\documentclass[coverpage,hyper,s5paper,11pt]{su_thesis}		
\usepackage{natbib_suthesismulticol}
\usepackage{pdfpages}
\usepackage{ifthen}
\usepackage{amssymb}
\usepackage[fleqn]{amsmath}
\usepackage{graphicx}
\usepackage{changepage}
\usepackage{calc}
\usepackage{cancel}
\usepackage{feynmf}
\usepackage{multicol}

\def\araa{ARA\&A}%
\def\apj{ApJ}%
\def\apjl{ApJ}%
\def\apss{Ap\&SS}%
\def\aap{A\&A}%
\def\jhep{JHEP}%
\def\jcap{JCAP}%
\def\mnras{MNRAS}%
\def\prd{Phys.\ Rev.\ D}%
\def\prl{Phys.\ Rev.\ Lett.}%

\newcounter{papercount}

\newlength{\paperboxheight}
\newlength{\paperboxwidth}
\newlength{\verticalpaperboxoffset}

\newcommand{\paperboxbackgroundcolor}{black}
\newcommand{\paperboxtextcolor}{Goldenrod}
\newcommand{\papersummary}[8]{

  \begin{description}
    \item[\bfseries\sffamily #8] {#1}. \emph{{#2}}, {#3}~\textbf{{#4}}, {#5} ({#6}) {#7}.
  \end{description}

}

\newcommand{\paper}[9]{

  \ifthenelse{\equal{#8}{}}{}{ 
    \ifodd \value{page}
    \else
      \newpage
      \null
      \newpage
    \fi
  }
  
  \addtocounter{papercount}{1}
  \vspace*{\verticalpaperboxoffset}

  \begin{adjustwidth*}{}{-6.2em}
    \begin{flushright}
      \huge\sffamily\bfseries \color{\paperboxtextcolor} \colorbox{\paperboxbackgroundcolor}
        {\parbox[c][\paperboxheight]{\paperboxwidth}{\hspace{1cm}Paper \Roman{papercount}}}
    \end{flushright}
  \end{adjustwidth*}

  \vfill
  
  \begin{flushleft}
    {#1}\\
    \emph{{#2}}\\
    {#3}~\textbf{{#4}}, {#5} ({#6}) {#7}.
  \end{flushleft}

  \phantomsection
  \addcontentsline{toc}{chapter}{Paper \Roman{papercount}: #2}
  \label{#9}

  \newpage  
  \ifthenelse{\equal{#8}{}}{}{ 
    \null
    \newpage
    \includepdf[pages=-]{#8}
  }

  \addtolength{\verticalpaperboxoffset}{1.2\paperboxheight} 
  
}

\title{Searches for Particle Dark Matter}
\subtitle{Dark stars, dark galaxies, dark halos and global supersymmetric fits}
\author{Pat Scott}
\date{May 2010}
\shortdate{2010}
\type{Doctoral Thesis in Theoretical Physics}

\division{Oskar Klein Centre for Cosmoparticle Physics\\\vskip 0.5em
  and\\\vskip 0.5em
  Cosmology, Particle Astrophysics and String Theory\\\vskip 0.5em}
\department{Department of Physics}
\address{SE-106 91 Stockholm}
\city{Stockholm}
\country{Sweden}
\cplogo{\includegraphics[height=4cm, trim = 0 0 250 510, clip=true]{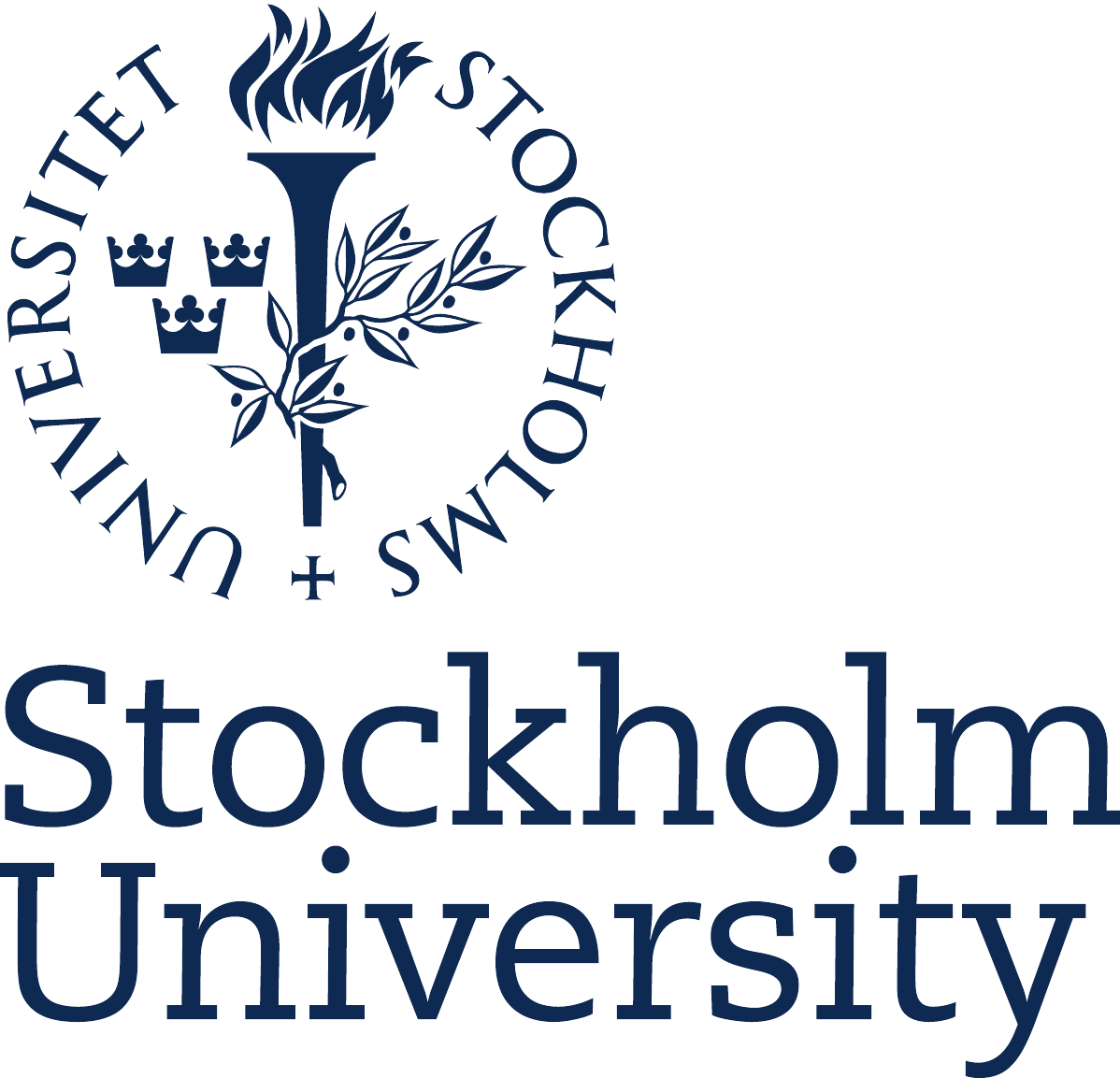}}

\publisher{\vspace{2mm}\\Printed by Universitetsservice US-AB, Stockholm, Sweden, 2010.\vspace{5mm}\\ Figures \ref{fig_rotcurve}, \ref{fig_bbn}, \protect{\ref{fig_lcdm}}, \protect{\ref{fig_relicdens}} and \protect{\ref{fig_sourcevols}} used with permission.\vspace{5mm}\\Typeset in pdf\LaTeX}
\copyrightline{pp.\ i--xiv, 1--84 \copyright\ Patrick Scott, 2010\vspace{2mm}}
\trita{xx}
\issn{aa}
\isrn{bb}
\isbn{978-91-7447-031-4 (pp.\ i--xiv, 1--84)}
\comment{\emph{Cover images, clockwise from top left:}  Artist's impression of a black hole in a globular cluster; G.~Bacon (STScI), NASA.  An `artistic simulation' of the collision which produced the Bullet Cluster (see Figure \ref{fig_bullet}); NASA/CXC/M.~Weiss.  A false-colour X-ray image of the Galactic Centre; D.~Wang (UMass) et al., CXC, NASA.  Computer-generated model image of the \emph{Fermi} Gamma Ray Space Telescope; General Dynamics C4 Systems/NASA.  \emph{Inset:} An artist's impression of a protostar, which dark stars at the Galactic Centre may resemble;  D.~Darling.}
\innerlogo{\includegraphics[height=3cm, trim = 0 0 250 510, clip=true]{sulogo}}
\extrainnerlogo{\includegraphics[height=2cm]{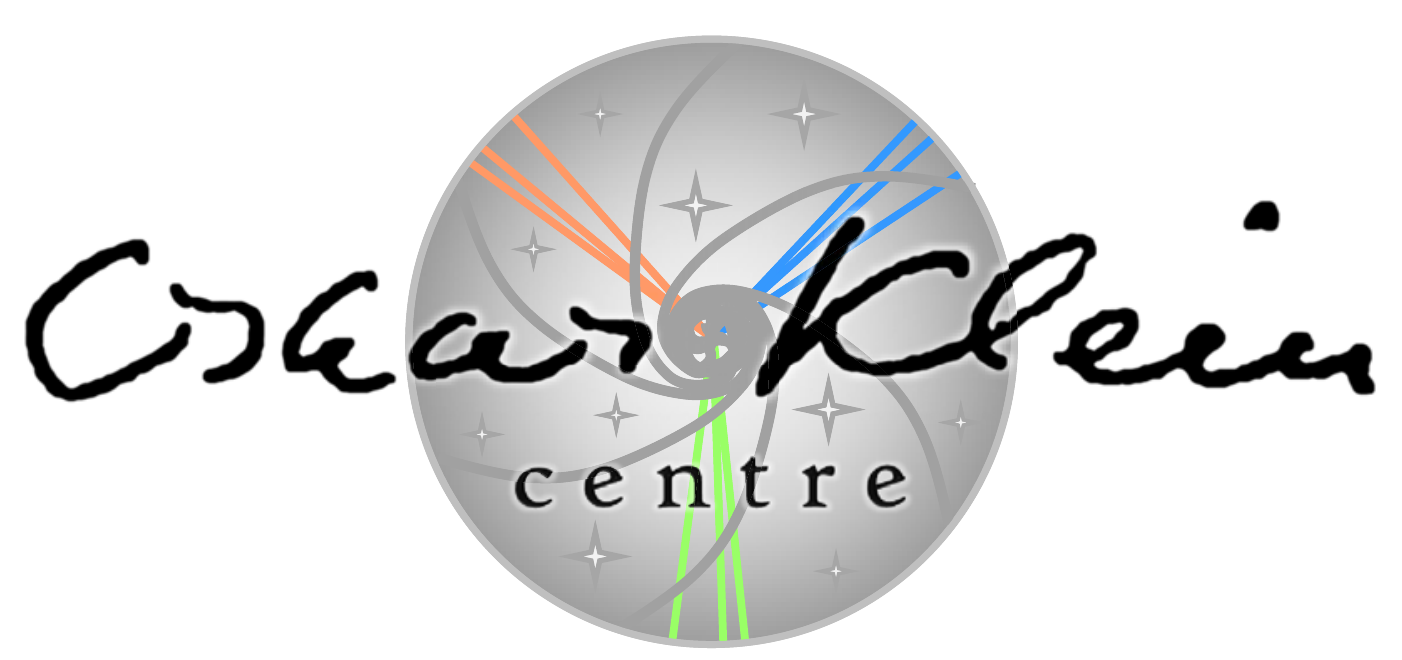}}

\bibpunct{(}{)}{;}{a}{}{,}
\newcommand\bibfont{\footnotesize}
\allowdisplaybreaks[1]

\newcommand{\Fairbairn}{Paper I}
\newcommand{\ScottII}{Paper II}
\newcommand{\SSiv}{Paper III}
\newcommand{\ScottIV}{Paper IV}
\newcommand{\Akrami}{Paper V}
\newcommand{\Zackrisson}{Paper VI}

\begin{document}

\begin{abstract}
The identity of dark matter is one of the key outstanding problems in both particle and astrophysics.  In this thesis, I describe a number of complementary searches for particle dark matter.  I discuss how the impact of dark matter on stars can constrain its interaction with nuclei, focussing on main sequence stars close to the Galactic Centre, and on the first stars as seen through the upcoming James Webb Space Telescope.  The mass and annihilation cross-section of dark matter particles can be probed with searches for gamma rays produced in astronomical targets.  Dwarf galaxies and ultracompact, primordially-produced dark matter minihalos turn out to be especially promising in this respect.  I illustrate how the results of these searches can be combined with constraints from accelerators and cosmology to produce a single global fit to all available data.  Global fits in supersymmetry turn out to be quite technically demanding, even with the simplest predictive models and the addition of complementary data from a bevy of astronomical and terrestrial experiments; I show how genetic algorithms can help in overcoming these challenges.
\\\noindent \strut  
\end{abstract}

\chapter*{List of Papers}
\pagestyle{plain}
\addcontentsline{toc}{chapter}{List of Papers}
\section*{Papers included in this thesis}
\papersummary{Malcolm Fairbairn, Pat Scott \& Joakim Edsj\"o}{The zero age main sequence of WIMP burners}{\prd}{77}{047301}{2008}{\href{http://arxiv.org/arXiv:0710.3396}{arXiv:0710.3396}}{\protect{\hyperref[papI]{Paper I}}}
\papersummary{Pat Scott, Malcolm Fairbairn \& Joakim Edsj\"o}{Dark stars at the Galactic Centre -- the main sequence}{\mnras}{394}{82--104}{2009}{\href{http://arxiv.org/arXiv:0809.1871}{arXiv:0809.1871}}{\protect{\hyperref[papII]{Paper II}}}
\papersummary{Pat Scott \& Sofia Sivertsson}{Gamma-rays from ultracompact primordial dark matter minihalos}{\prl}{103}{211301}{2009}{\href{http://arxiv.org/arXiv:0908.4082}{arXiv:0908.4082}}{\protect{\hyperref[papIII]{Paper III}}}
\papersummary{Pat Scott, Jan Conrad, Joakim Edsj\"o, Lars Bergstr\"om, Christian Farnier \& Yashar Akrami}{Direct Constraints on Minimal Supersymmetry from Fermi-LAT Observations of the Dwarf Galaxy Segue 1}{\jcap}{01}{031}{2010}{\href{http://arxiv.org/arXiv:0909.3300}{arXiv:0909.3300}}{\protect{\hyperref[papIV]{Paper IV}}}
\papersummary{Yashar Akrami, Pat Scott, Joakim Edsj\"o, Jan Conrad \& Lars Bergstr\"om}{A Profile Likelihood Analysis of the Constrained MSSM with Genetic Algorithms}{\jhep}{04}{057}{2010}{\href{http://arxiv.org/arXiv:0910.3950}{arXiv:0910.3950}}{\protect{\hyperref[papV]{Paper V}}}
\papersummary{Erik Zackrisson, Pat Scott, Claes-Erik Rydberg, Fabio Iocco, Bengt Edvardsson, G\"oran \"Ostlin, Sofia Sivertsson, Adi Zitrin, Tom Broadhurst \& Paolo Gondolo}{Finding high-redshift dark stars with the James Webb Space Telescope}{\apj}{717}{257}{2010}{\linebreak[4]\href{http://arxiv.org/arXiv:1002.3368}{arXiv:1002.3368}}{\protect{\hyperref[papVI]{Paper VI}}}

\section*{Papers and proceedings not included in this thesis}
\papersummary{Martin Asplund, Nicolas Grevesse, A.~Jacques Sauval \& Pat Scott}{The chemical composition of the Sun}{\araa}{47}{481--522}{2009}{\href{http://arxiv.org/arXiv:0909.0948}{arXiv:0909.0948}}{Paper VII}
\papersummary{Pat Scott, Martin Asplund, Nicolas Grevesse \& A.~Jacques Sauval}{On the solar nickel and oxygen abundances}{\apjl}{691}{L119--L122}{2009}{\href{http://arxiv.org/arXiv:0811.0815}{arXiv:0811.0815}}{Paper VIII}
\papersummary{Pat Scott, Martin Asplund, Nicolas Grevesse \& A.~Jacques Sauval}{Line formation in solar granulation. VII. CO lines and the solar C and O isotopic abundances}{\aap}{456}{675--688}{2006}{\href{http://arxiv.org/arXiv:astro-ph/0605116}{arXiv:astro-ph/0605116}}{Paper IX}
\papersummary{Nicolas Grevesse, Martin Asplund, A.~Jacques Sauval \& Pat Scott}{The chemical composition of the Sun}{\apss~special issue \emph{Proceedings of the 5th HELAS Asteroseismology Workshop: Synergies between solar and stellar modelling}}{328}{179}{2010}{\!\!}{Proceeding X}
\papersummary{Pat Scott, Joakim Edsj\"o \& Malcolm Fairbairn}{The \textsf{DarkStars} code: a publicly available dark stellar evolution package}{Proceedings of \emph{Dark Matter in Astroparticle and Particle Physics: Dark 2009} (eds.\ H.~K.~Klapdor-Kleingrothaus and I.~V.~Krivosheina)}{\!\!}{320--327, World Scientific Publishing, Singapore}{2010}{\linebreak[4]\href{http://arxiv.org/arXiv:0904.2395}{arXiv:0904.2395}}{Proceeding XI}
\papersummary{Pat Scott, Malcolm Fairbairn \& Joakim Edsj\"o}{Impacts of WIMP dark matter upon stellar evolution: main-sequence stars}{Proceedings of \emph{Identification of Dark Matter 08}}{PoS(idm2008)}{073}{2008}{\href{http://arxiv.org/arXiv:0810.5560}{arXiv:0810.5560}}{Proceeding XII}
\papersummary{Pat Scott, Joakim Edsj\"o \& Malcolm Fairbairn}{Low mass stellar evolution with WIMP capture and annihilation}{Proceedings of \emph{Dark Matter in Astroparticle and Particle Physics: Dark 2007} (eds.\ H.~K.~Klapdor-Kleingrothaus and G.~F.~Lewis)}{\!\!}{387--392, World Scientific Publishing, Singapore}{2008}{\href{http://arxiv.org/arXiv:0711.0991}{arXiv:0711.0991}}{Proceeding XIII}

\tableofcontents

\begin{acknowledgments}
My first and foremost thanks go to Joakim Edsj\"o for his encouraging and enthusiastic supervision, and for his energy and creativity; these have made working together a genuine pleasure.  Thanks also to my secondary supervisors Lars Bergstr\"om and Jan Conrad, for much sage advice on scientific matters and the practicalities of working as a physicist.  Thanks to all three for the confidence they have shown in me, and the scientific and organisational independence they have granted me here in Stockholm.

I have also had the benefit of many other highly-skilled and motivated collaborators in the work of this thesis; without the particular efforts of Yashar Akrami, Malcolm Fairbairn, Christian Farnier, Sofia Sivertsson and Erik Zackrisson, many of these projects would not have come to fruition.  I have also benefited greatly from helpful discussions on various topics with Marcus Berg, Ross Church, Melvyn Davies, Steen Hansen, Fabio Iocco, Paolo Gondolo, Ariel Goobar, Antje Putze, Are Raklev, Riccardo Rando, Joachim Ripken, Chris Savage, Aldo Serenelli, Roberto Trotta and the Dark Matter \& New Physics Group within the \emph{Fermi}-LAT Collaboration.  Although our work together is not part of this thesis, Martin Asplund, Nicolas Grevesse and Jacques Sauval have also been steady sources of inspiration and wisdom over the last six years.

My thanks also to the other students and postdocs of the OKC and its forerunner, HEAC, for the pleasure of their company, lunch table, Swedish lessons, couch or beer glass over the last three and a half years, and to Yashar, for managing to put up so stoically with sharing an office with me all this time.

Thanks to my parents and sister Adele for all their encouragement over the years, and to my wife Susan for her companionship and unwavering support, from the moment I suggested moving overseas to do my PhD to its ultimate completion.

I gratefully acknowledge travel funding from the IAU Commission 46 Exchange of Astronomers Program, the European Network of Theoretical Astroparticle Physics ILIAS/N6 under contract number RII3-CT-2004-506222, the Helge Axelsson Johnsons Foundation, the C.~F.\ Liljevalchs Scholarship Fund and the G \& E Kobbs Scholarship Fund, as well as general financial support from the Swedish Research Council (Vetenskapsr{\aa}det) through grants to HEAC and the OKC.
\end{acknowledgments}

\begin{preface}
This thesis deals with strategies for detecting and identifying dark matter.  Identification means somewhat more than simply establishing that dark matter is a particle with a certain mass, and belongs to one or another proposed phenomenological candidate class (WIMPs, axions, etc).  Although this alone would be a tremendous achievement, it would be just the first step.  In my opinion, it is the second that is actually the more interesting (and difficult) of the two.

Rather, true identification means establishing which particle (or particles) are responsible for dark matter, and characterising the actual field theory to which they belong.  The papers included in this thesis all work towards achieving the first of these goals, but some (\hyperref[papIV]{\ScottIV} and \hyperref[papV]{\Akrami} especially) also make contributions towards the second.

\subsubsection*{Thesis plan}

This thesis is divided into two parts: Part~\ref{intro} gives an introduction to the field and my work in particular, and Part~\ref{papers} provides the included papers.  Chapter~\ref{needfordm} of Part~\ref{intro} describes the observational evidence for dark matter, Chapter~\ref{models} details the various models proposed to explain it, and Chapter~\ref{dmsearches} discusses the strategies which may be used to search for it.  Chapter~\ref{nuisances} gives a brief exposition of some of the most important uncertainties entering dark matter searches, and Chapter~\ref{summary} summarises the results presented in the papers of Part~\ref{papers}.  

The papers included in this thesis span quite a broad range of topics within the theme of dark matter searches.  Many of them include introductory sections with specific background material on the physics, existing literature and computational techniques relevant to their respective topics.  Each paper also includes extensive discussion and interpretation of the results it presents; as one should expect, they are quite self-contained and largely self-explanatory.  In most cases, it was I who was responsible for actually writing those sections (or at least extensively revising them).  Instead of simply repeating material already included in the papers themselves, I have thus intentionally written the first four Chapters of Part~\ref{intro} with the character of a textbook-level introductory review, so as to explain and justify the broader scientific context in which the six papers have been written.  In this undertaking I have not aimed for comprehensive coverage of the literature.  Full historical references and further details on the topics covered in Chapters \ref{needfordm}--\ref{nuisances} can be found in extensive review articles by \citet*{Jungman96}, \citet{Bergstrom00} and \citet*{Bertone05}, as well as the recent reference volume edited by \citet{BertoneBook}.

Unlike in some theses, the included papers of Part~\ref{papers} should not be considered appendices.  They are in fact the main body of this thesis, equivalent to Results, Discussion and Conclusion chapters in a traditional monograph.  Their main results, figures and conclusions are summarised for the reader's convenience in Chapter~\ref{summary}, but the actual papers should be recognised as the definitive exposition of the work contained in this thesis.  In short, the ethos I have adopted is that Part~\ref{intro} should essentially contain information not provided already by Part~\ref{papers}; there should be as little regurgitation as possible of content which is already contained in the six included manuscripts.  Accordingly, Chapter~\ref{summary} also places the results of Part~\ref{papers} back into the broader context established by Chapters \ref{needfordm}--\ref{nuisances}, and presents an outlook for future work.

\subsubsection*{Contribution to papers}

My contribution to the included papers is as follows.  For \hyperref[papI]{\Fairbairn}, I wrote and ran the stellar evolution code (as opposed to the static stellar structure code), and revised the manuscript.  I discussed methods, results and conclusions together with the other authors.  For \hyperref[papII]{\ScottII}, I designed most of the study, wrote the code and ran it using orbital and velocity data provided by Malcolm, and the results of neutrino simulations performed by Joakim.  We discussed methods, results and conclusions together.  I wrote the paper.  I had the original idea for \hyperref[papIII]{\SSiv}, designed the bulk of the study, wrote most of the paper and performed all calculations except for adiabatic contractions, which were done by Sofia.  We interpreted results together.    I designed almost all aspects of the work in \hyperref[papIV]{\ScottIV}, with significant input on statistical matters from Jan.  I wrote the software and ran it using data reduced by Christian, and wrote the paper.  I interpreted results together with the other authors.  For \hyperref[papV]{\Akrami}, I made large contributions to the design and implementation of the study.  I wrote sections of the paper and revised the remainder.  I also organised the constituent physical codes into a consistent and functional starting point, from which Yashar added the genetic algorithm code and ran it.  We discussed and interpreted results together with the other authors.  For \hyperref[papVI]{\Zackrisson}, I helped design the study, wrote sections of the paper and computed \textsc{tlusty} model atmospheres.  I discussed methods and results primarily with Erik, Fabio and Sofia.

\hspace{0cm}\\
\noindent Pat Scott\\
Stockholm, March 2010

\end{preface}

\mainmatter

\part{Introduction}
\label{intro}

\defcitealias{Fairbairn08}{Paper I}
\defcitealias{Scott09}{Paper II}
\defcitealias{SS09}{Paper III}
\defcitealias{Scott09c}{Paper IV}
\defcitealias{Akrami10}{Paper V}
\defcitealias{Zackrisson10}{Paper VI}
\defcitealias{AGSS}{Paper VII}
\defcitealias{Scott09Ni}{Paper VIII}
\defcitealias{ScottVII}{Paper IX}
\defcitealias{Scott09b}{Proceeding XI}
\defcitealias{Scott08a}{Proceeding XIII}

\setlength{\unitlength}{1mm}
\begin{fmffile}{feyn}

\chapter{The need for dark matter}
\label{needfordm}

Dark matter has become such an established paradigm in modern astro- and particle physics that its existence is generally accepted with little explanation.  Such blithe and widespread adoption of a cosmology in which an unknown matter plays the pivotal role has led some to become a little sceptical about its existence.  It is worth reminding ourselves, from time to time, of the strong and compelling evidence on which this paradigm actually stands.  As it so happens, we will also pick up a good deal of information along the way about what properties dark matter must have.

\begin{figure}[tbp]
\centering
\includegraphics[width=0.7\textwidth]{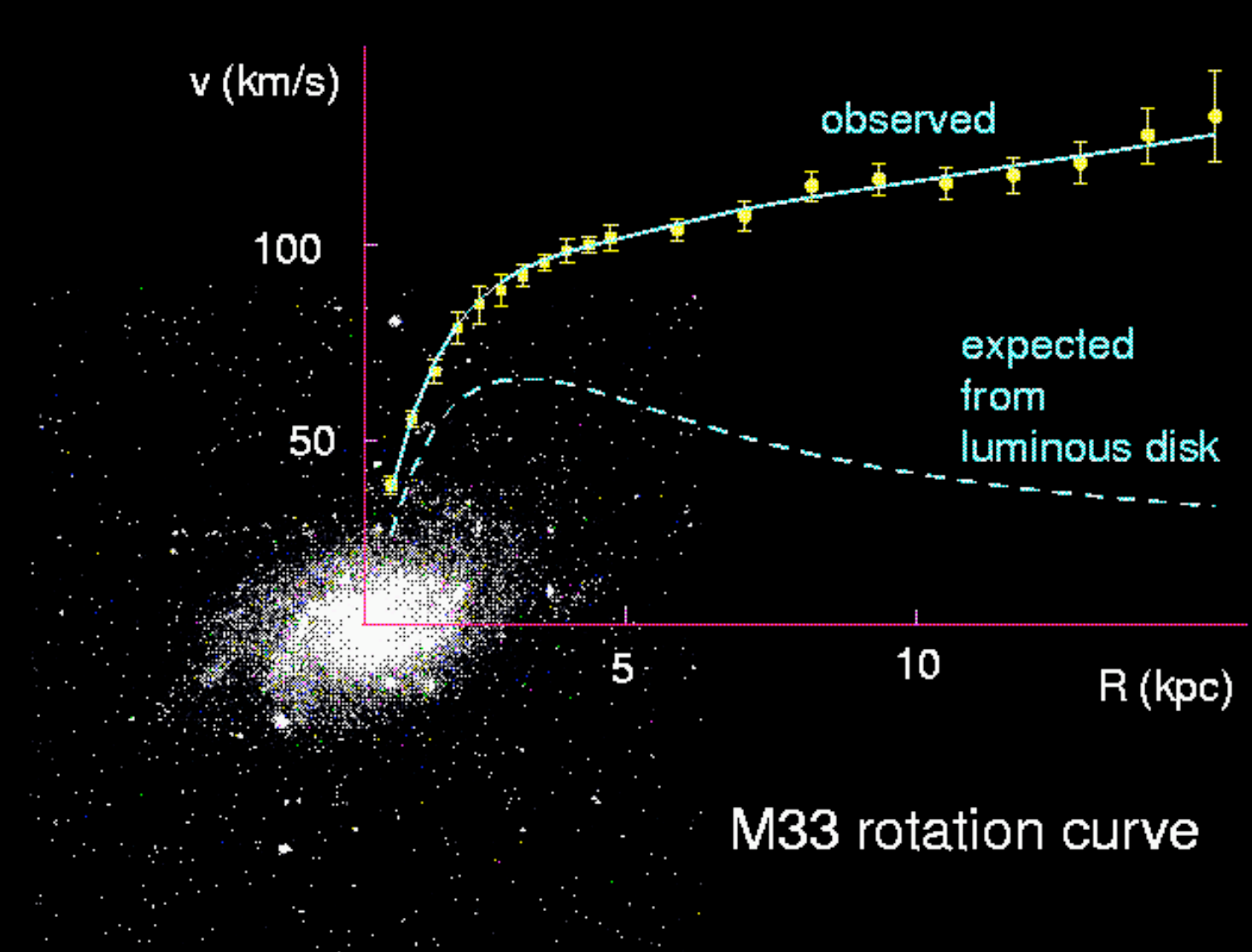}
\caption{The observed rotation curve of the dwarf spiral galaxy M33, and the curve one would predict purely on the basis of the luminous disc.  The fact that the curve is flatter than predicted suggests the presence of a halo of dark matter, extending to large galactic radii.  From \protect\citet{Bergstrom00}.}
\label{fig_rotcurve}
\end{figure}

\section{Kinematics}
\label{kinematics}

The earliest identification of dark matter came from the velocity dispersions of galaxies within clusters. \citet{Zwicky33} noticed that the outer members of the Coma cluster were moving far too quickly to be merely tracing the gravitational potential of the visible cluster mass.  The only way the observed velocities of the cluster members could be reconciled with the virial theorem was to postulate that the cluster also contained another large, but unseen, mass component: dark matter.  Today, an equivalent technique is used to weigh clusters in large-scale X-ray surveys \citep[e.g.][]{Chandra,XMM}.  In this case, because the hot, X-ray-emitting intracluster medium is entirely virialised, its temperature can be used to infer the total potential energy of the system, and therefore the cluster mass.  Such surveys consistently confirm \citeauthor{Zwicky33}'s finding that clusters contain far more dark than luminous matter.

It was not until similar effects were seen at the galactic level by Rubin and collaborators \citep{Rubin70, Rubin78, Rubin85} that the idea of dark matter really began to gain traction.  Like the galaxies in the outer region of the Coma cluster, stars in the outer reaches of spiral galaxies are seen to rotate far more quickly than would be expected if each galaxy consisted only of the matter visible in stars and gas.  An example of such anomalous circular velocities is shown in Fig.~\ref{fig_rotcurve} for a dwarf spiral galaxy, M33.  Here, the rotation curve expected purely on the basis of the stellar mass drops off markedly at long distances from the centre of the galaxy, whereas the observed curve is far flatter.  This can be explained by the presence of an additional halo of dark matter, far more extended than the observed stellar disk.

Similarly, the observed proportionality between the luminosity $L$ of spiral galaxies and the maximum circular velocity $V_\mathrm{max}$ of their members,
\begin{equation}
L \varpropto V_\mathrm{max}^\beta\ ,\hspace{1cm}\beta = 3\sim4,
\end{equation}
known as the Tully-Fisher relation \citep{TullyFisher}, can be explained by dark matter and the virial theorem \citep{Aaronson79}.  The virial theorem leads to a direct correspondence between the total galactic mass $M$ and $V_\mathrm{max}$.  The self-similarity of hierarchically-formed dark matter halos \citep*[e.g.][]{NFW} and their resident baryonic disks causes this to be a power-law dependency.  Assuming that for a given class of spiral galaxies (early-type, late-type, etc.) the mass-to-light ratio ($M/L$) is approximately constant, the relationship translates directly into a power-law dependency of $L$ upon $V_\mathrm{max}$.  A similar argument holds for the analogous relation for elliptical galaxies, the Faber-Jackson relation \citep{FaberJackson}.

\begin{figure}[t]
\centering
\includegraphics[height=0.4\textwidth]{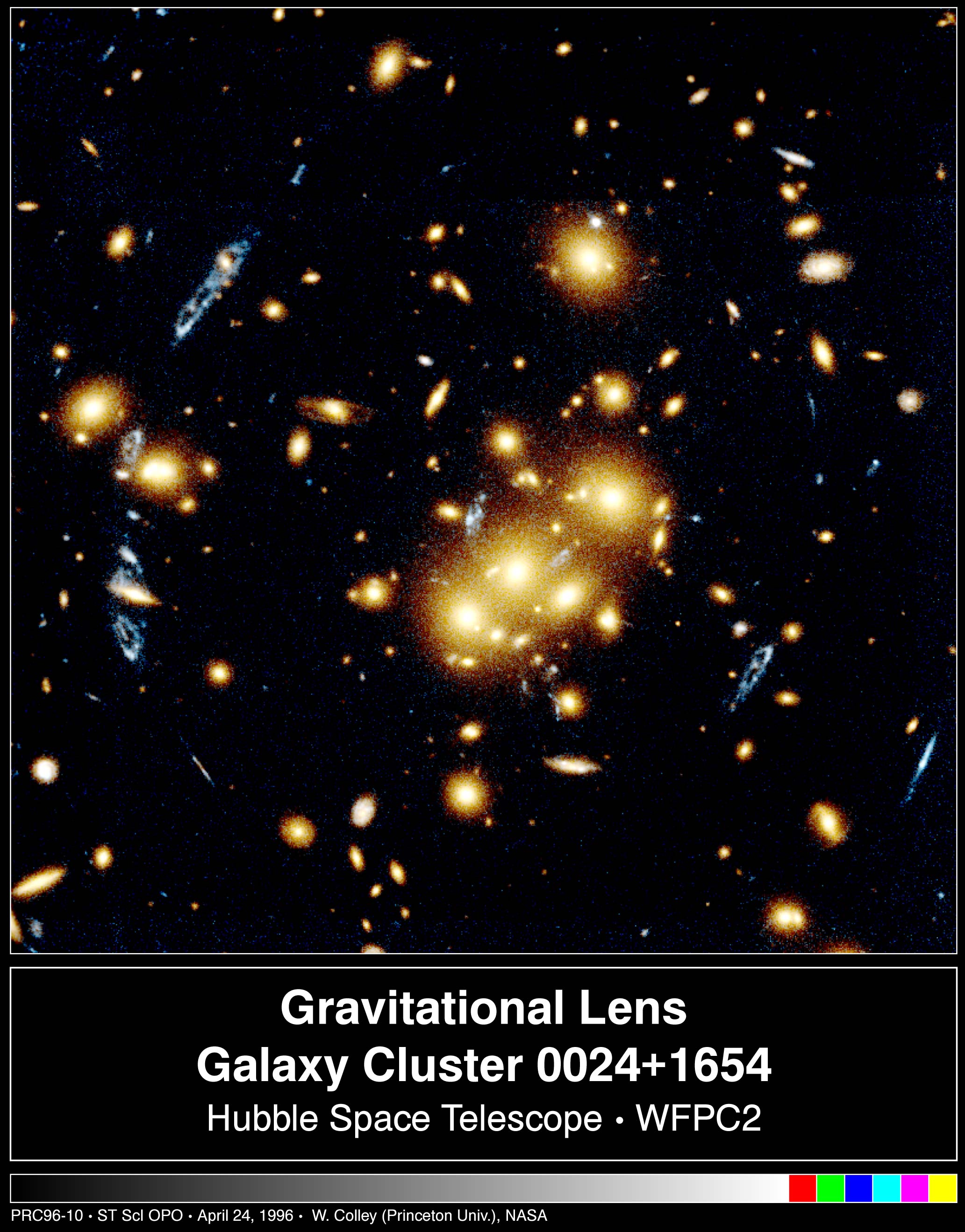}
\includegraphics[height=0.4\textwidth]{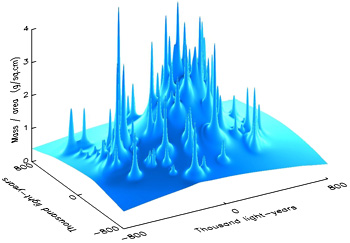}
\caption{An image of a gravitationally lensed cluster (left), along with the corresponding mass map of the foreground cluster.  In the image on the left, the foreground cluster can be seen in yellow, with the lensed image of the background cluster visible in blue.  The reconstructed mass map shows a large, broad peak around the centre of the cluster which is not visible in the optical image, indicating the presence a massive dark halo.  Credit: Greg Kochanski, Ian Dell'Antonio, and Tony Tyson (Bell Labs, Lucent Technologies).  See also \protect\citet{Tyson98}.}
\label{fig_gravlensing}
\end{figure}

\begin{figure}[t]
\centering
\includegraphics[width=0.7\textwidth]{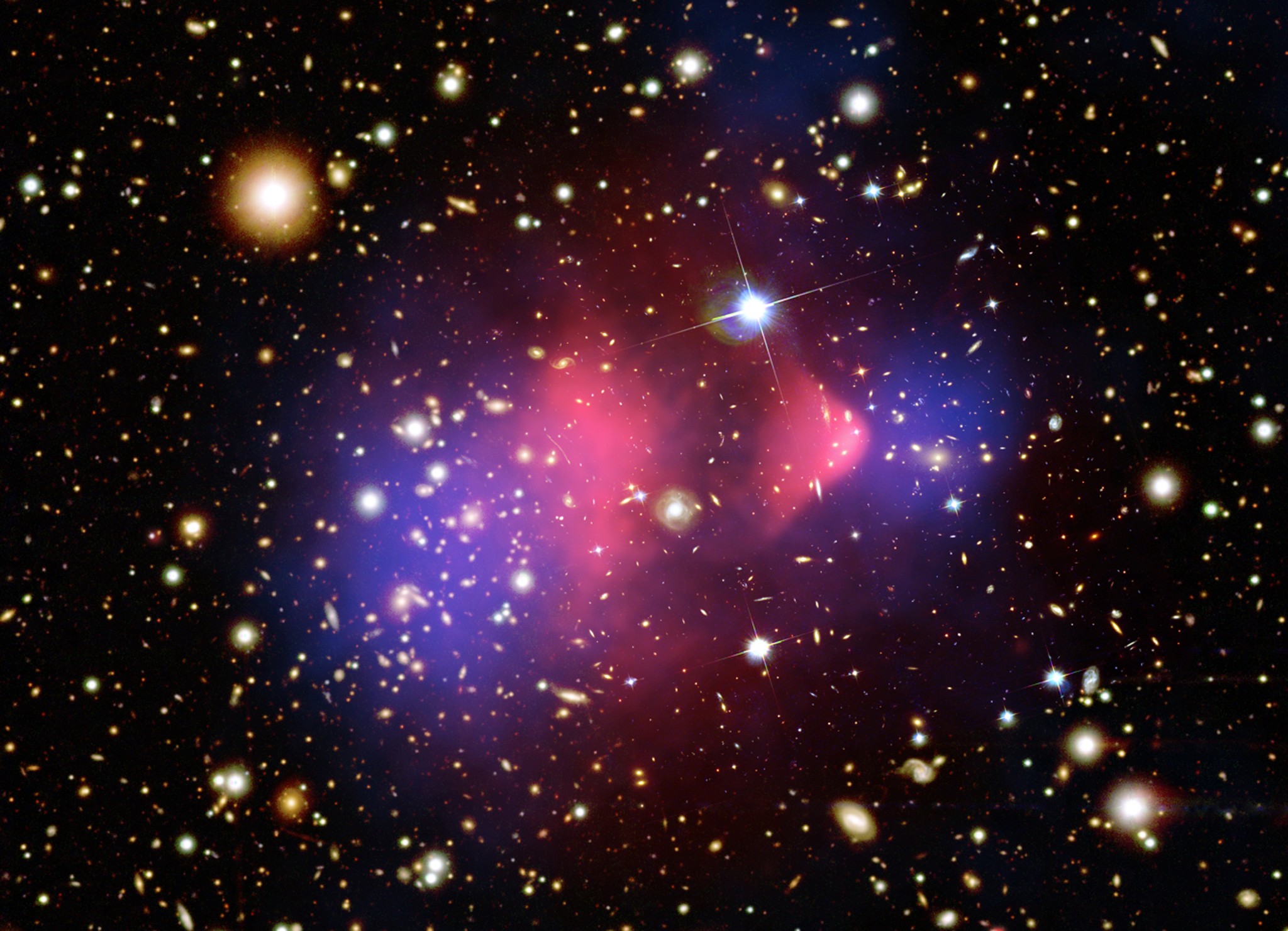}
\caption{Composite image of the colliding `Bullet Cluster', obtained with gravitational lensing.  The lensing mass map is shown in blue, and the X-ray observations tracing the gas component are shown in pink.  The offset of the mass and gas maps indicates that the mass of each cluster does not consist predominantly of gas, nor does it track the gas.  Whilst the gas has collided with and exerted friction on the other cloud, leading to the characteristic `punch-through' shape of the cluster to the right, the bulk of the mass has simply passed through the collision unimpeded.  This proves that the majority of the matter in each cluster is collisionless dark matter.  X-ray data: NASA/CXC/CfA/\citet{Markevitch06}. Lensing Map: NASA/STScI; ESO WFI; Magellan/U.Arizona/\citet{bullet}.  Optical data: NASA/STScI; Magellan/U.Arizona/\citet{bullet}.}
\label{fig_bullet}
\end{figure}

\section{Gravitational lensing}
\label{gravlensing}

According to General Relativity, the presence of any mass causes the space in its vicinity to curve.  Curvature of space implies curvature of geodesics, which results in bending of light rays around massive bodies.  In this way, light from background objects can be `lensed' by massive foreground objects, as the light rays passing through the foreground objects' gravitational field are bent towards a distant observer.  At the simplest level, this causes a background object to appear brighter than it otherwise would.  This phenomenon provides a clean and unambiguous means by which to probe the mass distribution in a foreground object.

Distant clusters lensed by closer ones show evidence of extensive gravitational lensing, far more than can be explained by the observed distribution of luminous matter in the foreground cluster \citep[e.g.][]{Tyson98, Massey07}.  Fig.~\ref{fig_gravlensing} gives an example of such a cluster, exhibiting a large, smooth mass component in the lensing map to the right, which is not accounted for by the distribution of luminous matter shown in the optical image to the left.  

Probably the most famous example of a dark matter lens is the so-called Bullet Cluster \citep{bullet}, shown in Fig.~\ref{fig_bullet}.  This object is in fact two clusters that have recently collided.  In this case, not only does the lensing map (in blue) exhibit large amounts of dark matter not evident in the X-ray gas map (in pink), but the distribution of the two components also provides crucial insight into the properties of the dark matter.  That is, the dark matter halos have passed straight through both the gas clouds and each other, and appear essentially undisturbed after the collision.  The gas clouds, on the other hand, have clearly exerted friction on each other as they have collided, leading to the characteristic ballistic shape of the rightmost cloud.  This shows that dark matter does not necessarily have to track luminous matter in any way, and that it does not interact strongly with either gas or itself; this means that it is effectively collisionless.

\section{The large scale structure of the Universe}
\label{lss}

On large scales, the Universe shows a wealth of structure: galaxies are gathered into clusters, clusters are part of superclusters, and superclusters are arranged into large-scale sheets, filaments and voids.  This cosmic scaffolding has been revealed by large-scale surveys such as 2dFGRS \citep[the 2-degree Field Galaxy Redshift Survey;][]{Colless01} and SDSS \citep[the Sloan Digital Sky Survey;][]{Tegmark04}.  Presumably, the pattern of galactic superstructure reflects the history of gravitational clustering of matter since the Big Bang.  If dark matter were present during structure formation, it should have influenced the pattern of large-scale structure we see today.

Large-scale cosmological `N-body' simulations \citep[e.g.][]{NFW, Millenium, vialactea, Aquarius, vialacteaII} demonstrate that the observed large-scale structure of luminous matter could only have been formed in the presence of a substantial amount of dark matter.  Furthermore, the bulk of dark matter must be both cold and non-dissipative for the correct structures to be produced.  `Cold' in this context means that it moves very non-relativistically, and so has a short free-streaming length (less than the size of a gas cloud undergoing gravitational collapse, for example).  Being cold means that the dark matter can gather gravitationally on small scales and so seed galaxy formation, but being non-dissipative prevents it from cooling and collapsing with the luminous matter, which would produce larger and more abundant galactic disks than are observed.  Hot (highly relativistic) and warm (borderline relativistic) dark matter could still make up a fraction of the total dark matter, though just how large a fraction depends critically upon how warm that fraction is.

Large-scale galaxy surveys not only provide information as to the amount and pattern of structure present in the Universe, but also a handle on the total mass contained within it \citep[e.g.][]{Tegmark04,Cole05,Percival10}.  To a first approximation, the characteristic size of baryonic density perturbations which can survive until matter-radiation equality (and therefore collapse to form galaxies and clusters) is set by the mean density of baryons in the Universe \citep{Silk68}; smaller perturbations are destroyed by radiative and neutrino damping, larger ones fragment.  In the presence of an additional matter component which gravitates but does not couple radiatively to the baryons, the power spectrum of perturbations which ultimately survive to form galaxies is modified by the enhanced gravitational clustering.  The resulting large-scale galaxy power spectrum is thus strongly dependent upon the total matter density of the Universe, and weakly dependent upon the fraction of matter contained in baryons.  Recent surveys indicate a total matter (i.e.\ dark plus luminous matter) density of $\Omega_\mathrm{m} \equiv \rho_\mathrm{m} / \rho_\mathrm{c} \approx 0.29$ \citep{Percival10}, where $\rho_\mathrm{c}$ is the critical density required to close the Universe.

\section{Big bang nucleosynthesis}
\label{bbn}

\begin{figure}[t]
\centering
\includegraphics[width=0.7\textwidth]{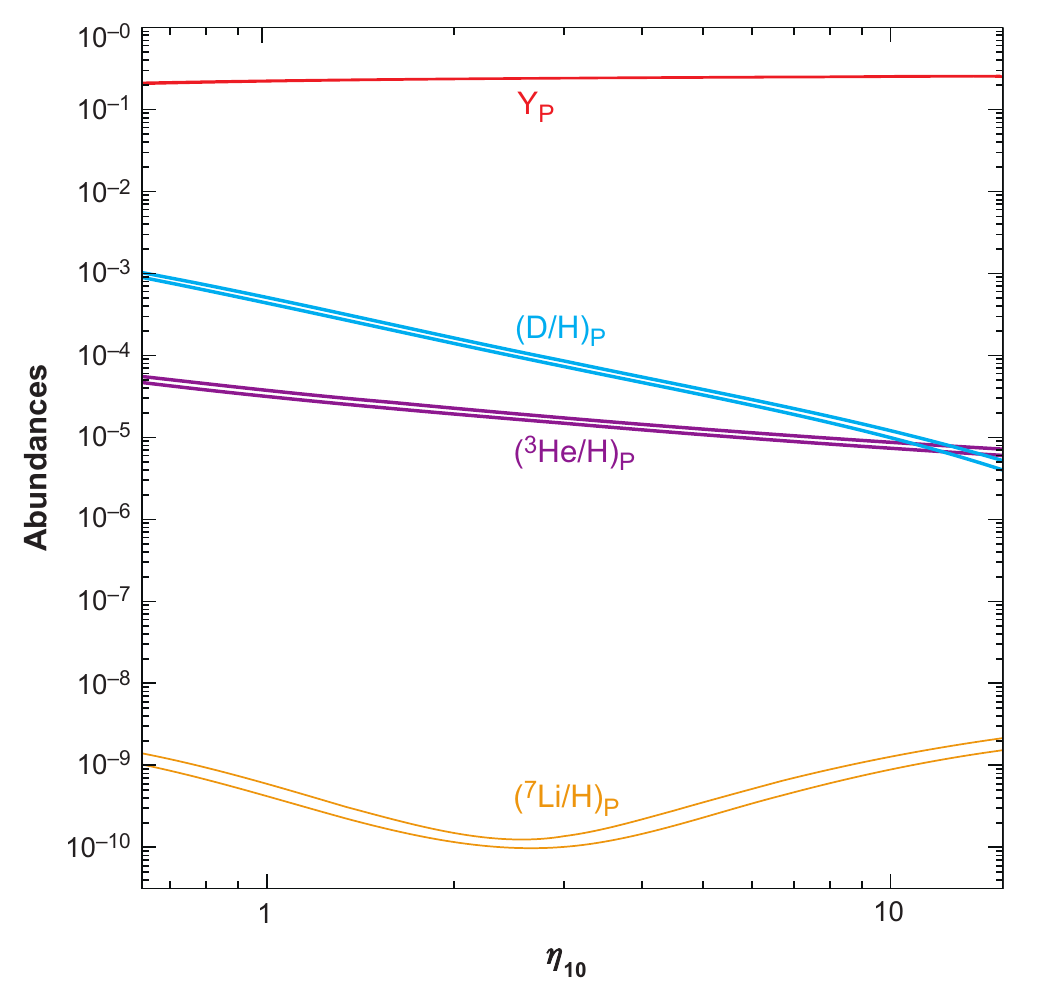}
\caption{Abundances of the light isotopes produced in Big Bang nucleosynthesis (BBN), as a function of the baryon-to-photon ratio $\eta_{10}\equiv 10^{10}\eta$, which directly corresponds to the baryon fraction of the universe ($\eta\varpropto\Omega_\mathrm{b}h^2$).  By comparing such predictions with observations of very old astronomical systems, one can use BBN to derive $\Omega_\mathrm{b}$.  Abundances are given as mass fractions.  The bands between the two curves for each isotope indicate the degree of uncertainty arising from nuclear reaction rates and other theoretical sources.  $\mathrm{Y}_\mathrm{P}$ refers to primordial $^4$He.  From \protect\citet{Steigman07}.}
\label{fig_bbn}
\end{figure}

A critical prediction of the hot Big Bang cosmology is that protons and neutrons were fused in the primordial fireball to create the light elements, as it cooled to temperatures of the order of an MeV.  Modulo the effect of neutrinos, the resulting elemental abundances depend only on the nuclear reaction rates and the baryon-to-photon ratio ($\eta$) at the time.  Laboratory and theoretically-calculated nuclear reaction rates \citep[e.g.][]{NACRE} can therefore be used to derive the primordial abundances of the elements as a function of $\eta$ \citep[e.g][]{Coc04,Steigman07,IoccoBBN}, as shown in Fig.~\ref{fig_bbn}.

The parameter $\eta$ is completely equivalent to $\Omega_\mathrm{b}h^2$ (up to a constant of proportionality), where $\Omega_\mathrm{b}$ is the baryon density of the Universe and $h\equiv H_0/100$\,km\,s$^{-1}$\,Mpc$^{-1}$ is the dimensionless Hubble parameter.  $H_0\equiv v/d$ is the Hubble constant, describing the speed $v$ at which galaxies at a distance $d$ appear to be receding due to the expansion of the Universe.  With observations of the true primordial abundances of the elements and an independent measurement of $h$, Big Bang Nucleosynthesis (BBN) therefore allows us to measure the primordial value of $\Omega_\mathrm{b}$.  Measurements of the Hubble parameter are abundant; the most widely-used comes from the Hubble Key Project \citep[$h=0.72\pm0.08$;][]{Freedman01}.  The primordial abundances of the light isotopes are somewhat more difficult to come by, requiring direct observations of extremely unevolved systems.  

Because deuterium (D) is easily destroyed in stellar interiors and not produced in substantial amounts by other processes, its abundance shows a roughly steady decrease with time.  The D/H ratio is often obtained by observing absorption lines in very old hydrogen clouds backlit by high-redshift quasars, and extrapolating backwards to nucleosynthesis.  $^3$He, on the other hand, is created and destroyed in stellar interiors, the interstellar medium and the Earth's atmosphere, so its primordial abundance is very difficult to estimate reliably.  The best upper limits to its primordial abundance come from emission line observations of the least evolved Galactic H\,\textsc{ii} regions, or the protosolar nebula.  Similarly, $^4$He is produced ubiquitously in hydrogen fusion via the $pp$-chain in stars, so its abundance is substantially polluted relative to BBN; its higher abundance than $^3$He at least allows determinations using emission from \emph{extra}galactic H\,\textsc{ii} regions.

$^7$Li is observed in absorption in the atmospheres of extremely metal-poor halo stars, where its abundance may eventually plateau with decreasing metallicity \citep{SpiteSpite}.  $^6$Li has also been seen in such stars \citep{AsplundLi6}, though its presence can only be inferred from a very careful treatment of the isotopic broadening of spectral lines, along with the related three-dimensional effects of convection and departures from local thermodynamic equilibrium (LTE) on the line shapes.  The heavier elements are rarely considered because their tiny primordial abundances make any reliable observational determination virtually impossible.

The abundances of D and the He isotopes point very consistently to a baryon density of $\Omega_\mathrm{b}\approx0.04$.  The corresponding prediction for the lithium isotopic abundances do exhibit some tension with observation: $^7$Li is seen to be under-abundant relative to the BBN prediction, and $^6$Li is not expected to have been produced in BBN at all.  Both lithium isotopes are intrinsically unstable however, and subject to assorted processing in different stellar layers and the interstellar medium.  The stellar determinations are also extremely difficult, depending crucially upon the input atomic data required for isotopic shifts and non-LTE calculations, and the fine points of detailed convective line shapes in stars known to be heavily effected by 3D corrections to their inferred abundances.  Although it remains to be seen if the discrepancies will indeed be borne out by future observations, there is speculation that they might be explained by non-standard BBN caused by particle physics beyond the standard model (BSM; beyond the SM), including some specific dark matter models \citep[e.g.][]{Jedamzik09}.

The independent measurements of $\Omega_\mathrm{b}$ from BBN and $\Omega_\mathrm{m}$ from large-scale structure together provide incontrovertible evidence for the existence of dark matter.  Since all luminous matter essentially consists of baryons, $\Omega_\mathrm{m}\approx0.29$ and $\Omega_\mathrm{m}\approx0.04$ together imply that the remaining $\Omega_\mathrm{leftover}\approx0.25$ must be dark matter.  Furthermore, we have another invaluable piece of information about its nature: we see immediately that dark matter must be predominantly non-baryonic.  Being non-baryonic allows the possibility that the dark matter is not capable of interacting with photons electromagnetically (something baryons clearly \emph{are} capable of), which sits very well with the fact that it does indeed appear dark.  This is also consistent with dark matter being dissipationless; were it able to absorb and re-emit photons, it would obscure stars within distant halos, as well as radiate away angular momentum and collapse with baryons to form stellar and galactic disks.  

It is worth mentioning that some amount of \emph{baryonic} dark matter also remains unaccounted for today in the figure $\Omega_\mathrm{b}\approx0.04$, in that it has not been directly observed in surveys of gas or galaxies.  Nevertheless, we know this to be far less than what exists in non-baryonic dark matter.  For the purposes of this thesis, `dark matter' is taken as shorthand for the dominant, non-baryonic component.

\begin{figure}[tbp]
\centering
\begin{minipage}[c]{0.5\textwidth}
  \textbf{a}\\\includegraphics[width=\textwidth]{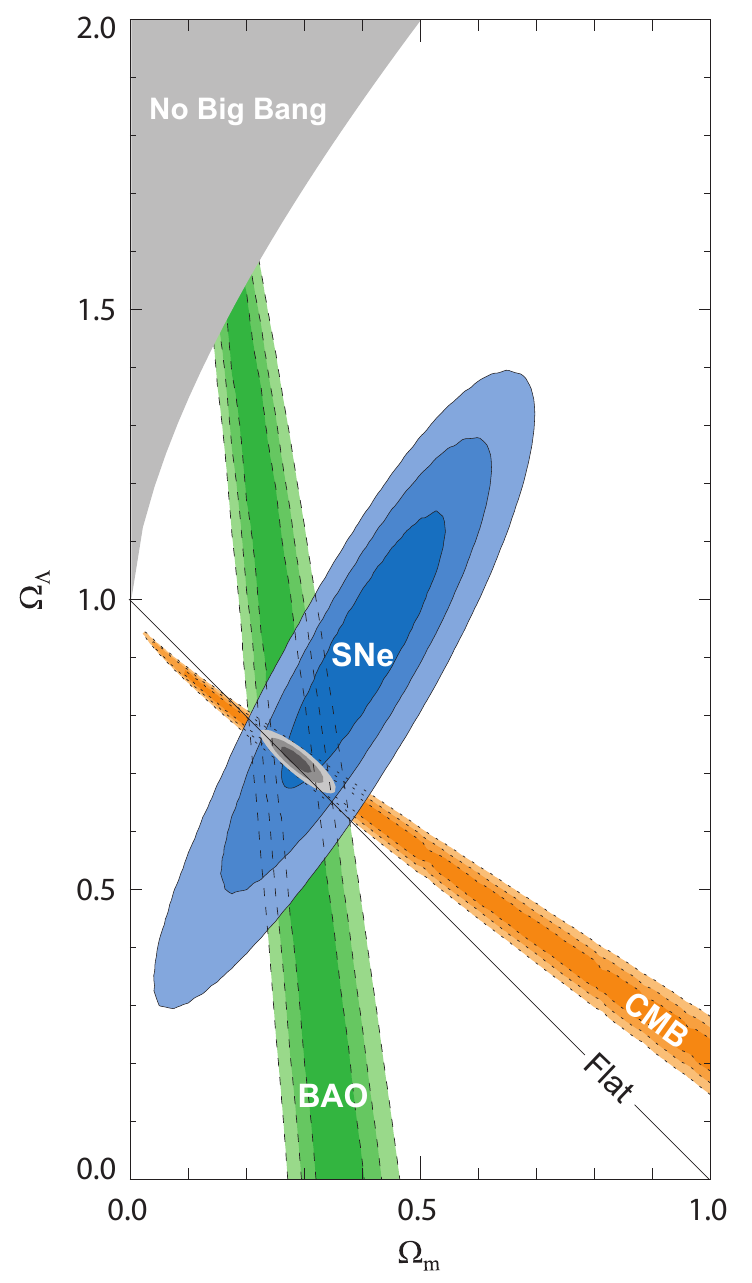}
\end{minipage}%
\begin{minipage}[c]{0.5\textwidth}
  \textbf{b}\\\includegraphics[width=\textwidth]{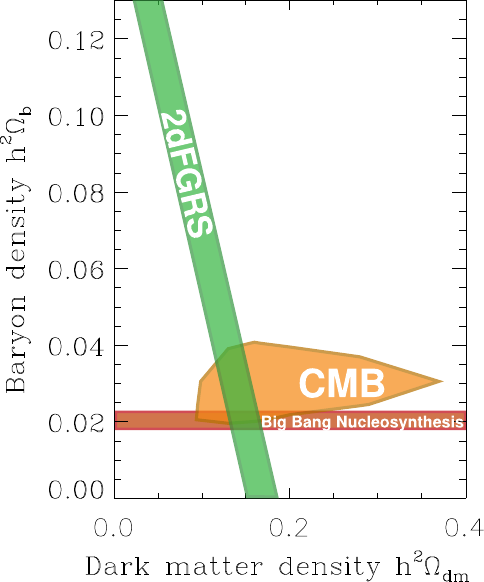}
\end{minipage}
\caption{The current concordance model of cosmology, indicating that the Universe is dominated by dark energy and dark matter, and essentially flat.   These figures illustrate the agreement between cosmological fits to BBN, observations of the cosmic microwave background (CMB), large scale structure (BAO/2dFGRS) and Type 1a supernovae.  The fit is so good that any one of the data sets could be removed and essentially the same cosmology would be inferred from the remainder.  Contours in the figure on the left \emph{(a)} indicate 1-, 2- and $3\sigma$ plausible ranges, whilst areas plotted on the right \emph{(b)} are $1\sigma$ ranges.  Left figure from \protect\citet[][reproduced by permission of the AAS]{Kowalski08}, right figure adapted from \citet{Tegmark01} with additional 2dF results from \protect\citet{Cole05}.}
\label{fig_lcdm}
\end{figure}

\section{Concordance cosmology}
\label{cosmology}

In a sense, the evidence presented in Sects.~\ref{kinematics}--\ref{bbn} is already sufficient to conclude that dark matter definitely exists, and must be neither baryonic nor able to interact substantially with baryons, photons or itself.  Final confirmation of this comes from observations of the cosmic microwave background (CMB).

The large-scale distribution of galaxies shows density variations on scales which reflect the situation at matter-radiation equality.  Similarly, the temperature inhomogeneities of the CMB exhibit characteristic scales which reflect the situation shortly after, at the time of recombination \citep[see e.g.][]{HuDodelson,CMBpedestrian}.  Unlike the galaxy power spectrum however, the angular temperature anisotropies of the CMB exhibit the primordial density perturbations of the coupled baryon-photon fluid directly; following recombination, the angular distribution of the massless photons is held nearly uniform by radiation pressure, whereas the matter collapses due to gravity.  The angular power spectrum of the CMB observed today is therefore sensitive to the full range of cosmological parameters which play a role in the evolution of the baryon-photon oscillations: the total energy density of the Universe, the baryon fraction and the spectral shape of the primordial perturbations.  It is also sensitive to the large-scale geometry of the Universe \emph{since} recombination, as the observed angular diameter of the characteristic scales frozen into the CMB at recombination depends upon the geometry of the space through which they have travelled to reach us.

Thanks to the extremely high resolution of recent CMB missions, fits to the microwave background provide accurate measurements of the matter density of the Universe, as well as the baryon fraction.  The 7-year Wilkinson Microwave Background Probe (WMAP) results \citep{WMAP7Larson} give posterior mean values of $\Omega_\mathrm{m}=0.267\pm0.026$, $\Omega_\mathrm{b}=0.0449\pm0.0028$, $\Omega_\mathrm{DM}=0.222\pm0.026$.  These results are in excellent agreement with those obtained from BBN and large-scale structure, confirming the need for non-baryonic dark matter beyond any doubt.

The CMB fits also indicate that the Universe is approximately flat, requiring some form of `dark energy' to make up the remaining $\Omega_\mathrm{\Lambda}=0.734\pm0.029$.  Although difficult to explain theoretically, the need for this component was confirmed by observations that distant supernovae are apparently dimming, indicating that the expansion of the Universe is accelerating \citep{Riess98,Perlmutter99}.  The CMB, Type Ia supernovae, large scale structure and BBN together paint an entirely self-consistent picture of the Universe in which we live: baryons make up little more than a few percent of the energy budget of the Universe, non-baryonic dark matter about a quarter, and dark energy the remaining three quarters.  The remarkable level of consistency between these different data sets can be seen in Fig.~\ref{fig_lcdm}; were any one of the data sets to be removed, the model would stand perfectly well constrained with the remaining three.  A combined fit to all data \citep{WMAP5} gives $\Omega_\mathrm{m}=0.274\pm0.013$, $\Omega_\mathrm{b}=0.0456\pm0.0015$, $\Omega_\mathrm{DM}=0.228\pm0.013$ and $\Omega_\mathrm{\Lambda}=0.726\pm0.015$.

\chapter{Models for dark matter}
\label{models}

As we have seen in the previous chapter, the identity of dark matter is not entirely unconstrained.  Firstly, it must be massive, because we see its gravitational influence.  We know that it must of course also be dark, which implies that it can neither absorb nor emit photons at rates comparable to normal matter.  This means that it must have a very small effective coupling to photons, either because it exists in some very special geometry that prevents it coming into contact with them very often, or due to its intrinsic particle properties.  In this case, it must be electrically neutral or possess some extremely small fractional charge, as any particle with charge $q\ne0$ couples to photons via tree-level QED processes, leading to cross-sections proportional to $q^4\alpha^2$ (where $\alpha$ is the fine-structure constant).

We know that dark matter must be non-dissipative in order to produce the correct structures in the early Universe.  This also implies that there can be no substantial (i.e. electromagnetic-strength or stronger) interaction between dark matter and normal matter.  If there were, energy would be efficiently transferred from the dark matter to the normal matter and radiated away as photons, allowing the dark matter to form disks along with normal matter.  Dark matter is thus effectively collisionless with respect to all normal matter: baryons, leptons and photons.  We know then that it must be non-baryonic, because it does not couple to photons or baryons; this is spectacularly confirmed by the anisotropies in the CMB.  Observations of systems such as the Bullet Cluster provide even more direct proof that dark matter does not couple substantially to baryons, and even show that dark matter must be effectively collisionless with respect to itself.

Dark matter must also be cold in order for structure formation to proceed correctly, which places limits on the mass(es) of its constituent particle(s)  in some cases.  Finally, any dark matter candidate must be produced with the correct relic abundance to plausibly explain observations that $\Omega_\mathrm{DM}=0.23$ today.  This also means that it must be stable, or at least very long-lived, in order to have persisted in significant numbers to the present day. 

\section{The good, the bad and the ugly}
\label{goodbadugly}

A great many theories have been put forward for the identity of dark matter.  A good candidate theory fulfils all the requirements above, fits all other experimental constraints (most of which are described below or in Chapter \ref{dmsearches}), requires minimal arbitrary choices in its parameter values, and makes some testable predictions.  The last two points are debatable; one is the requirement that the candidate presents a subjectively `natural' solution to the problem, and the other simply makes the theory practically useful.  Mainly due to the non-baryonic requirement, all theories which qualify as `good' today fall into the realm of particle dark matter: namely, that dark matter consists of some as-yet undiscovered particle.  The most notable theories of this kind are outlined in Sects.~\ref{axions}--\ref{others}. 

A theory that was rather popular in astronomical circles for a time was that of MACHOs \citep{Paczynski86}, MAssive Compact Halo Objects.  These were hypothesised to be dim but otherwise quite normal astrophysical objects, which populated the outer halos of galaxies.  Known candidate objects were red, white and brown dwarfs, black holes and substellar-mass objects (in effect, hostless planets).  These systems of condensed baryons would appear dark because they would be so dim, but would also be effectively transparent to background light because their compact nature would cause the effective cross-section for light absorption per unit mass across an entire halo to be very small.  The obvious trouble with MACHOs was that they would have been baryonic dark matter, violating the non-baryonic requirement.  There was also never any convincing argument for why such objects should preferentially congregate in the outer halo, whereas other more luminous stars should not.  Nonetheless, MACHOs were popular because they were testable (not because they were logical): microlensing searches towards the Magellanic Clouds \citep{Alcock00,Tisserand06,OGLE09} promptly ruled MACHOs out as the dominant contributor to dark matter some years ago.

Primordial black holes (PBHs) are another dark matter candidate with significant problems.  These would be formed from small-scale primordial density perturbations which are so strong that they cause the entire horizon mass at the time (or at least a very substantial fraction of it) to collapse directly into a black hole.  The fluctuations would be produced before matter-radiation equality, either during inflation or by phase transitions in the early Universe.  Because PBHs form before BBN or the CMB, the baryons of which they consist are effectively sequestered from the rest of the Universe, allowing them to act like non-baryonic dark matter.  Like MACHOs, they would escape detection via shadowing effects today simply by virtue of their compactness.  The difficulty with this scenario is that the density contrast 
\begin{equation}
\delta \equiv \frac{|\rho_\mathrm{perturbation} - \rho_\mathrm{ambient}|}{\rho_\mathrm{ambient}}
\end{equation}
required to form a PBH is of the order of $\delta\gtrsim30$\% \citep[e.g.][]{Green04c}.  In comparison, the initial density perturbations from inflation were about $\delta\sim10^{-5}$.  A very bottom-heavy spectrum of perturbations is then required in order to provide enough power on small scales to produce a substantial number of PBHs without violating the level of large-scale anisotropy seen in the CMB.

Another problem is that PBHs can be produced with virtually any primordial abundance, creating a major fine-tuning problem.  Their relic density is obtained by integrating the probability distribution of primordial density perturbations between their formation threshold ($\delta\sim30\%$) and $\delta=1$.  Assuming e.g.~a Gaussian power spectrum of perturbations for simplicity \citep[e.g.][]{GreenLiddle}, this produces a density at matter-radiation equality of
\begin{equation}
\Omega_\mathrm{PBH}(M_\mathrm{H}) = \int^{1}_{0.3} \frac{\delta}{\sqrt{2\pi}\sigma(M_\mathrm{H})} \exp\left(-\frac{\delta^2}{2\sigma(M_\mathrm{H})^2}\right)\mathrm{d}\delta.
\label{pbhrelicdens}
\end{equation}
Here $\sigma(M_\mathrm{H})^2$ is the variance of perturbations at the time when the horizon mass is $M_\mathrm{H}$.  With a scale-independent perturbation spectrum of index $n$,
\begin{equation}
\sigma(M_\mathrm{H}) \varpropto M_\mathrm{H}^{-n/4}.
\label{sigma}
\end{equation}
On CMB scales, $n$ is close to 1 \citep{wmap7}, but at small scales, it could be very different.  The resultant relic density from Eq.~\ref{pbhrelicdens} is extremely sensitive to $n$, even for a single epoch of formation (i.e.\ a single value of $M_\mathrm{H}$).  The same is essentially true for other reasonable choices of the power spectrum and scale-dependence.  See \hyperref[papIII]{\SSiv} for a more detailed discussion.  To the best of my knowledge, there is no known mechanism which naturally produces exactly the values of $n$ and $M_\mathrm{H}$ required to reproduce the observed cosmological density of dark matter with PBHs.  Even more disastrously though, PBHs are mostly disfavoured now by experiment.  More massive PBHs should have been picked up by microlensing searches, and the Hawking radiation produced in the evaporation of low mass PBHs should have created effects on the extragalactic diffuse gamma-ray background, BBN, the CMB or cosmic ray data that have not been seen \citep[see][for an extensive compilation of limits]{Josan09}.

SM neutrinos were also considered viable dark matter candidates for a time \citep[e.g.][]{Steigman78}, since they are massive, stable, non-baryonic and do not interact electromagnetically with other matter.  With the current limits on their masses however \citep[e.g.][]{Schwetz08}, they are far too light to meet the definition of cold dark matter, and can only have a quite small cosmological abundance \citep[$\Omega_\nu \lesssim 0.03$;][]{Bergstrom09}.

For a time, the notion of a heavy fourth-generation SM-like neutrino was popular as a candidate for dark matter.  In order to achieve the correct relic abundance, its mass would have to be either 1--50\,eV, or more than about a few GeV \citep{Jungman96,Bergstrom00}; the former is ruled out for SM-like neutrinos by the Tremaine-Gunn limit \citep{TremaineGunn}, and the latter is excluded by direct searches for dark matter (Sect.~\ref{directdetection}).

To some, the MOND paradigm \citep[MOdified Newtonian Dynamics;][]{Milgrom83,Bekenstein04} is appealing.  This approach involves modifying gravity in an ad hoc fashion at long distances in order to fit galactic rotation curves without the need for dark matter.  In this sense, one could say that it is quite a successful empirical fitting formula for parameterising the effects of dark matter on the kinematic observables discussed in Sect.~\ref{kinematics}.  As a physical theory however, it has no basis to speak of, and fails entirely to explain any of the additional evidence for dark matter presented in Sects.~\ref{gravlensing}--\ref{cosmology}.

\section{Axions}
\label{axions}

Because gluons should be pure gauge fields at spatial infinity, the QCD vacuum possesses a rather complex structure \citep{Peccei08}.  Being a pure gauge field implies only that these boundary values must be drawn from the set of field configurations which can be obtained by gauge transformations on the null field (i.e.\ 0).  The vacuum structure arises because the pure gauge boundary condition introduces a freedom in the choice of boundary field, which translates into similar freedom in the choice of QCD vacuum.  Because of this vacuum structure, the QCD Lagrangian picks up an additional effective term
\begin{equation}
\mathcal{L}_\theta = \theta\frac{g_\mathrm{s}^2}{32\pi^2}G_a^{\mu\nu}\tilde{G}_{a\,\mu\nu}.
\label{laxion}
\end{equation}
Here $g_\mathrm{s}$ is the strong coupling constant, $G_a^{\mu\nu}$ is the QCD field strength tensor for the $a$th gluon, and $\tilde{G}_{a\,\mu\nu} = \frac12 \epsilon_{\mu\nu\alpha\beta}G_a^{\alpha\beta}$ is its dual.  Unlike the rest of the QCD Lagrangian, $\mathcal{L}_\theta$ does not conserve CP.  This is a problem, as the strong interaction is known empirically to conserve CP rather well.  This additional term would induce effects which have not been observed, like a non-zero electric dipole moment for the neutron.  The limits on such a moment to date constrain the vacuum angle $\theta$ to be less than $10^{-9}$ \citep{Peccei08}.  Understanding why $\theta$ is so small is known as the `strong CP problem'.

One widely held suspicion is that the solution to the strong CP problem lies in the existence of an additional spontaneously-broken, global chiral $U(1)$ symmetry of the SM Lagrangian, known as the Peccei-Quinn symmetry \citep{PQ}.  The axion is the Goldstone boson of this broken symmetry \citep[see e.g.][for a review]{Kim08}.  It possesses a potential with a minimum that naturally sets the field to a value that cancels $\mathcal{L}_\theta$.  This is true for essentially any value of the Peccei-Quinn symmetry breaking scale $v_a$, relieving the naturalness problem posed by the smallness of $\theta$.

The low-energy phenomenology of axions is set almost entirely by their decay constant $f_a = v_a/N$.  Here $N$ is an integer which gives the degree of the colour chiral anomaly of the Peccei-Quinn symmetry (a corresponding electromagnetic chiral anomaly also exists, and is typically indexed with another integer $E$).  The axion mass for example is given by \citep{Sikivie08}
\begin{equation}
m_\mathrm{a} \approx 6\,\mu\mathrm{eV} \left(\frac{10^{12}\,\mathrm{GeV}}{f_a}\right).
\end{equation}
Axions possess a vertex with two photons \citep{Raffelt08},
\begin{equation}
\mathcal{L}_{a\gamma\gamma} = \frac{\alpha}{2\pi f_a}\left(\frac{E}{N} - \frac23\frac{4+z}{1+z}\right)\mathbf{E}\cdot\mathbf{B}a,
\end{equation}
where $z\equiv m_\mathrm{u}/m_\mathrm{d}$ is the ratio of up and down quark masses, $\mathbf{E}$ and $\mathbf{B}$ are the electric and magnetic fields, and $a$ is the axion field itself.  The two photon vertex not only allows axion decay to two photons, but allows axion conversion to photons (and vice versa) in the presence of electromagnetic fields.  This feature is used as the prime phenomenological means for searching for axions, as discussed in Sect.~\ref{othersearches}.

Certain classes of axions, dubbed `invisible axions', make very good dark matter candidates because they interact extremely weakly with normal matter.  This class of axion involves a very high Peccei-Quinn breaking scale, and therefore very light axions.  Such axions have virtually no kinematically-accessible decay channels, so are stable on cosmological timescales.  These axions constitute cold dark matter despite being so light, because they are never in thermal equilibrium in the early Universe, so are never heated to relativistic temperatures along with the other forms of matter.  

There are two commonly discussed mechanisms for the primordial production of dark matter axions: vacuum misalignment and string decay.  Depending on the timing of Peccei-Quinn breaking relative to inflation, the correct relic density of dark matter axions can be obtained either only for a very specific value of $f_a$ ($\sim$$7\,\times10^{10}$\,GeV), or for a tightly correlated corridor of values in the $f_a$--$\varphi_\mathrm{i}$ plane \citep{Visinelli09}.  The variable $\varphi_\mathrm{i}$ is the initial value of the misalignment angle between the axion field and its minimum; in this model, it is the initial offset from the minimum which causes the field to oscillate and create particles.  In this sense axions have some fine-tuning issues of their own as dark matter candidates, though they are not nearly as severe as for e.g.~primordial black holes.

\section{WIMPs}
\label{wimps}

The most widely-studied dark matter candidates are the Weakly Interacting Massive Particles (WIMPs).  WIMPs interact with SM fields only via the weak nuclear force, making them non-baryonic and electrically neutral by definition.  WIMPs must carry some sort of conserved quantum number to keep them stable on cosmological timescales.  This is usually achieved by making the WIMP the lightest member of a matter sector which is charged under some discrete symmetry.  More often than not, this symmetry needs to be essentially imposed by hand on the underlying theory.  Being members of $SU(2)_\mathrm{L}$ multiplets, in the absence of any other symmetries which might force their masses down, WIMPs should naturally acquire masses $m_\chi$ within a few orders of magnitude of the $SU(2)_\mathrm{L}\times U(1)_\mathrm{Y}$ electroweak symmetry-breaking scale (i.e.~a few GeV or TeV).  This makes them sufficiently heavy to constitute cold dark matter even if they have been produced thermally in the early Universe.  Examples of WIMPs include the lightest neutralino in supersymmetry (Sect.~\ref{susywimps}), the lightest Kaluza-Klein (KK) particle \citep{ServantTait} and an additional inert Higgs boson \citep{Barbieri06}.

\subsection{Thermal relics}
\label{thermalrelics}

A thermal relic is a particle whose cosmological abundance is set by thermal production of the particle in the early Universe.  `Thermal production' actually refers to \emph{chemical} decoupling of a species from the primordial particle soup created in the Big Bang.  Here `chemical' has the meaning of particle creation and destruction by quantum processes, not the usual meaning of creation and destruction of molecules and free atoms by atomic interactions.  True \emph{thermal} decoupling on the other hand, otherwise known as kinetic decoupling, refers to the time when the velocities of a relic species cease to reflect the temperature of the Universe.

Chemical decoupling, or `freeze out', happens when the expansion of space eventually overcomes the rate at which the species is created and annihilated in interactions with other particles.  The primordial fireball necessarily begins in chemical (and thermal) equilibrium at some very high temperature $T_\mathrm{i}$.  The populations of different particle species are set by the equilibrium rates of particle creation and annihilation at $T_\mathrm{i}$, and their velocities follow a Maxwellian distribution with temperature $T_\mathrm{i}$.  As the Universe expands, it cools, and for a time, the equilibrium populations of the particles adjust accordingly as chemical equilibrium is maintained.  At some point in the expansion, when $T=T_\mathrm{c}$, the creation and annihilation of a particular species will cease to be able to keep pace with the expansion, and the particle will fall out of chemical equilibrium.  If it is stable, at this point the comoving density of the species will become essentially fixed.\footnote{This is analogous to the transition from LTE to non-LTE populations of atomic energy levels in astrophysical plasmas as one considers progressively lower gas densities or temperatures.}  Similarly, when elastic collisions between particles become sufficiently infrequent, both due to the increasing scale and decreasing temperature of the Universe, it will fall out of kinetic equilibrium.

The time of chemical freeze out therefore sets the final population of any stable relic, and the time of kinetic decoupling sets its temperature.  In general, kinetic decoupling of WIMPs happens after chemical decoupling \citep[see e.g.][]{Bringmann09_clumps}.  With weak-scale masses and creation/anni\-hilation cross-sections, $T_\mathrm{c} \approx 0.05m_\chi$ \citep{Jungman96}.  As kinetic decoupling happens at comparable or even lower temperatures than this, WIMPs freeze out at sub-relativistic energies.  They are therefore guaranteed to be moving non-relativistically at the time of structure formation, so qualify as cold dark matter.

The resultant abundance of the relic species depends upon the history of the expansion rate up until freeze out, and the rate at which the species annihilates into lighter particles.  The expansion rate is $H\equiv \dot{a}/a$ (where $a$ is the scale factor of the Universe and the dot denotes time differentiation), and is obtained from the cosmological model.  To a first approximation, the evolution of the population is thus described by the Boltzmann equation \citep{Jungman96}
\begin{equation}
\frac{\mathrm{d}n_\chi}{\mathrm{d}t} + 3Hn_\chi = \langle\sigma v\rangle (n_{\chi,\,\mathrm{eq}}^{\phantom{\chi,\,\mathrm{eq}}2} - n_\chi^{\phantom\chi2})
\label{boltzmann}
\end{equation}
where $n_{\chi,\,\mathrm{eq}}$ and $n_\chi$ are the equilibrium and actual number densities of the species $\chi$, and $\langle\sigma v\rangle$ is the thermally-averaged product of the relative velocity and the total annihilation cross-section for $\chi\bar{\chi}\to$ other particles.  Eq.~\ref{boltzmann} holds whether $\chi$ is Majorana or Dirac, though in the Dirac case it is only valid if there is no particle-antiparticle asymmetry to begin with, meaning that the total particle population is given by $2n_\chi$.

\begin{figure}[t]
\centering
\includegraphics[width=0.7\textwidth]{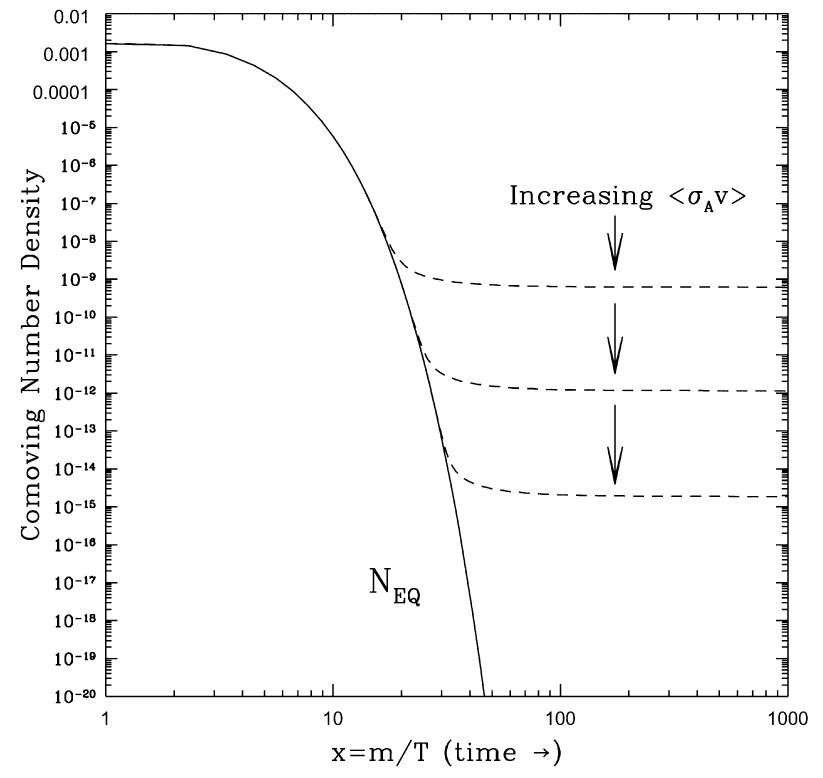}
\caption{Freeze out of a stable thermal relic.  At early times, the relic particle is in chemical and thermal equilibrium with other particles, with its abundance set by the equilibrium rates of particle creation and annihilation.  Because of thermal coupling, these are set entirely by the temperature of the cooling Universe.  At later times, the expanding Universe cools to the point where the equilibrium density is too low to maintain chemical contact with other particles.  At this point, the particle falls out of chemical equilibrium with the rest, and its comoving density becomes fixed.  The density at which this occurs depends upon the annihilation cross-section of the particle; larger cross-sections allow equilibrium to be maintained for longer, resulting in lower relic abundances.  From \protect\citet{Jungman96}, modelled on a figure from \protect\citet{KolbTurner}.}
\label{fig_relicdens}
\end{figure}

Eq.~\ref{boltzmann} can be solved numerically in order to obtain the relic density of any species of interest; Fig.~\ref{fig_relicdens} gives solutions for the chemical freeze out of some example WIMPs.  To a good approximation \citep{Jungman96,Bergstrom00}, for particles with masses in the relevant range for WIMPs 
\begin{equation}
\Omega_\chi h^2 = 0.1\frac{3\times 10^{-26}\,\mathrm{cm}^3\,\mathrm{s}{-1}}{\langle\sigma v\rangle}.
\label{relicdens}
\end{equation}
As seen in Eq.~\ref{relicdens} and Fig.~\ref{fig_relicdens}, larger self-annihilation cross-sections cause the particle to freeze out later, and so exhibit a lower relic density.  In practice, co-annihilations between the WIMP and other particles nearly degenerate with it in mass, as well as effects such as annihilation thresholds and resonances, can make the calculation a little bit trickier than Eqs.~\ref{boltzmann} and \ref{relicdens} might suggest.  Dedicated computer packages have been developed to solve Eq.~\ref{boltzmann} in detail, for a variety of specific WIMP models \citep[e.g.][]{DarkSUSY,MicrOmegas}.

Eq.~\ref{relicdens} shows that the canonical annihilation cross-section implied by the cosmological observations discussed in Chapter~\ref{needfordm} is $\langle\sigma v\rangle\approx3\times10^{-26}\,\mathrm{cm}^3\,\mathrm{s}^{-1}$.  The natural value of the annihilation cross-section for a weakly-interacting particle can be estimated as $\langle\sigma v\rangle \approx \alpha^2 / E_\mathrm{weak}^{\phantom{\mathrm{weak}}2}\approx10^{-8}$\,GeV$^{-2}=10^{-25}$\,cm$^3$\,s$^{-1}$, where $E_\mathrm{weak}\sim100$\,GeV is the electroweak scale.\footnote{That this is approximately true can be verified by e.g.~considering $Z$-mediated $\chi$ self-annihilation, such that $|\mathcal{M}|^2\sim g^4E_\chi^{\phantom{\chi}4}/m_Z^{\phantom{Z}4}$, and using Eq.~4.84 of \protect\citet{Peskin} to arrive at
\begin{equation}
\sigma v \approx \frac{|\mathcal{M}|^2}{128\pi^2E_\chi^{\phantom{\chi}2}} \approx \frac{G^2E_\chi^{\phantom{\chi}2}}{4\pi^2}, 
\label{wimpmiracle}
\end{equation}
where $g$ is the electroweak coupling and $G$ is the Fermi coupling constant.  In the case where the incoming $\chi$s are moving non-relativistically (as WIMPs do), $E_\chi \approx m_\chi \approx E_\mathrm{weak}$, giving the zero-velocity limit $\langle\sigma v\rangle_0 \approx 10^{-8}\,\mathrm{GeV}^{-2}=10^{-25}$\,cm$^3$\,s$^{-1}$.}
Here we have an amazing coincidence: purely by postulating that dark matter is a stable weakly-interacting particle, we have predicted the relic density to within an order of magnitude.  The naturalness with which WIMPs can provide the correct relic density has lead this to be dubbed `the WIMP miracle'.  Whilst far from definitive, it is a very strong hint that WIMPs might be the correct solution to the dark matter puzzle.

\subsection{The hierarchy problem}

Indeed, the process of chemical freeze out is so generic that if a stable neutral particle exists around the electroweak scale, then we really should \emph{expect} it to be dark matter.  But what reason do we actually have for believing that such stable particles might exist around or just above the electroweak scale?  This is a question of why we think there might be new physics at such an energy scale, as new physics generically means new particles (though not necessarily stable ones).  There are a number of reasons to expect that new physics exists beyond the SM, since the SM cannot explain phenomena such as the net baryon asymmetry of the Universe, neutrino masses, the imperfect unification of the known gauge couplings, and the spectrum of particle masses -- but the one compelling reason to believe that new physics exists specifically at the TeV scale is the mass of the Higgs boson.

\begin{figure}
\begin{center}
  \begin{fmfgraph}(60,30)
   \fmfleft{i1}
   \fmfright{o1}
   \fmf{fermion,left,tension=0.4}{v1,v2,v1}
   \fmf{dashes}{i1,v1}
   \fmf{dashes}{v2,o1}
   \fmfdot{v1,v2}
  \end{fmfgraph}
\caption{An example one-loop contribution to the Higgs mass from a fermionic loop.  The amplitude of this diagram is quadratically divergent, so the Higgs mass is sensitive to some large mass scale (which may be expressed as the cut-off used to regularise the diagram's amplitude, amongst other presentations; see e.g.~\protect\citealt{LutyTASI}).}
\label{fig_fermionloop}
\end{center}
\end{figure}
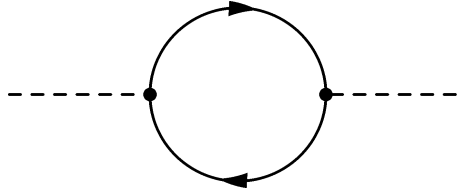

As a fundamental scalar, the mass of the Higgs is not protected by any symmetry; quadratic divergences generated by fermionic loop diagrams such as Fig.~\ref{fig_fermionloop} have no apparent cancellation or suppression, nor can they be absorbed into the renormalisation group running of any coupling constant.  For fermions, the case is different: similar mass corrections are controlled by the approximate chiral symmetry of the fermionic part of the SM Lagrangian \citep{LutyTASI}.  For the gauge bosons, gauge symmetry suppresses the divergences and shifts them into the running of the gauge couplings.  For the SM Higgs however, the quadratic divergences remain bare, causing extreme corrections to its mass.  There is therefore no reason to believe that the SM Higgs should have a mass any lower than the GUT scale.  That is, except for the pesky detail of its fundamental role in electroweak symmetry breaking and the generation of mass for the entire spectrum of SM fermions, which it needs an electroweak-scale vacuum expectation value (VEV) to achieve.  How then is the Higgs mass stabilised at the weak scale?  This problem is known as the gauge hierarchy problem, and strongly suggests that some new physics must kick in at energies just above the electroweak scale, allowing us to calculate the effective Higgs mass below this scale by regularising the amplitudes of the divergent diagrams with a cut-off at $\Lambda\sim1$\,TeV.

The hierarchy problem is a very generic motivation for new physics at or just above the electroweak scale.  That is to say, it is independent of what the nature of this new physics might be, so there is no \emph{a priori} reason for us to naively expect that it necessarily has anything to do with dark matter.  The fact that its resolution would appear to require new physics at exactly the right scale for the associated particles to be WIMPs is thus another amazing coincidence, and a second very strong hint that WIMPs constitute the majority of dark matter.

\section{Supersymmetry}
\label{susy}

Although this chapter is primarily intended as something of a bestiary of dark matter models, here it is worth detouring slightly into the theory of supersymmetry (SUSY).  This is because SUSY not only provides multiple examples of WIMPs, but because it is also exactly the sort of `new physics' discussed in the previous section as necessary to solve the hierarchy problem.  

\subsection{General features}

The basic idea of SUSY is to give every SM particle one (or more) superpartner(s), which differ from their SM counterparts only by their spins \citep[see e.g.~][for reviews; the bulk of the treatment in this section follows \protect\citeauthor{BaerTata}]{Martin,BaerTata,Aitchison}.  Every SM fermion has a sfermionic superpartner (which is a boson) and every SM boson has a bosino superpartner (which is a fermion).  The symmetry is a continuous extension of the Poincar\'e group to fermionic (i.e.~anti-commuting) degrees of freedom.  Supersymmetry is thus essentially an extension of special relativity into a four-dimensional fermionic `superspace'; the fermionic dimensions can be viewed either as four additional anti-commuting dimensions, or simply as fermionic extensions of the existing four.  Correspondingly, the generators of SUSY transformations are themselves spinorial, in contrast to the vector and scalar generators associated with traditional spacetime and internal symmetries.  It is entirely possible for there to exist multiple distinct supersymmetry generators, and therefore multiple supersymmetries (known as $N>1$ supersymmetries, where $N$ is the number of generators).  $N=1$ supersymmetry is the only version which permits chiral fermions, so known phenomenology dictates that supersymmetry can only be of the $N=1$ variety at low energies.

It is often said that SUSY provides a non-trivial way of unifying spacetime and internal symmetries, since the anti-commutation relation for the spinor components of the SUSY generator $Q$
\begin{equation}
\{Q_a,\bar{Q}_b\}= (\gamma_\mu)_{ab}P^\mu
\end{equation}
gives the generator of spacetime translations, $P^\mu$.\footnote{Here I am working with 4-component spinor notation as per \protect\citet{BaerTata}.  The bar denotes the spinor adjoint $\bar{Q}\equiv Q^\dagger\gamma^0$, and $\gamma_\mu$ are the standard Dirac gamma matrices, forming a representation of the Clifford algebra of Minkowski space.}  This means that the combination of any two SUSY transformations is equivalent to a spacetime translation.  So, the supersymmetry algebra mixes spacetime transformations with boson-fermion exchange, which would seem to be something of an `internal' operation because it directly modifies particle properties.  This interpretation seems unreasonable though, as the swapping of bosons for fermions or vice versa involves modifying only the particle's spin, which is arguably no less of a spacetime property than momentum or position.  Even if supersymmetry were gauged (more on this later), the resulting gauge theory of gravity would still seem to be an entirely spacetime symmetry, and all the other gauge symmetries of particle physics still entirely internal ones.  This is effectively what was said by the SUSY version \citep{Haag75} of the celebrated Coleman-Mandula `no-go' theorem \citep{Coleman67}: it is not possible to have a larger spacetime symmetry than the super-Poincar\'e algebra, so there are no non-trivial ways to unify spacetime and internal gauge symmetries into the same group.

\begin{figure}
\begin{center}
  \begin{fmfgraph}(50,20)
   \fmfleft{i1,i2}
   \fmfright{o1,o2}
   \fmf{dashes}{i1,v1,o1}
   \fmf{phantom}{i2,v2,o2}
   \fmffreeze   
   \fmf{scalar,right}{v1,v2,v1}
   \fmfdot{v1}
  \end{fmfgraph}
\caption{An example one-loop contribution to the Higgs mass from a sfermionic loop.  The amplitude of this diagram has exactly the same sensitivity to the high-scale cut-off as the fermionic loop diagram of Fig.~\protect\ref{fig_fermionloop}, except for a factor of $-\frac12$.  In unbroken supersymmetry, two of these diagrams appear for each fermionic one, exactly cancelling the fermionic loop's contribution to the Higgs mass.  The same occurs to all orders in perturbation theory.  This removes the sensitivity to the high scale, stabilising the Higgs mass against higher-order corrections.}
\label{fig_bosonloop}
\end{center}
\end{figure}
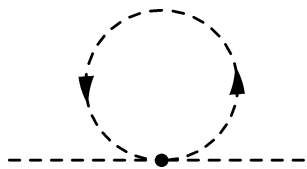

SUSY solves the hierarchy problem in a very simple and elegant way: because every fermion has a bosonic superpartner and vice versa, every divergent loop diagram containing a SM fermion like Fig.~\ref{fig_fermionloop} is matched by corresponding scalar sfermion loop diagrams like Fig.~\ref{fig_bosonloop}.  In this case there are actually two scalar loop diagrams, because chiral fermions have two degrees of freedom, so their scalar superpartner fields must be complex, resulting in two oppositely charged sfermions for each chiral fermion.  Because closed fermionic loops introduce a factor of $-1$ to their diagram's amplitude (whereas bosonic loops do not), and the particles are otherwise identical, the (regularised) contributions of the two copies of the scalar loop diagram cancel that of the fermionic one.  The same is so to all orders in perturbation theory, so the Higgs mass receives no extreme corrections, and is stabilised somewhere near the SUSY-breaking scale.

SUSY must clearly be a broken symmetry, as we have not observed any of the superpartners to date (known particles cannot be the superpartners of each other, as they do not fit correctly into the supermultiplet structure required by the supersymmetry algebra).  Except for some very specific scenarios where particular superpartners happen to be especially sneaky, we can infer that they must all be considerably more massive than the SM particles, as they have yet to show up in e.g.~collider experiments.  We would expect then that the SUSY particles should generally have masses of around the SUSY-breaking scale.  The cancellation of divergences thus cannot be exact following SUSY-breaking, as particles and sparticles must have different masses.  We can then conclude that if SUSY is a viable solution to the hierarchy problem, it must be broken at about a TeV, otherwise the cancellation of the different loop contributions would again require an unexplained fine-tuning.  In fact, another appealing feature of SUSY is that it can actually \emph{cause} electroweak symmetry-breaking (EWSB), in that the Higgs mass term in the $SU(2)_\mathrm{L}\times U(1)_\mathrm{Y}$ Lagrangian can be driven negative by SUSY renormalisation group evolution -- but only if the SUSY-breaking scale is about a TeV.  So, we might expect EWSB to have been induced by SUSY breaking, and hence taken place at only a slightly lower energy than SUSY-breaking itself.  Since we know the electroweak scale to be a few hundred GeV, this gives further credence to the idea that SUSY-breaking should occur at around a TeV.

Actually, the cancellation of quadratic divergences depends a lot more crucially upon the gauge coupling constants in the SM and SUSY sector being identical than the particle masses.  It is possible to introduce explicit mass terms to the SUSY Lagrangian which break SUSY `softly', in that they give the sparticles different masses to the SM particles, but do not introduce any quadratic divergences.

The simplest prescription then for writing down a minimal supersymmetric theory is to take the supersymmetrised version of the SM Lagrangian, 
$\mathcal{L}_\textrm{SUSY-SM}$, and augment it with all possible renormalisable soft SUSY-breaking terms \citep{BaerTata}
\begin{subequations}
\label{lsoft}
\begin{align}
\mathcal{L}_\mathrm{soft} =& -\frac12\left[M_1\bar{\tilde{B}}^0\tilde{B}^0 + M_2\bar{\tilde{W}}_A\tilde{W}_A + M_3\bar{\tilde{g}}_B\tilde{g}_B\right] \label{gauginos}\\
& -\frac{i}{2}\left[M_1'\bar{\tilde{B}}^0\gamma_5\tilde{B}^0 + M_2'\bar{\tilde{W}}_A\gamma_5\tilde{W}_A + M_3'\bar{\tilde{g}}_B\gamma_5\tilde{g}_B\right] \label{CP-violating gauginos}\\
& + \Big[b{H}^a_u{H}_{da} + \mathrm{h.c.}\Big] + m^2_{H_u}|H_u|^2 + m^2_{H_d}|H_d|^2 \label{Higgses}\\
& +\textstyle\sum_{i,j=1,3} \left\{ -\left[\tilde{Q}_i^\dagger(\mathbf{m}^\mathbf{2}_Q)_{ij}\tilde{Q}_j +
\tilde{d}_{\mathrm{R}i}^\dagger(\mathbf{m}^\mathbf{2}_d)_{ij}\tilde{d}_{\mathrm{R}j} \right. \right. \nonumber\\ 
& + \tilde{u}_{\mathrm{R}i}^\dagger(\mathbf{m}^\mathbf{2}_u)_{ij}\tilde{u}_{\mathrm{R}j} 
  + \tilde{L}_{i}^\dagger(\mathbf{m}^\mathbf{2}_L)_{ij}\tilde{L}_{j} 
  + \tilde{e}_{\mathrm{R}i}^\dagger(\mathbf{m}^\mathbf{2}_e)_{ij}\tilde{e}_{\mathrm{R}j} \Big] \label{sfermion masses}\\
& + \left[(\mathbf{A}_u)_{ij}\epsilon_{ab}\tilde{Q}^a_iH^b_u\tilde{u}^\dagger_{\mathrm{R}j} 
  + (\mathbf{A}_d)_{ij}\tilde{Q}^a_iH_{da}\tilde{d}^\dagger_{\mathrm{R}j} \right. \nonumber\\
& \hspace{1cm} + \left. (\mathbf{A}_e)_{ij}\tilde{L}^a_iH_{da}\tilde{e}^\dagger_{\mathrm{R}j} + \mathrm{h.c.}\right] \label{trilinear couplings}\\
& + \left[(\mathbf{C}_u)_{ij}\epsilon_{ab}\tilde{Q}^a_iH^{*b}_d\tilde{u}^\dagger_{\mathrm{R}j} 
  + (\mathbf{C}_d)_{ij}\tilde{Q}^a_iH^*_{ua}\tilde{d}^\dagger_{\mathrm{R}j} \right. \nonumber\\
& \hspace{1cm} + \left.\left. (\mathbf{C}_e)_{ij}\tilde{L}^a_iH^*_{ua}\tilde{e}^\dagger_{\mathrm{R}j} + \mathrm{h.c.}\right] \right\}. \label{trilinear C couplings} 
\end{align}
\end{subequations}
Here the generation indices $i$ and $j$ are explicitly summed over, whilst summation is implied over the $SU(2)_\mathrm{L}$ indices $a,b=1,2$ and the gauge generator indices $A=1\,..\,3$ and $B=1\,..\,8$.  Tilded operators denote superparticle fields ($\tilde{Q}_j$, $\tilde{u}^\dagger_{\mathrm{R}j}$, etc.), where $\tilde{B}^0$, $\tilde{W}_A$ and $\tilde{g}_B$ are the superpartners of the familiar SM gauge bosons.  Other capitalised fields are $SU(2)_\mathrm{L}$ doublets, whilst lowercase fields are singlets: $Q_i\equiv(u_{\mathrm{L}i},d_{\mathrm{L}i})^\mathrm{T}$, $L_i\equiv(e_{\mathrm{L}i},\nu_{e_i\mathrm{L}})^\mathrm{T}$, $H_u\equiv(H_u^+,H_u^0)^\mathrm{T}$ and $H_d\equiv(H_d^-,H_d^0)^\mathrm{T}$.  The doublet definitions for the superpartner fields are simply tilded versions of the same.  Notice that there are now two Higgs doublets instead of the normal one from the SM.  After EWSB, this gives rise to two additional charged Higgs bosons and a further two neutral Higgs compared the SM (recall that three real degrees of freedom are lost to the gauge bosons in EWSB, so two doublets means $8-3=5$ real degrees remaining, instead of $4-3=1$ real degree in the SM).  The two doublets give masses separately to the up-type and down-type quarks, hence the subscripts $u$ and $d$; the up-type doublet corresponds to the SM Higgs.  The down-type doublet is required because the up-type Higgs cannot couple to the down-type quarks without breaking the SUSY-invariance of the action.  Unless multiple Higgs doublets are present, Higgsinos also ruin certain anomaly cancellations which exist in the SM.

The first two terms in Eq.~\ref{lsoft} (\ref{gauginos} and \ref{CP-violating gauginos}) give explicit masses to the gauginos via the real parameters $M_1,M_2,M_3$ and $M'_1,M'_2,M'_3$.  The second of these terms violates $CP$, so $M'_1,M'_2$ and $M'_3$ should be small or zero.  The Higgs sector (\ref{Higgses}) includes explicit mass terms with real parameters $m^2_{H_u}$ and $m^2_{H_d}$, as well as a bilinear coupling with complex parameter $b$.  Explicit sfermion masses (\ref{sfermion masses}) are given by five $3\times3$ Hermitian mass-squared matrices $\mathbf{m}^\mathbf{2}_Q$, $\mathbf{m}^\mathbf{2}_u$, $\mathbf{m}^\mathbf{2}_d$, $\mathbf{m}^\mathbf{2}_L$ and $\mathbf{m}^\mathbf{2}_e$.  The final two terms (\ref{trilinear couplings} and \ref{trilinear C couplings}) give trilinear couplings between the Higgs and squarks or sleptons, with general Yukawa-type complex $3\times3$ matrices $\mathbf{A}_u, \mathbf{A}_d, \mathbf{A}_e$ and $\mathbf{C}_u, \mathbf{C}_d, \mathbf{C}_e$.  The latter terms, proportional to the $\mathbf{C}$ matrices (\ref{trilinear C couplings}), tend to be left out of most low-energy effective models because they are suppressed in many SUSY-breaking schemes.

$\mathcal{L}_\textrm{SUSY-SM}$ is itself straightforward to write down, but lengthy and not especially phenomenologically illuminating (see for example Eq.~6.44 of \citealt{BaerTata}).  For the purposes of this general description, the most important aspect is that it contains derivatives of the `superpotential'
\begin{equation}
\begin{split}
\hat{W} = \mu\hat{H}^a_u\hat{H}_{da} + \textstyle\sum_{i,j=1,3} & \left[(\mathbf{Y}_u)_{ij}\epsilon_{ab}\hat{Q}^a_i\hat{H}^b_u\hat{U}^c_j \right. \\ 
          & + (\mathbf{Y}_d)_{ij}\hat{Q}^a_i\hat{H}_{da}\hat{D}^c_j + \left. (\mathbf{Y}_e)_{ij}\hat{L}^a_i\hat{H}_{da}\hat{E}^c_j\right].
\end{split} 
\label{superpot}
\end{equation}
Here the indices $i$ and $j$ are again generation number, and $a$ and $b$ are $SU(2)_\mathrm{L}$ indices.  The carets indicate superfields, containing both the SM fields (and where applicable, $SU(2)_\mathrm{L}$ doublets) such as $u_{\mathrm{R}i}, L_i$, etc.~and their superpartners $\tilde{u}_{\mathrm{R}i}, \tilde{L}_i$, etc.  The terms $\hat{U}^c_j$, $\hat{D}^c_j$ and $\hat{E}^c_j$ are the left chiral superfields containing the charge conjugates of the right-handed $SU(2)_\mathrm{L}$ singlets: up-type (s)quarks, down-type (s)quarks and (s)electrons, respectively.  Derivatives of $\hat{W}$ with respect to its scalar fields give rise to all non-gauge interaction terms in $\mathcal{L}_\textrm{SUSY-SM}$.  In this sense, it is somewhat analogous to the non-gauge part of the scalar potential seen in non-supersymmetric quantum field theories.  Here it plays a similar role, specifying the Higgs potential via the complex parameter $\mu$, as well as the Higgs-fermion interactions and ultimately, the fermion masses, via the general complex $3\times3$ Yukawa coupling matrices $\mathbf{Y}_u$, $\mathbf{Y}_d$ and $\mathbf{Y}_e$.

In Eqs.~\ref{lsoft} and \ref{superpot}, only terms which conserve baryon ($B$) and lepton number ($L$) have been retained.  $B$ and $L$ are known to be broken non-perturbatively in the SM \citep[][see also e.g.~the introductory remarks of \protect\citealt{Morrissey05}]{t'Hooft76a,t'Hooft76b}.  This happens as the $B$ and $L$ currents become anomalous (non-conserved) due to a freedom in the electroweak vacuum, similar to that discussed for the QCD vacuum in the context of the axion in Sect.~\ref{axions}.  `Instanton' transitions between these vacua can then cause simultaneous violation of $B$ and $L$ by 3 units.  Nevertheless, $B$ and $L$ are observed to be very good (albeit accidental) symmetries of the SM at tree level; the electroweak vacuum doesn't seem to have done much tunnelling recently.  Exactly why the SM contains no $B$ or $L$-violating operators is an open question.  Certain other terms which explicitly violate $B$ and $L$ are allowed in $\mathcal{L}_\mathrm{soft}$ and the superpotential by all other considerations, but in the interests of constructing a model with minimal new interactions, and which agrees with experimental constraints, they are generally excluded.

Together, $\mathcal{L}_\textrm{SUSY-SM}$ and the soft terms in Eq.~\ref{lsoft} give the Lagrangian for the the Minimal Supersymmetric Standard Model (MSSM)
\begin{equation}
\mathcal{L}_\mathrm{MSSM} = \mathcal{L}_\textrm{SUSY-SM} + \mathcal{L}_\mathrm{soft},
\label{lmssm}
\end{equation}
after SUSY breaking but prior to EWSB.  The model is minimal in that it contains the least additional field content possible for a supersymmetrised version of the SM.  Note that the only new parameter actually introduced in initially supersymmetrising the SM is $\mu$, the Higgs parameter in the superpotential.  Once we demand a low-energy theory for broken supersymmetry, the soft terms introduce more than a hundred additional parameters.

\subsection{Supersymmetric WIMPs}
\label{susywimps}

In order for a SUSY particle to realistically constitute dark matter, it must somehow be stabilised against decay into lighter SM states.  The most common way this is achieved is to postulate that aside from supersymmetry, a discrete $\mathbb{Z}_2$ symmetry exists between the SM particles and their SUSY partners.  The corresponding conserved quantum number is known as $R$-parity, and has the form
\begin{equation}
R = (-1)^{3(B-L)+2s},
\label{rparity}
\end{equation}
where $s$ is the particle's spin.  All SM particles thus have $R$-parity $+1$ and all their SUSY partners $R$-parity $-1$.  Clearly any theory which simultaneously conserves $L$ and $B$ automatically conserves $R$; the MSSM superpotential and soft terms (Eqs.~\ref{lsoft} and \ref{superpot}) hence conserve $R$-parity by construction.  Note that the converse is emphatically not true: $R$-parity conservation does not imply $B$ or $L$ conservation.  For example, the most interesting of the $B$ and $L$-violating terms discussed in the previous subsection (the ones which are arguably valid additions to the MSSM Lagrangian), include some which conserve $R$-parity.

Although $R$-parity is generally imposed by hand in most SUSY variants (either via $B$ and $L$ conservation, or directly), it is not plucked entirely from thin air.  Some grand unified theories (GUTs) give rise to $R$-parity conservation naturally in their group and/or representation structure; examples are certain supersymmetric $SO(10)$ GUTs \citep[e.g.][]{LeeMohapatra}.  Indeed, $R$-parity conservation in a higher theory might be part of the reason $B$ and $L$ are approximate symmetries of the SM today.  If $R$-parity is conserved, then the lightest SUSY particle (LSP) is absolutely stable.  If it is also weakly-interacting and electrically neutral, then it is a viable WIMP.

The quintessential SUSY WIMP is the lightest neutralino, the lightest linear combination of the two neutral Higgsinos, the neutral wino and the bino,
\begin{equation}
\tilde{\chi}^0_1 = V^*_{1l} \tilde{H}_u^0 + V^*_{2l} \tilde{H}_d^0 + V^*_{3l} \tilde{W}_3^0 + V^*_{4l} \tilde{B}^0,
\label{neutralino}
\end{equation}
which mix following EWSB because they share quantum numbers.  There are thus four neutralinos in the MSSM, corresponding to the different linear combinations; only the lightest is ever stable, and it is always this particle that is meant when people talk about `the neutralino' as a WIMP.  Their masses are given by the eigenvalues of the mass mixing matrix which appears in the post-EWSB Lagrangian
\begin{equation}
\mathcal{M}_{\chi^0} = \left( \begin{array}{cccc}
  0 & \mu & -\frac{gv_u}{\sqrt{2}} &  \frac{g'v_u}{\sqrt{2}} \\
  \mu & 0 &  \frac{gv_d}{\sqrt{2}} & -\frac{g'v_d}{\sqrt{2}} \\
  -\frac{gv_u}{\sqrt{2}} & \frac{gv_d}{\sqrt{2}} & M_2 & 0   \\
  \frac{g'v_u}{\sqrt{2}} & -\frac{g'v_d}{\sqrt{2}} & 0 & M_1 \end{array}\right),
\label{neutralinomass}
\end{equation}
in the basis $\vec{G} \equiv (\tilde{H}_u^0, \tilde{H}_d^0, \tilde{W}_3^0, \tilde{B}^0)^\mathrm{T}$.  Here $v_u$ and $v_b$ are the VEVs obtained by the up- and down-type Higgs during EWSB, and $g'=g\tan\theta_\mathrm{W}$, with $\theta_\mathrm{W}$ the weak mixing angle.  The coefficients $V^*_{1l}, V^*_{2l}, V^*_{3l}, V^*_{4l}$ of the individual gaugino/Higgsino components of each mass eigenstate are the entries in the inverse of the unitary matrix $V$ which diagonalises $\mathcal{M}_{\chi^0}$,
\begin{equation}
\vec{\chi} = V^\dagger\vec{G},
\label{neutralinoeigenvectors}
\end{equation}
where $\vec{\chi} \equiv (\tilde{\chi}^0_A,\tilde{\chi}^0_B, \tilde{\chi}^0_C, \tilde{\chi}^0_D)^\mathrm{T}$.  The index $l$ in the $V^*_{1l}, V^*_{2l}, V^*_{3l}, V^*_{4l}$ refers to the column of $V$ which applies to the lightest mass eigenstate.  The usual labelling for mass eigenstates is to sort $\tilde{\chi}^0_A, \tilde{\chi}^0_B, \tilde{\chi}^0_C, \tilde{\chi}^0_D$ in increasing mass and assign them the numbers from 1 to 4, i.e.~$\tilde{\chi}^0_1$, $\tilde{\chi}^0_2, \tilde{\chi}^0_3$, $\tilde{\chi}^0_4$.

Sneutrinos, the spin-0 partners of neutrinos, are also weakly-interacting and neutral, so qualify as SUSY WIMPs when the lightest of their number is the LSP.  Unfortunately, this is not usually the case, because in most models one of the sleptons has a slightly lower mass than the lightest sneutrino \citep{Jungman96}.  In any case, constraints on the sneutrino-nucleon scattering cross-section from direct detection (Sect.~\ref{directdetection}) all but rule it out as the dominant component of dark matter, as the nuclear scattering cross-section for sneutrinos is even higher than for Dirac neutrinos \citep{Falk94}.

It also is worth pointing out that in certain (generally non-minimal) SUSY scenarios, the dark matter particle might not be absolutely stable, but metastable on cosmological timescales.  In some cases, this removes the need for $R$-parity conservation.  This applies to some types of gravitinos (the SUSY partner of the graviton, the gauge quantum of gravity), an example of non-WIMP SUSY dark matter discussed in Sect.~\ref{gravitinos}.

\subsection{SUSY breaking and parameterisations}

As a low-energy effective theory, the MSSM involves no specification of the mechanism by which SUSY is broken.  In its full form (Eqs.~\ref{lsoft}--\ref{lmssm}), the MSSM possesses a vast number of free parameters that are not present in the SM.  All but one ($\mu$) come from the soft terms in the SUSY-breaking sector (Eq.~\ref{lsoft}).  Counting parameters, in the gauge sector we have the normal 3 gauge couplings $(e, g, g_s)$ and the QCD vacuum angle $\theta$ (Eq.~\ref{laxion}) of the SM, as well as 6 gaugino masses (\ref{gauginos}, \ref{CP-violating gauginos}).  One of the CP-violating masses can be removed following a field transformation, leaving 9 free parameters.  From the Higgs sector (\ref{Higgses}) there are two real squared masses $m^2_{H_u}$ and $m^2_{H_d}$, and one complex coefficient $b$.  Together with the complex parameter $\mu$ and the fact that one degree of freedom can again be removed by a field definition, this leaves 5 real parameters.  In the fermion sector, there are 5 Hermitian $3\times3$ mass-squared matrices (\ref{sfermion masses}), as well as 9 general complex $3\times3$ trilinear coupling matrices (\ref{trilinear couplings}, \ref{trilinear C couplings}, \ref{superpot}), giving $5\times9+9\times18 = 207$ real parameters.  Careful treatment of field redefinitions \citep{BaerTata} reduces this number by 43, leaving $5+9+207-43=178$ free parameters in the full $B$ and $L$-conserving MSSM, and $178-3\times18=124$ in the version where the `$\mathbf{C}$-terms' (\ref{trilinear C couplings}) are excluded.  Given 19 parameters in the SM, there are therefore either 159 or 105 more parameters in the full MSSM than the SM.

SUSY-breaking itself is a jungle, and even the zoo of viable breaking schemes is vast and varied \citep[see e.g.][for an introduction]{LutyTASI}.  Spontaneous SUSY breaking is hard to achieve at tree level, because none of the MSSM supermultiplets can easily acquire an appropriate VEV \citep[this is related to a particular sum rule for the tree-level masses; see e.g.][]{Martin,BaerTata}.  The solution seems to be for SUSY to be broken in a particle sector which has only very weak or indirect interactions with the MSSM, and then communicated to the MSSM by some heavy mediator particle $X$.  The mass at which SUSY is broken in the other sector is related to the mass of the mediator by
\begin{equation}
\frac{M^2_{\cancel{\mathrm{SUSY}}}}{M_X} \sim m_\mathrm{soft} \sim 1\,\mathrm{TeV}.
\end{equation}

The decoupled matter sector is often referred to by the unfortunate sobriquet of `the hidden sector', which (at least for this author) evokes notions of fine-tuning and careful sequestering away of a whole matter sector because experimentally, it should be `seen and not heard'.  In reality though, most hidden sectors required in SUSY-breaking scenarios have a very high characteristic mass scale.  We should therefore expect them to decouple from the low-energy theory as the influence of the heavy particles is integrated out, into mass dimension $d>4$ operators suppressed by $d-4$ powers of the large mass scale.  In that case, the `hidden sector' is then simply one which is phenomenologically disjoint (effectively hidden) from the SM because of its high energy scale.

Three scenarios have received the bulk of attention to date: gravity mediation \citep{Chamseddine82}, gauge mediation \citep{Giudice99} and anomaly mediation \citep{Randall99}.  In gravity-mediated scenarios, SUSY-breaking occurs in the matter sector associated with new physics around the Planck scale $M_\mathrm{P}$.  The breaking is hence mediated by particles with $M_X\sim M_\mathrm{P}$, leading to hidden sector SUSY-breaking at about 10$^{11}$ GeV.  The coupling between the hidden and observable sectors is naturally very weak because gravity is so weak, and in this case can actually even occur at tree level.  This is because SUSY-breaking in the gravitational sector actually implies that we are talking about supergravity (SUGRA), where SUSY has been made a local rather than a global symmetry, and gauged.  This happens to modify the tree-level mass sum rule in such a way that an appropriate VEV is possible to come by from within the MSSM.  It is important to note that gravity-mediation must always occur at some level if all interactions are supersymmetric, the issue is just whether or not it dominates over other contributions to SUSY breaking.  Aside from string theory, $N=8$ SUGRA is also the only known potentially-viable theory of quantum gravity \citep{Bern09}.

Gauge-mediated SUSY breaking (GMSB) takes an entirely orthogonal approach, where SUSY breaking is mediated by a particle which has both hidden sector and SM gauge couplings.  Actually, the mediator gets an entire sector of its own in many GMSB scenarios, known as the `messenger sector'; here the hidden sector breaks SUSY, passes it on to the messenger sector, which then communicates it to the observable sector.  The mediator conveys SUSY breaking radiatively from the messenger sector to the observable sector, coupling to the MSSM only at loop level.  In this case the mediator can be of almost any mass; if it is light, SUSY breaking in the messenger sector can occur at much lower masses than in the gravity-mediated scenario.  Anomaly-mediated SUSY breaking (AMSB) relies on extra dimensions and somewhat contrived brane geometries and dynamics to suppress gravity mediation to a level where loop contributions from a particular anomaly dominate the mediation. 

For the purposes of phenomenology, the full 124 or 178-parameter MSSM is not really very practical to work with.  Indeed, many of the terms should be suppressed anyway in order to agree with experiment: the $M'$ gaugino masses should be small or absent in order to prevent excessive $CP$-violation, as should the various $CP$-violating phases arising from complex MSSM parameters.  The off-diagonal entries in the mass matrices are constrained to be small by the level of flavour-changing neutral currents (FCNCs) they cause.  One approach to SUSY phenomenology is thus to work with the effective low-energy theory, but approximate all these dangerous terms to zero.  One can go a step further and assume reality and universality in the mass and trilinear coupling matrices, such that each is just a real constant times the identity matrix.  One could even be so bold as to assume that some of \emph{those} constants are equal to one another.  Many different low-energy parameterisations have been used in the literature: 6, 7, 8, 19, 24-parameter versions and so on, depending upon the individual authors' computational resources and the nuances of the specific observable under investigation.

The alternative is to take one of the scenarios for SUSY breaking, which invariably predict universality and reality in the structure of their soft terms at the scale of the mediator particles $M_X$.  Investigating the phenomenology then requires evolving the soft terms down to the weak scale using the renormalisation group equations (RGEs).  In the minimal supergravity (mSUGRA) model, all scalar masses, gaugino masses and trilinear couplings are unified at respective values $m_0$, $M_\mathrm{\frac12}$ and $A_0$, either at $M_\mathrm{P}$ or more commonly, $M_\mathrm{GUT}$.  The only remaining non-SM parameters are $\tan\beta$, the ratio of up-type to down-type Higgs VEVs at the weak scale, and the sign of the $\mu$ parameter.  The bilinear coupling $B\equiv b/\mu$ and the magnitude of $\mu$ are set by the condition that SUSY-breaking radiatively generates EWSB.  The parameter space is thus\footnote{Strictly speaking, what I describe here is the `constrained MSSM' (CMSSM).  According to some definitions, the mSUGRA model differs from the CMSSM, in that the condition of EWSB does not necessarily have to be imposed, making $\mu$ a parameter, $B$ is used instead of $\tan\beta$, and the relation $A_0=B+m_0$ is sometimes used to eliminate $B$ as a free parameter (or equivalently, $\tan\beta$ in versions where EWSB \emph{is} imposed).}
\begin{equation}
{ m_0, M_\frac12, A_0, \tan\beta, \mathrm{sgn}(\mu)}.
\label{mSUGRA}
\end{equation}

The minimal gauge-mediated model (mGMSB) is characterised by the parameter set
\begin{equation}
{ \Lambda, M, n_5, \tan\beta, \mathrm{sgn}(\mu), C_\mathrm{grav}}, 
\label{mGMSB}
\end{equation}
where 
\begin{equation}
\Lambda\equiv \frac{M^2_{\cancel{\mathrm{SUSY}},\mathrm{mess.}}}{\langle\mathcal{S}\rangle},
\end{equation}
with $M_{\cancel{\mathrm{SUSY}},\mathrm{mess.}}$ the SUSY-breaking scale in the messenger sector and $\langle\mathcal{S}\rangle$ the VEV of the scalar part of a gauge singlet superfield $\hat{\mathcal{S}}$ which also couples to the messenger sector.  $M$ is the mass scale of the messenger sector, which consists of $n_5$ generations of vector-like multiplets of quark and lepton superfields carrying SM charges.  $C_\mathrm{grav}$ is the gravitino mass parameter
\begin{equation}
C_\mathrm{grav} = \frac{M^2_{\cancel{\mathrm{SUSY}},\mathrm{hidd.}}}{\lambda M^2_{\cancel{\mathrm{SUSY}},\mathrm{mess.}}},
\end{equation}
where $M_{\cancel{\mathrm{SUSY}},\mathrm{hidd.}}$ is the SUSY-breaking scale in the hidden sector and $\lambda$ is the common messenger-sector Yukawa coupling between $\hat{\mathcal{S}}$ and the messenger quark and lepton supermultiplets.  The trilinear couplings are essentially absent in this model, so $\mathbf{A}_u=\mathbf{A}_d=\mathbf{A}_e=0$.

The minimal version of AMSB (mAMSB) is specified by
\begin{equation}
{ m_0, m_\frac32, \tan\beta, \mathrm{sgn}(\mu)}, 
\label{mAMSB}
\end{equation}
where $m_\frac32$ is the gravitino mass, which is also ultimately responsible for determining the gaugino masses and the trilinear couplings.  The other three parameters are defined identically to their mSUGRA counterparts.

\subsection{SUSY scanning}
\label{susyscanning}

In order to compare any particular version of SUSY with experimental data, one generally needs to consider a number of different points in the model parameter space, covering a large range of each of the free parameters.  One then scans over the parameters using some sort of search algorithm, evaluating the values of the relevant observables at each point.  These can then be compared with existing data to determine if the point is consistent with experiment, or to make predictions for future experiments.

With such high-dimensional parameter spaces, and a very nonlinear mapping from some of the model parameters to actual observables like cross-sections, it is no surprise that scanning SUSY parameter spaces is not a simple exercise.  This is especially so if one wants to make meaningful statements about which regions of the parameter space are allowed or disallowed, and to what degree of statistical significance.  Traditional analyses simply assigned a status of either `allowed' or `excluded' to a point if it lay within or beyond a certain confidence level of the measured data.  These were often based on brute-force random scans, with little concern for convergence or any sort of statistical interpretation.  The modern approach is to consider the full likelihood for each point, and analyse the resultant map in terms of either a Bayesian or frequentist statistical framework.  These scans have often also employed sophisticated scanning techniques like Markov Chain Monte Carlos (MCMCs), nested sampling or genetic algorithms.  The reader is referred to p.~3 of \hyperref[papV]{\Akrami} for a detailed background on the differences between the Bayesian and frequentist approaches in this context.  Details of the differences between scanning techniques and a review of significant work in SUSY scanning can be found on p.~3 of \hyperref[papIV]{\ScottIV} and pp.~4--5 of \hyperref[papV]{\Akrami}.

In particular, the most well-known products of SUSY scanning are the various regions of mSUGRA compatible with the relic density from the CMB.  These regions, as well as benchmark points drawn from them, have been widely used for collider studies and design over the last 20 years.  A short review of these regions is given on p.~3 of \hyperref[papIV]{\ScottIV}.

\section{Gravitinos and Axinos}
\label{gravitinos}

If supersymmetry is made local and extended to supergravity, the graviton acquires a spin-$\frac32$ superpartner, the gravitino.  In gravity-mediated SUSY-breaking scenarios, the gravitino obtains a similar mass to the other sparticles, i.e.~$m_\mathrm{soft}$, whereas in gauge-mediated schemes it has a mass of the order of a few keV \citep{Martin}.  It can therefore be the LSP in some cases, the next-to-lightest SUSY particle (NLSP) in others, or just another more massive member of the SUSY spectrum.  

If it is the LSP and $R$-parity is conserved, the gravitino is stable, so it can constitute a viable but depressing dark matter candidate.  This is because it interacts only gravitationally with the rest of the spectrum, easily fulfilling the collisionless, non-dissipative and electrically-neutral requirements of a good dark matter candidate -- but the very weakness of the interaction means that it is however virtually undetectable.  If the gravitino is the NLSP, the weakness of its interactions can make it meta-stable, potentially leading to non-thermal LSP dark matter production by decay.  The caveat is that the long lifetime can interfere with later-time processes like BBN, depending upon which SM particle the gravitino predominantly decays into along with the LSP.  The same is also true if the gravitino is the LSP, as in this case the weakness of its interactions again make the NLSP long-lived.  Their long lifetimes mean that gravitinos can be viable dark matter candidates even if $R$-parity is violated, since they would again be meta-stable.  Because thermal gravitinos decouple from the primordial bath around the Planck mass, their cosmology is particularly complicated and uncertain; they could constitute either warm or cold dark matter, depending upon their mass and cosmological history.

If one extends the SM to include the axion (cf.~Sect.~\ref{axions}) as a solution the strong CP problem, then the supersymmetrised version of the theory of course also contains a spin-$\frac12$ axino \citep[see e.g.][]{CoviKim}.  Like the gravitino, the axino is extremely weakly interacting, so would be perfectly viable as dark matter if it were the LSP.  In contrast to the gravitino though, the axino should at least be detectable indirectly though the observation of the axion itself.  Because of its ultra weak coupling to other particles, if it is the LSP or NLSP the axino might have any of the same features and issues as the gravitino with regard to long-lived states.

\section{Sterile neutrinos}
\label{neutrinodm}

The fact that neutrinos have been observed to oscillate indicates that they have masses, though experimental limits constrain them to be sub-eV.  This presents quite a fine-tuning problem, as the masses must be many orders of magnitude less than any other known massive particles.  The favoured mechanism for producing such small masses is known as the see-saw, which operates by introducing heavy singlet (right-handed) neutrino states of mass $\sim$$M$.  The observed neutrino masses become inversely proportional to $M$ when the mass matrix is diagonalised.  The singlet neutrinos do not interact with other particles because they carry no charges under any of the SM gauge groups.  Being massive fermions, the actual neutrino mass eigenstates differ from the weak eigenstates; the splitting is sufficiently large that the light neutrinos are almost entirely left-handed in character, and the heavy ones almost entirely singlet.  This makes the heavy states very weakly interacting, or `sterile', and sufficiently long-lived to constitute dark matter.

$M$ can in principle take on almost any value (there are valid naturalness arguments for it being around $M_\mathrm{GUT}$, or for being rather small).  If it is in the keV mass range, then the sterile neutrino can be a viable warm dark matter candidate, with the correct relic abundance \citep[see e.g.][for a recent review]{Kusenko09}.  Depending on the production mechanism in the early Universe, it is also possible to make sterile neutrinos cold(ish) dark matter.  

\section{Other candidates}
\label{others}

I will not attempt to detail the myriad other viable dark matter candidates which exist; this subsection merely provides a summarising `sound bite' on some of the more notable ones, and some references for further reading.

A range of `WIMP-derivative' models build upon the basic idea of WIMP dark matter, or at least make use of a number of its desirable features. WIMPless dark matter \citep{FengKumar08,Feng10} employs the WIMP miracle in a hidden sector, by tuning the WIMP mass and interaction strength to achieve the desired relic density, \`a la Eq.~\ref{wimpmiracle}.  WIMPzillas \citep{Kolb99} are superheavy, weakly-interacting particles formed out of thermal equilibrium, arising from e.g.~the GMSB messenger sector.  Minimal Dark Matter \citep{Cirelli06,Cirelli09} involves the trial-and-error addition of any colourless $SU(2)_L$ multiplet with a neutral lightest member to the SM, in the search for a WIMP.  Inelastic Dark Matter \citep[iDM;][]{SmithWeiner01,TuckerSmithWeiner05} and eXciting Dark Matter \citep[XDM;][]{FinkbeinerWeiner07} are WIMPs with excited states, nowadays often paired with models exhibiting Sommerfeld-enhanced annihilation \citep{AHDM,NTDM}.  These models are all designed to explain specific observational anomalies.  

SIMPs \citep{Starkman90} are Strongly Interacting Massive Particles which could form colourless bound states \citep{Kang08} and hide their strong interactions, whilst milli-charged particles \citep{Holdom86,Davidson00} might manage to appear dark because they carry only a very small fractional electric charge.  Both these options are very strongly constrained at the present time \citep{Taoso08_10pt}.

\chapter{Searches for dark matter}
\label{dmsearches}

Dark matter can be sought in a number of complimentary ways.  In Sects.~\ref{directdetection}--\ref{solarneutrinos} I give an overview of the main techniques used to search for WIMPs, followed in Sect.~\ref{othersearches} by a quick exposition of other tests, including those for non-WIMP models.  It is unlikely that any single search technique will ultimately be enough for us to confidently characterise or completely exclude a particular dark matter model; combining results from different searches is precisely the strategy adopted in \hyperref[papIV]{\ScottIV} and \hyperref[papV]{\Akrami}, and discussed in Sect.~\ref{susyscanning} in the context of SUSY.

\section{Direct detection}
\label{directdetection}

Because WIMPs interact with SM particles by the weak force, they should have weak-scale scattering cross-sections with normal nuclei.  One of the most promising ways to detect WIMPs is thus to look for nuclear recoils in large-volume target materials on Earth \citep{GoodmanWitten85}.  

The expected number of WIMP-nucleon scattering events $dN$ per nuclear recoil energy window $dE_\mathrm{r}$ is given \citep{Gaitskell04} by 
\begin{equation}
\frac{dN}{dE_\mathrm{r}} = \frac{\sigma\rho}{2\mu^2m_{\chi}}F^2\int_{v_\mathrm{min}(E_\mathrm{r})}^{v_\mathrm{esc}}\frac{f(v)}{v}dv,
\end{equation}
where $\sigma$ is the WIMP-nucleus cross-section, $\rho$ the local WIMP density, $m_\chi$ the WIMP mass and $F$ the nuclear form factor.  The WIMP-nucleus reduced mass is $\mu \equiv (m_\chi m_\mathrm{nuc})/(m_\chi+m_\mathrm{nuc})$.  The distribution of WIMPs in the halo with velocities $v$ is given by $f(v)$, which is integrated over all possible velocities.  These range from $v_\mathrm{min}(E_\mathrm{r})$, the minimum velocity necessary to produce a recoil of energy $E_\mathrm{r}$, to the halo escape velocity $v_\mathrm{esc}$.  Nuclear recoil searches attempt to measure $\frac{dN}{dE_\mathrm{r}}$ directly.  By assuming a certain halo model, this then allows a relation to be inferred between the WIMP mass and cross-section.  The standard halo model assumes a Maxwellian velocity distribution with $v_\mathrm{RMS}\approx220$\,km\,s$^{-1}$ and a local density of 0.3\,GeV\,cm$^{-3}$; alternative halo models and their influence upon direct detection results have also been discussed \citep[e.g.][]{Belli02, Stiff03, Green07, Green08, Read09}.

The cross section has two parts: a spin-independent interaction $\sigma_\mathrm{SI}$ between WIMPs and all nucleons, and a spin-dependent component $\sigma_\mathrm{SD}$ coupling only to nucleons with net spin.  The kinematics of WIMP-nucleon collisions mean that both parts are proportional to the square of the WIMP-nucleon reduced mass.  This dependence is generally suppressed in practice however, as direct detection experiments are never sensitive to the full WIMP recoil spectrum.  The spin-independent part has a further dependence on the square of the nuclear mass.  The spin-dependant component instead depends upon the nuclear spin $J$ and the spins of the individual proton and neutron subsystems in the nucleus (see e.g.\ Eq.\ 2.8 of \hyperref[papII]{\ScottII}, or \citealt{Jungman96,DarkSUSY,CerdenoGreen10}).  Target nuclei with different isotopic compositions can be chosen to optimise an experiment for spin-dependent or spin-independent searches.

Three physical consequences of nuclear recoils are used to search for evidence of WIMP scattering.  One is ionisation of target atoms caused by energy transfer from the recoiling nucleus.  Another is fluorescent radiation given off by electrons of target atoms, as they decay after having been excited by energy transfer from the recoiling nucleus.  Materials known as scintillators, which are transparent at their fluorescent wavelength and have a very fast decay time, are used extensively for this purpose; in this case, the process is referred to as scintillation.  The third technique is to measure phonon excitations generated in crystals by the nuclear recoils, where minute heat changes corresponding to the absorption of individual vibrational quanta are measured.  Various combinations of these detection methods are employed in different experiments.

Due to the proper motion of the solar system within the galactic halo, there should be a mean net velocity between WIMPs and the Earth.  This should result in both a diurnal modulation of WIMP-nucleus collisional directions, and an annual modulation of the total detection rate due to the non-perpendicularity of the solar ecliptic and galactic planes.

\begin{figure}[t]
\centering
\begin{minipage}[t]{0.59\textwidth}
  \vspace{0mm}
  \centering
  \includegraphics[width=\linewidth]{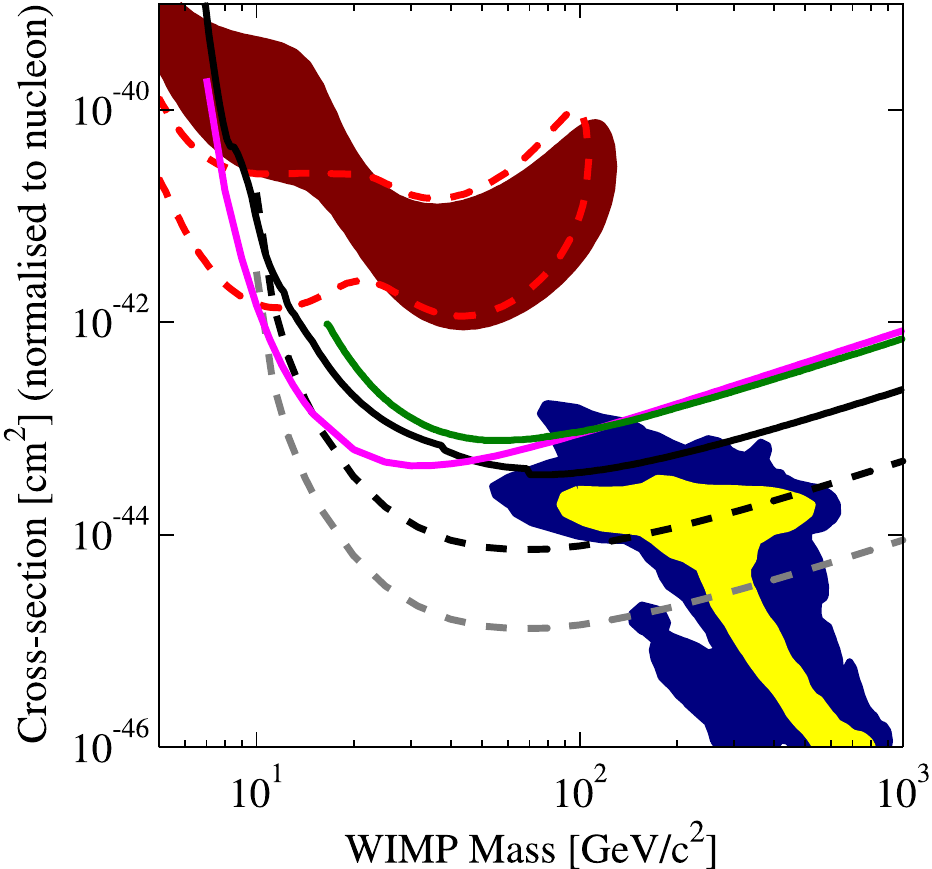}
\end{minipage}%
\begin{minipage}[t]{0.1\textwidth}
  \vspace{0mm}
  \centering
  \includegraphics[width=\linewidth]{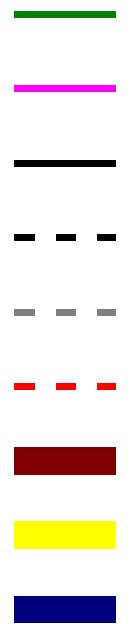}
\end{minipage}%
\begin{minipage}[t]{0.31\textwidth}
  \vspace{0mm}
  \raggedright
  \tiny
  ZEPLIN III \protect\citep[Dec 2008;][]{zeplinDec08}\\\vspace{1.8mm}
  XENON 10 \protect\citep[2007 result, 136\,kg\,d exposure;][]{Angle08a}\\\vspace{1.8mm}
  CDMS II \protect\citep[Soudan 2004-2009, Ge target;][]{cdms2event}\\\vspace{1.8mm}
  SuperCDMS, projected \protect\citep[2 towers, Soudan;][]{supercdms}\\\vspace{1.8mm}
  SuperCDMS, projected \protect\citep[7 towers, SNOLab;][]{supercdms}\\\vspace{1.8mm}
  DAMA/LIBRA, with ion channelling \protect\citep[$5\sigma$;][Fig.~11]{Savage09b}\\\vspace{0.5mm}
  DAMA/LIBRA, no ion channelling \protect\citep[$5\sigma$;][Fig.~10]{Savage09b}\\\vspace{1.8mm}
  CMSSM Bayesian posterior \protect\citep[68\% contour;][]{Trotta08}\\\vspace{1.8mm}
  CMSSM Bayesian posterior \protect\citep[95\% contour;][]{Trotta08}
\end{minipage}%
\caption{Exclusion curves, fit regions and SUSY predictions for WIMP masses and spin-independent nuclear scattering cross-sections.  The areas above each curve are excluded by the respective experiments, listed to the right.  The red regions are claimed detections of an annual modulation signal by the DAMA collaboration, under different assumptions about channeling by the detector crystal.  The yellow and blue regions are Bayesian posterior predictions within the CMSSM, taking into account all other experimental constraints.}
\label{fig_DDresults}
\end{figure}

Exclusion limits placed upon the WIMP mass and spin-independent nuclear scattering cross-sections by various direct detection experiments are shown in Fig.~\ref{fig_DDresults}.  The plot also shows a theoretical prediction from a Bayesian CMSSM parameter scan \citep[blue and yellow solid region;][]{Trotta08}, as well as the area compatible \citep[red solid region]{Savage09b} with the claimed detection of an annual modulation signal by the DAMA collaboration \citep{Bernabei00,Bernabei08}.  Other experiments appear to disfavour a dark matter interpretation for this signal.  However, the solid compatible region on this plot has been drawn assuming standard WIMP, halo and target models; the annual modulation signal can be made marginally compatible with other experiments if these assumptions are relaxed \citep[e.g.][]{Savage09a,Savage09b,Fairbairn09}.  An example of how the compatibility region changes when channelling in the detector crystals is included \citep{Savage09b} is outlined in red dashes.  A similar story holds for interpretation of the DAMA signal in terms of spin-dependent WIMP scattering (not shown).  Predicted sensitivity curves for upcoming experiments are also shown, showing that a substantial part of the CMSSM parameter space should become accessible in the near future.

\section{Indirect detection}
\label{indirectdetection}

If WIMPs are thermal relics, they should generically possess weak-scale self-annihilation cross-sections.  This is true whether the particles are Majorana or Dirac.  In the Dirac case, being a thermal relic implies that there is no net matter-antimatter asymmetry, since any excess of WIMPs or anti-WIMPs would cause the relic abundance to be determined by the asymmetry rather than thermal production.  If dark matter is produced non-thermally, it could also have a comparable, or even higher, self-annihilation cross-section.  Except in the case of a non-thermally produced Dirac particle with an initial matter-antimatter asymmetry, WIMPs should therefore annihilate at a non-vanishing rate today.

Indirect detection methods aim to detect the primary or secondary products of these annihilations, in the form of photons, neutrinos or other cosmic rays.  The same techniques can also be used to search for the products of decaying dark matter, such as the LSP in versions of SUSY which weakly violate $R$-parity.  Searches place constraints on annihilation cross-sections in annihilating models (and therefore the relic density, in ones where the particle is thermally-produced), and on the lifetime in decaying models.  As a two-body process, the annihilation rate is proportional to the square of the dark matter density, whereas the single-body decay process is proportional to the first power of the density.  

Whether searching for annihilations or decays, the most promising targets are those with large dark matter densities and/or low astrophysical backgrounds.  The Galactic Centre (GC) would seem the most obvious target given its distance and dark matter concentration, but it is also one of the most difficult areas to work with because of its complex and poorly-understood background \citep{Vitale09,HESS10_GC}, and uncertain dark matter profile \citep{Stoehr03,Merritt10}.  Better prospects might be had just outside the GC, in a broken annulus which excludes the galactic disc \citep{Stoehr03, Serpico08}.  Dwarf galaxies are a good option because they are extremely dark-matter-dominated, leading to a low background, but have the disadvantage of also having rather low predicted fluxes \citep[e.g.][]{Bringmann09,Pieri09,Martinez09}.  Smaller, previously unidentified clumps of dark matter with no association to any known astrophysical sources (e.g.~\citealt{Green04b}; \citealt{Green05}; \citealt{Kuhlen08}; \citealt{Bringmann09_clumps}; \hyperref[papIII]{\SSiv}) would make for a spectacular, background-free discovery if any were detected, but there is significant theoretical uncertainty in their expected number, mass and proximity to Earth.  Galaxy clusters and the total integrated contribution to the extragalactic background have also been considered competitive targets.

The relative merits of the different targets depend upon whether limits are sought on annihilation or decay, and the observation channel.  Massive charged particles like electrons are attenuated and deflected by magnetic fields, causing them to almost always arrive isotropically.  Further discussion of the uncertainties in dark matter density profiles and the astrophysical production and propagation of cosmic rays can be found in Chapter \ref{nuisances}.

The expected differential primary gamma-ray flux per unit solid angle \citep[e.g.][]{Bergstrom98} from WIMP annihilations is  
\begin{equation}
\frac{\mathrm{d}\Phi}{\mathrm{d}E\mathrm{d}\Omega} = \frac{1 + BF}{8\pi m_\chi^2} \sum_f \frac{\mathrm{d}N^\gamma_f}{\mathrm{d}E}\sigma_f v \int_\mathrm{l.o.s.} \rho_\chi^2(l) \mathrm{d}l,
\label{fluxeqn}
\end{equation}
where $BF$ is the boost factor from unresolved substructure in the source, $f$ labels different final states, $\mathrm{d}N^\gamma_f/\mathrm{d}E$ is the differential photon yield from any particular final state, $\sigma_f$ is the cross-section for annihilation into that state, $v$ is the WIMP relative velocity, and the integral runs over the line of sight to the source.  In the absence of any bound states (i.e. Sommerfeld enhancements), WIMPs move so slowly that they can effectively be considered to collide at rest, allowing $\sigma_f v$ to be replaced with the velocity-averaged term in the zero-velocity limit, $\langle\sigma_f v\rangle_0$.

In the case of neutralino annihilation, three main channels contribute to the gamma-ray spectrum.  Through loop processes, annihilation can proceed directly into two photons \citep{Bergstrom88,Bergstrom97}
\begin{equation}
\frac{\mathrm{d}N^\gamma_{\gamma\gamma}}{\mathrm{d}E} = 2\delta(E - m_\chi),
\end{equation}
or into a Z boson and a photon \citep{Ullio98}
\begin{equation}
\frac{\mathrm{d}N^\gamma_{Z\gamma}}{\mathrm{d}E} = \delta(E - m_\chi + \frac{m_Z^2}{4m_\chi}).
\end{equation}
This gives a monochromatic gamma-ray line.  Because no other known process produces such a line, this would be a smoking gun signal for WIMP dark matter.  Unfortunately, the loop suppression means that very few models actually have substantial branching fractions into monochromatic photons.  A hard spectrum can also be produced by internal bremsstrahlung (final-state radiation plus virtual internal bremsstrahlung), generated when a photon is emitted from a virtual particle participating in the annihilation \citep{Bergstrom89, Bringmann08}.  Continuum gamma-rays can also be produced by annihilation into quarks, leptons and heavy gauge bosons (including the $Z$ from the $Z\gamma$ line), which subsequently decay via $\pi^0$ and the emission of bremsstrahlung to softer photons.

Indirect detection with photons is currently dominated by large air-\v{C}erenkov gamma-ray telescopes (ACTs) and the \emph{Fermi} Large Area Telescope \citep[LAT;][]{Atwood09}, a pair-conversion gamma-ray space telescope.  Limits on the total cross-section from dwarf galaxies (\citealt{Lombardi09,Essig09}; \hyperref[papIV]{\ScottIV}; \citealt{LATdwarfpaper}), the isotropic diffuse background \citep{LATcosmowimp,Abazajian10} and galaxy clusters \citep{LATclusterpaper} are approaching the canonical thermal cross-section ($3\times10^{-26}$\,cm$^3$\,s$^{-1}$; cf.~Sect.~\ref{thermalrelics}), but cannot yet conclusively exclude it for any range of WIMP masses.  LAT limits on annihilation into gamma-ray lines also exist \citep{LATlinepaper}.  Claims have been made from outside the collaboration of excesses in public \emph{Fermi} data in the inner Galaxy \citep{Goodenough09,Dobler09}, which some have interpreted as signals of WIMP annihilation.  Their status as true excesses rather than instrumental foibles or systematics arising from overly simplified background modelling \citep[e.g.][]{Linden10} have yet to be confirmed or disproved by the LAT collaboration.

Gamma-ray and longer-wavelength observations also make it possible to hunt for secondary photons from WIMP annihilation or decay.  In particular, inverse Compton scattering of CMB and interstellar radiation field photons by primary leptons injected in annihilations or decays can lead to substantial signals in gamma rays and X-rays \citep{Regis08,Cirelli09,Profumo09,Belikov10,LATcosmowimp,LATclusterpaper}.  Synchrotron emission from primary leptons in regions with significant magnetic fields can also be a competitive probe \citep{Regis08,Bertone09,Bergstrom09_MW}, as can modifications of the CMB by particle injection from dark matter annihilation at early times \citep{Slatyer09,Galli09}.

Recent electron and positron cosmic ray data from \emph{Fermi} \citep{Fermielectron}, the PAMELA satellite \citep[Payload for Anti-Matter Exploration and Light-nuclei Astrophysics;][]{pamelapositron} the ATIC balloon mission \citep[Advanced Thin Ionisation Calorimeter;][]{Chang08} and the HESS ACT \citep[High Energy Stereoscopic System;][]{HESSelectron} indicate a slight excess in the total number of events at $\sim$10\,GeV--10\,TeV over that expected purely from background.  A substantial excess is also seen in the fraction of events due to positrons.  This has lead to a flood of papers explaining the data in terms of dark matter annihilation or decay \citep[e.g.][]{Bergstrom08,AHDM,NTDM,Chen09,Donato09}.  Almost all annihilation scenarios require a substantial boost to the annihilation cross-section above that expected for a typical thermal relic, either from substructure or specific particle models (these are the Sommerfeld-enhanced models mentioned in Sect.~\ref{others}).  Such boosted models are put under rather severe pressure by limits from recent gamma-ray observations \citep[e.g.][]{LATdwarfpaper, LATcosmowimp, LATclusterpaper, Abazajian10}.  The positron excess can also be explained in terms of conventional astrophysical sources like pulsars \citep{Profumo08,Yuksel09,Hooper09} and supernova remnants \citep{Blasi09,Piran09,Fujita09}, and possibly even standard secondary production \citep{Katz09}.

In many models, WIMP annihilation produces a substantial number of antiprotons.  Observations of cosmic ray antiprotons by PAMELA \citep{PAMELAantiproton} show no excess above the expected astrophysical background.  This places rather severe limits on the types of models which can be invoked to explain the PAMELA positron excess, as they must not overproduce antiprotons.  Together with the steepness of the rise observed in the positron fraction, this suggests that such models should annihilate predominantly to lepton-antilepton pairs \citep[so-called `leptophilic' dark matter;][]{Cholis09,Bergstrom09,Bergstrom09b}, which is somewhat difficult to achieve in the MSSM \citep{Bergstrom08}.

Antideuterons are another promising indirect detection channel.  Although predicted yields from dark matter annihilation \citep{Donato00,Baer05} and decay \citep{Ibarra09} are certainly smaller than those for antiprotons, the very low backgrounds expected \citep{Donato08} could make antideuterons a useful detection method.  This is especially true for low WIMP masses, where kinematics would prevent background antideuteron formation from spallation.  Future experiments, such as the imminent Alpha Magnetic Spectrometer space shuttle mission \citep[AMS-02;][]{Choutko08} and the dedicated balloon mission GAPS \citep[General AntiParticle Spectrometer;][]{Hailey09}, should make interesting inroads into model parameter spaces.

In certain models, such as Kaluza-Klein dark matter in Universal Extra Dimensions (UED), WIMPs can annihilate directly into neutrinos.  Other models produce neutrinos in secondary decays and/or cascade interactions with baryonic matter.  Whilst their weak interactions, small masses and atmospheric background make neutrinos a difficult indirect detection prospect, they have the advantage of pointing directly back to a source in the same way photons do.  With observations of the GC \citep{Bertone04} and the inner Galactic halo \citep{Yuksel07}, the SuperKamiokande neutrino telescope has been used with some success to constrain the annihilation cross-section of traditional WIMP models.  Upcoming cubic-kilometre telescopes such as IceCube should also prove powerful enough to essentially rule out Sommerfeld-enhanced leptophilic models, through observations of the central regions of our own Galaxy \citep{Hisano09,Liu09,Buckley10,Mandal10} or dwarfs \citep{Sandick09}.

Indirect detection could also include dark stars and searches for neutrinos from WIMP annihilation in the Sun.  Unlike the indirect searches just described, stellar probes depend more upon nuclear scattering than the annihilation cross-section, so I will describe them separately (in Sects.~\ref{darkstars} and \ref{solarneutrinos}, respectively).

\section{Accelerator searches}
\label{acceleratorsearches}

Finding dark matter at accelerators is unlikely to be straightforward.  The most obvious collider WIMP signature is expected to be missing transverse energy (missing $E_\mathrm{T}$), which refers to an apparent missing component of the total final-state momentum in the direction transverse to a collider beam.  If the outgoing transverse momenta of a reaction do not sum to zero (as they do in the original beam), this can be attributed to the production and escape of a massive particle with a very small interaction cross-section with the detector material.  The catch is that whilst the particle must be stable enough to travel beyond the detector, there is no way of to know how stable it is on cosmological timescales.  There would hence be no way to infer its relic abundance.

By looking at kinematic endpoints in the momentum distributions of observed particles, one can derive the masses of intermediate states in the decay chain leading to the missing $E_\mathrm{T}$, including the mass of the missing particle itself \citep{Battaglia04,White07}.  By carefully examining the shapes of the distributions, in principle one can also distinguish between different theories giving rise to the same (observable) decay products.  The difficulty with this prospect is that in a hadron collider like the LHC (Large Hadron Collider), collisions take place between bound-state quarks.  It is therefore impossible to ever know exactly where the rest frame of each interaction is.  Whilst there should be a known distribution of rest frames across all collisions, this still introduces a significant uncertainty to the distributions for rare processes like those that would produce WIMPs, and also makes efforts to disentangle the spins of the new particles very difficult \citep{Baltz06}.

The main alternative is to attempt to reconstruct the whole underlying theory responsible for the new TeV-scale physics, by using other channels (than missing $E_\mathrm{T}$) to constrain the masses and couplings of other new particles.  In this way, one might infer the mass, spin and cross-section of the dark matter particle(s) without measuring them directly.  A more reliable option would be to employ a lepton collider like the proposed ILC (International Linear Collider), which should allow direct pair production of WIMPs if they exist in the appropriate mass range.  This would allow direct measurements of the spin, couplings and mass of the particle, though the issue of proving stability would remain \citep{Baltz06}.  The couplings could however be measured well enough to determine a particle's relic density to a similar level of accuracy as from the CMB \citep{Battaglia09}; if the two matched, stability would be strongly implied.

Another interesting alternative allowed by a lepton collider would be to search for initial state internal bremsstrahlung in dark matter pair production \citep{Birkedal04}.  If dark matter were a thermal relic, the event rate would be related to the relic density via the production/annihilation cross-section.  Some signal should thus be expected for any thermally-produced WIMP, assuming a non-vanishing annihilation branching fraction into the particular leptons used.

In general, an unequivocal identification and characterisation of the particle responsible for dark matter will require verification from a range of different searches.  Accelerators will help most in characterising the theory to which it belongs, whilst direct and indirect searches should provide information about its stability and cosmological abundance.  All can contribute to determining the mass and couplings.

\section{Dark stars}
\label{darkstars}

If dark matter annihilates at a non-negligible rate today, we might expect that the energy from those annihilations could impact nearby baryonic systems.  This might contribute to heating of the intergalactic medium \citep{Ripamonti07}, reionisation \citep{Natarajan08a}, the CMB temperature fluctuations \citep{Slatyer09,Galli09}, or the energy budget of stars \citep{Steigman78,BouquetSalati89a,SalatiSilk89}.  Stars whose structure or evolution are affected in this way are often referred to as `dark stars'.

Stars might obtain a substantial amount of dark matter in their cores by nuclear scattering, gravitational contraction, or both.  In the same way that we expect WIMPs to collide with nuclei in direct detection apparatuses (Sect.~\ref{directdetection}), we should also expect them to scatter on nuclei in astrophysical objects.  Thus, in the first case WIMPs could scatter on nuclei in stars, lose sufficient energy to become gravitationally bound, and return to repeat the process \citep{Press85,Gould87b} until they either thermalise with the stellar core or annihilate with each other.  In the second case, steepening of the gravitational potential caused by dissipative collapse of a baryonic gas cloud could draw dark matter into a star during its formation \citep{Spolyar08,Natarajan08b,Freese08b}.  In either case, conductive heat transport by WIMP-nucleon scattering could in principle also affect the stellar structure or evolution, but current direct detection bounds on the scattering cross-sections make this unlikely (see e.g. \citealt{Bottino02}; \hyperref[papII]{\ScottII}).

The effects upon different types of stars in a variety of locations have been considered in recent years: white dwarfs at the GC \citep{Moskalenko07,Hooper10} and in globular clusters \citep{Bertone07,McCullough10}, population III stars (\citealt{Spolyar08,Iocco08a,Iocco08b,Natarajan08b,Freese08c,Freese08a,Freese10}; \hyperref[papIII]{\Zackrisson}) and main sequence stars near the GC (\hyperref[papI]{\Fairbairn}; \citetalias{Scott08a}; \hyperref[papII]{\ScottII}; \citealt{Casanellas09}).  A review of the historical and recent developments in this field can be found in Sect.~1 of \hyperref[papII]{\ScottII}.  The two different dark star simulation strategies employed by different groups are described in Sect.~2.1 of \hyperref[papIII]{\Zackrisson}.  A technical treatment of WIMP capture by scattering, and the subsequent impacts upon stellar evolution can be found in Sect.~2 of \hyperref[papII]{\ScottII}.  For constraining dark matter properties, dark stars are most useful for putting limits on either nuclear scattering cross-sections or the growth of dark matter halos.

\section{High-energy solar neutrinos}
\label{solarneutrinos}

In the same way that WIMPs would scatter on nuclei in dark stars and become gravitationally captured, so they should in the Sun as well.  The primary difference in this case is that the expected capture rates are far lower in the Sun than at the GC or in early proto-halos, mainly due to the comparatively low dark matter density in the solar system.  Although annihilation of WIMPs in the solar core should have essentially no impact on the Sun's structure or evolution, some fraction of the annihilation energy might escape as neutrinos.  

These $\mathcal{O}$(GeV) neutrinos would be much more energetic than the $\mathcal{O}$(MeV) solar neutrinos from nuclear fusion \citep[e.g][]{Ahmad01}, so would be clear evidence of WIMP dark matter.  The only known solar neutrino background at such high energies comes from cosmic ray interactions with the corona, and is expected to be low \citep{Seckel91,Ingelman96}.  The main background concerns are atmospheric muons and neutrinos, produced by interactions of cosmic rays with the Earth's atmosphere.  This background can be mostly avoided by triggering only on upwards-going events, at times when a terrestrial neutrino telescope is pointing away from the Sun \citep[e.g.][]{IceCube09, IceCube09_KK}.

The capture and annihilation of WIMPs in the Sun has been used together with telescopes such as SuperKamiokande \citep{Desai04}, AMANDA \citep{AMANDA06}, ANTARES \citep{Lim09} and IceCube \citep{IceCube09, IceCube09_KK, Flacke09, Blennow10} to place limits on WIMP-nucleon scattering cross-sections.  Because the Sun consists predominantly of hydrogen, the most competitive limits from neutrino telescopes are on the spin-dependent cross section.  Spin-dependent limits from IceCube are already stronger than those from direct detection \citep{IceCube09, IceCube09_KK}, and should improve significantly as the DeepCore section of the detector is added \citep{Ellis09}.

Neutrino telescope limits from the Sun are considerably more model-dependent than those from direct detection, as the only way to calculate the expected neutrino yields is to have a particular model for the annihilation branching fractions of the WIMP \citep[see e.g.][]{Wikstrom09}.  Some channels give rise to more neutrinos than others, producing different limits depending upon the assumed annihilation channel.  KK dark matter for example can annihilate directly into monochromatic neutrinos, whilst neutralinos produce neutrinos from a series of secondary interactions between stellar nuclei and daughter quarks and gauge bosons \citep[e.g.][]{Blennow08}.  The expected signal from neutrino telescopes also depends critically on the dark matter velocity distribution in the solar neighbourhood \citep[e.g.][]{Bruch09}; capture is far more efficient for WIMPs which arrive at the Sun with lower velocities, because they need not lose as much energy in the scattering process as faster WIMPs to become gravitationally bound.  Perturbations of WIMP orbits by planets \citep{Gould91,Lundberg04,Peter09} adds further uncertainty to the expected capture rates for a given cross-section.

\section{Searches for other dark matter candidates}
\label{othersearches}

Because axions in electromagnetic fields can convert into photons and vice versa (Sect.~\ref{axions}), the best places to search for them are in objects with strong magnetic fields.  The Sun is a good target, as some fraction of the photons created in the photosphere will convert to axions as they pass through the strong magnetic fields of the chromosphere and corona.  The CAST \citep[CERN Axion Solar Telescope,][]{CAST09} experiment searches for these axions by attempting to observe the Sun through an opaque screen and a very strong magnet, in the hope that some of the solar axions will convert back to photons in the local magnetic field, after having passed through the screen (which the solar photons could not).

A similar idea is pursued in the `light shining through a wall' experiments \citep[e.g.][]{Robilliard07,Pugnat08}, where a photon source is shone against a barrier after passing through a strong magnetic field.  A photodetector and a second magnet on the other side of the wall look for photons regenerated in the magnetic field from axions that have passed through the wall.  The same strategy could be used to detect axions by looking at background X-ray sources through the Sun \citep{Fairbairn07}. Cavity experiments like ADMX \citep[Axion Dark Matter eXperiment,][]{ADMX10} attempt to detect halo axions, by tuning the frequency of a magnetic field to the axion mass and searching for microwaves from resonant conversion.  Astrophysical limits on axions can be obtained by considering the maximum amount of energy that they could carry out of supernovae and stellar cores without exceeding observed cooling rates \citep[see e.g.][for an up-to-date review]{Raffelt08}.

One of the more promising ways to search for sterile neutrinos is via X-ray line emission produced in the loop decay $\nu_\mathrm{s} \to \gamma \nu$.  Such a signal has been claimed in \emph{Chandra} observations of the ultra-faint dwarf galaxy Willman I \citep{Lowenstein10}, but at less than the $2\sigma$ level.

Microlensing searches might also still have some distance to run in determining the identity of dark matter.  This is because ultracompact dark matter minihalos (\citealt{Ricotti09}, \hyperref[papIII]{\SSiv}) could constitute non-baryonic MACHOs.  These would apparently have escaped microlensing searches because they are slightly more extended than traditional MACHOs, but might appear in future surveys \citep{Ricotti09}.

\chapter{Nuisances}
\label{nuisances}

When searching for dark matter, uncertainties in a range of necessary input data can influence results.  The most relevant of these are the assumed distribution of dark matter, and the backgrounds to which each search is subject.  Backgrounds are comparatively well controlled and understood in terrestrial direct detection and accelerator experiments, but not in indirect searches in astrophysical targets.

In the types of parameter-scanning exercises discussed in Sect.~\ref{susyscanning}, uncertainties in input data can be fully included in the analysis if they have been statistically quantified.  In this case the uncertain quantities are referred to as `nuisance' parameters, because they are parameters one is not actually interested in, but that effect results nonetheless.  Here I give a very brief synopsis of the two most dangerous nuisances in dark matter searches.  I also provide a brief summary of the most important nuisance parameters in particle physics, which necessarily arise from the SM.

\section{The distribution of dark matter}
\label{dmdistribution}

Simulations of structure formation using just cold dark matter indicate that halos develop self-similar density profiles, approximately following an NFW \citep*{NFW} or Einasto \citep{NFWsmooth} profile.  Both these profiles include a steep cusp in the central region.  This would seem to be at odds with kinematic data, which often indicate that the central parts of halos posses smooth cores of dark matter \citep{deBlok01,Gentile04,DelPopolo09}.  New simulations which include baryons \citep{Governato10} show that feedback from star formation and supernovae might solve this problem, resulting in cores very much like those observed.  This mechanism would not be feasible for such low-mass objects as ultra-faint dwarf galaxies, as they do not posses enough baryons to form sufficient stars and supernovae to make the process efficient.  On the other hand, kinematic data from the ultra-faint dwarfs is so sparse \citep[e.g.][]{Geha09} that they show no real preference so far for either cores or cusps anyway, so the smallest galaxies may still actually possess the cusps predicted by simulation.

Because annihilation rates depend on the square of particle densities, the predicted rates for indirect detection depend very sensitively upon the lower end of the halo mass distribution.  This is the source of $BF$ in Eq.~\ref{fluxeqn}, the `boost factor'.  The minimum possible mass of a dark matter halo depends upon the nature of the dark matter particle.  In particular, this is set by the particle's mass and couplings, as well as the history of its kinetic decoupling and free streaming in the early Universe \citep{Hofman01,Green04b,Green05}, and how those might have been influenced by events like the QCD phase transition \citep[e.g][]{Bringmann09_clumps}.  For thermal WIMPs for example, the most likely minimum mass is $\sim$$10^{-6}\,M_\odot$ \citep{Green04b,Green05}, though values from $10^{-4}$ to $10^{-9}\,M_\odot$ are also possible \citep{Martinez09}.  Smaller halos are expected to be far more numerous than larger ones in the picture of hierarchical structure formation governed by cold dark matter, although just how much more numerous is still a matter of considerable debate between $N$-body groups \citep{vialacteaII,Kuhlen08,Aquarius,Springel08}.

A standard assumption is that the velocities of WIMPs in halos follow an isotropic, spherically-symmetric, isothermal (Gaussian) distribution, with the width determined by the Keplerian velocity at the solar position.  In fact none of the details of this assumption turn out to be correct if one looks at $N$-body simulations: velocity distributions differ in the radial and angular directions, vary with galactocentric radius, and are neither Gaussian nor spherically-symmetric (\citealt{Hansen:2005yj}; \citealt{Fairbairn09}; \hyperref[papII]{\ScottII}).  In any case, the distribution must be truncated at the local Galactic escape velocity, which also has some uncertainty attached to it.  Local inhomogeneities in the phase space distribution further complicate matters, due to the presence of streams and cool clumps.  If this were not enough, recent simulations \citep{Read09} also suggest that a fraction of Galactic dark matter might exist in a disk that co-rotates with the baryonic disk.  The presence of such a `dark disk' would substantially boost the bottom end of the velocity distribution, resulting in greater solar capture rates and spectacularly improved limits on the nuclear scattering cross-sections \citep{Bruch09}.  Other uncertainties in the velocity distribution tend to impact direct detection more than solar capture, as the former is sensitive to WIMPs of any energy above threshold, whereas the Sun only really responds to the low-velocity part of the distribution.

\section{Production and propagation of cosmic rays}
\label{cosmicrays}

Cosmic rays (CRs) include protons and heavy nuclei, as well as most of the interesting species for indirect detection of dark matter: electrons, positrons, antiprotons, antideuterons and gamma rays.  CRs are produced when particles are accelerated in primary sources such as supernova remnants, interstellar shocks, pulsars and active galactic nuclei (AGN), and in secondary interactions or decay of other CRs.  Secondary production mechanisms include spallation, radioactive decay of unstable nuclei (including by electron capture), pion production and decay in collisions of primary CRs with diffuse gas, inverse Compton scattering of diffuse radiation, bremsstrahlung and synchrotron emission \citep{Strong07}.

\begin{figure}[tbp]
\centering
\includegraphics[width=\linewidth, trim=0 90 0 150, clip=true]{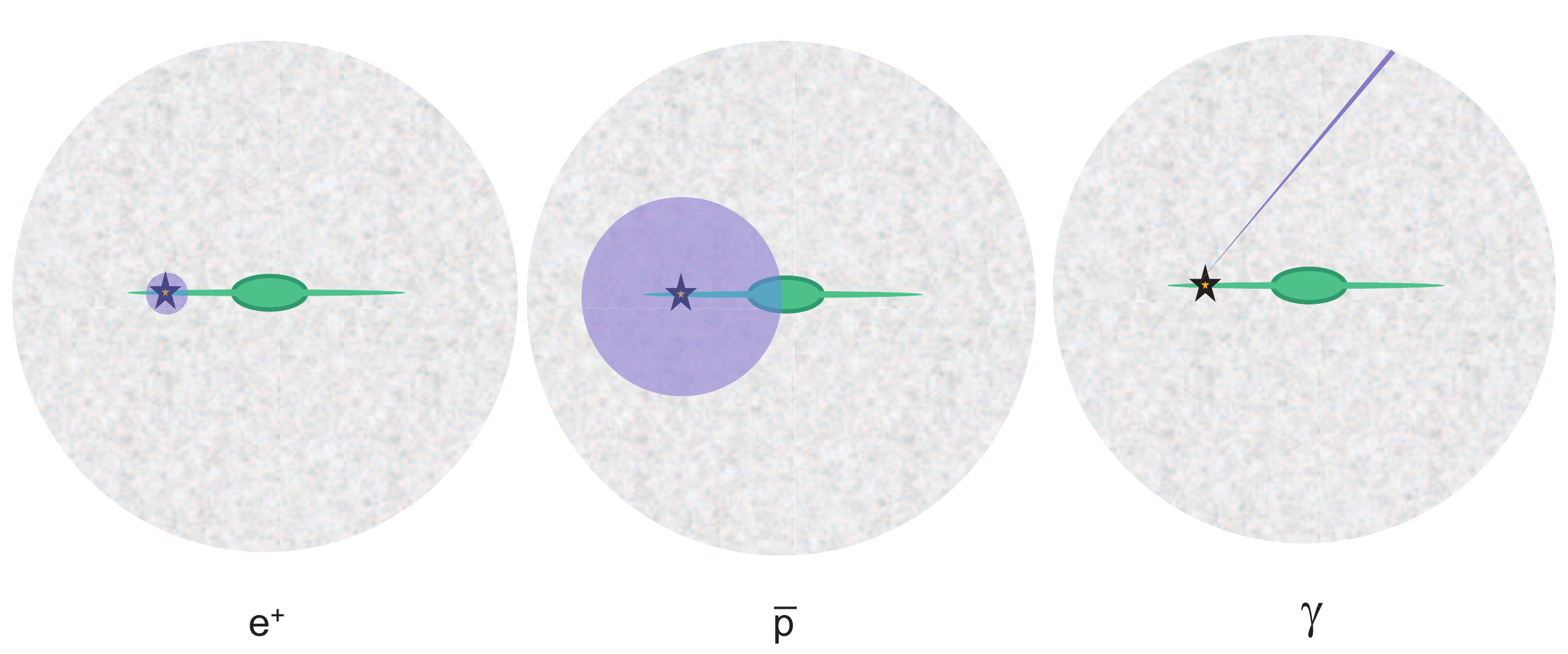}
\caption{Propagation volumes for direct observations of cosmic rays.  Positrons sample only the local environment, as they lose energy very efficiently via synchrotron emission, bremsstrahlung and inverse Compton scattering, and are easily deflected by magnetic fields because they are so light.  Antiprotons suffer a similar fate, but are not as strongly affected because they are much more massive.  Gamma rays are neutral and massless, so point directly back to their production site.  From \citet{Bergstrom09}.}
\label{fig_sourcevols}
\end{figure}

The propagation of CRs depends strongly upon the particle in question, as illustrated in Fig.~\ref{fig_sourcevols}.  Gamma-rays are neutral and massless, so arrive at Earth essentially undeflected and unattenuated.  Positrons and antiprotons are massive and charged, so suffer deflection in magnetic fields and energy losses from bremsstrahlung, synchrotron and inverse Compton emission \citep{Moskalenko98,Strong98}.  At the energies relevant to dark matter searches, positrons are much more strongly effected than antiprotons.  These energy loss processes all produce photons, resulting in a diffuse Galactic gamma-ray emission centred on the GC and Galactic disk \citep{Strong00}.  Similar processes produce gamma rays in external galaxies, especially those actively engaged in star formation \citep{LATstarburst}.  Unresolved external galaxies and AGN could be responsible for the near-isotropic diffuse gamma ray emission observed by \emph{Fermi} \citep{LATEGBGpaper} and its predecessors.

Modelling the propagation of CRs essentially requires solving a diffusion equation in the full phase space of the particles \citep[see e.g.][]{Strong07}, and adding additional terms for particle convection and decay.  The diffusion is sometimes solved for analytically in 1- or 2D \citep[e.g.][]{Maurin01, Putze10}, or numerically in 2- or 3D \citep[e.g.][]{Strong98,Strong07}.  The numerical approach allows the inclusion of much more realistic physics, source distributions and geometries.  At the present time, the analytic approach is the only one which allows a full statistical estimate of the relevant parameters \citep[e.g.][]{Putze09}, and also arguably gives a clearer understanding of the physics involved.  The parameters of the propagation model can be constrained by comparing with primary-to-secondary ratios such as boron to carbon, which are not strongly dependent upon the primary injection spectrum.

Given a set of propagation parameters, a theoretical source distribution and injection spectrum, and interstellar gas and magnetic field maps, one can then derive a predicted phase space distribution of charged CRs.  The local form of the distribution can then be used as a background prediction in e.g.~positron or antiproton indirect dark matter searches.  To obtain a similar estimate of gamma-ray backgrounds, the charged CR distribution is combined with the gas map to predict yields from bremsstrahlung and pion decay, with interstellar radiation maps (and less importantly, the CMB) to predict inverse Compton yields, and with the magnetic field map to produce synchrotron yields.

\section{Standard Model nuisances}
\label{smnuisances}

In principle, uncertainties in all 19 SM parameters constitute nuisances for SUSY scans, and could impact dark matter analyses.  In practice though, some parameters dominate over others.  The most important experimental uncertainties in the SM are those attached to the measured masses of the particles.  As the heaviest fermions, the masses of the top and (to a lesser extent) bottom quarks are the worst constrained of all the SM particles except the Higgs.  Unfortunately, the heaviest particles also have the greatest impacts upon SUSY phenomenology.  The masses of the $Z$ and $W$ bosons are not substantial nuisances, as they are known very accurately \citep{PDG}.

The strengths of the gauge coupling constants also have a substantial effect, and can be difficult to constrain accurately because they run with renormalisation group evolution.  Only the strong and electromagnetic couplings are significant nuisances, as the weak coupling is known as a function of the electromagnetic coupling and the masses of the $W$ and $Z$ bosons.  See e.g.\ \citet{Allanach06}, \citet{Ruiz06} or \citet{Trotta08} for the importance of considering gauge couplings and third generation quark masses as nuisance parameters in SUSY scans.

\chapter{Summary of results}
\label{summary}

\begin{figure}[tb]
\centering
\includegraphics[width=0.5\linewidth]{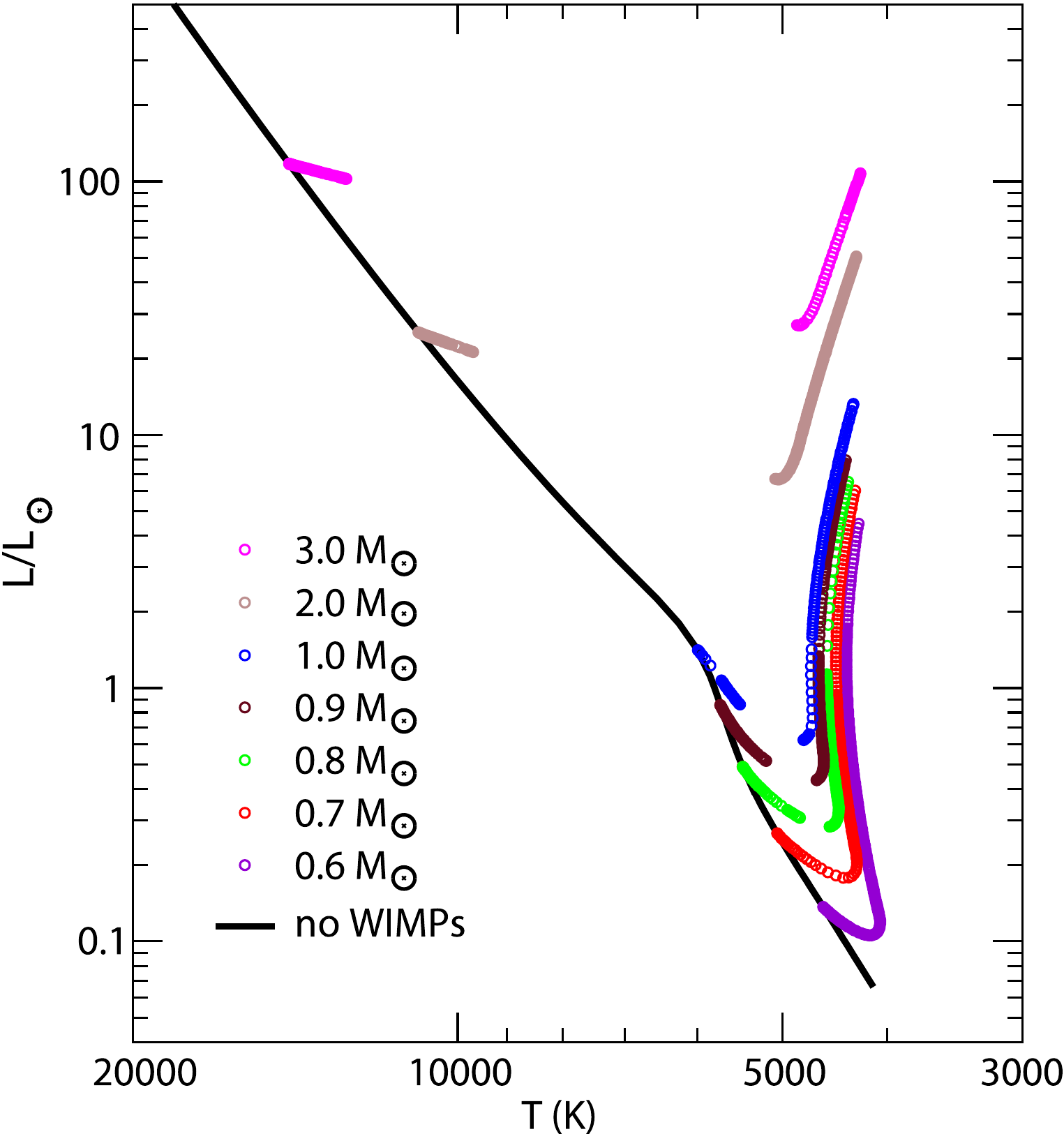}
\caption{HR-diagram showing steady-state solutions for main-sequence stars powered by differing amounts of dark matter annihilation.  The solid black line indicates the zero-age main sequence (ZAMS), where standard hydrogen-burning stars reside.  Coloured tracks indicate solutions for different stellar masses.  Stars situated further along the respective tracks (i.e.\ further from the ZAMS) have been provided with more dark matter.  The tracks strongly resemble the Hayashi track, along which protostars cool as they condense to form main sequence stars.  From \protect{\hyperref[papI]{\Fairbairn}}.}
\label{fig_I}
\end{figure}

\hyperref[papI]{\Fairbairn}, \hyperref[papII]{\ScottII} and \hyperref[papVI]{\Zackrisson} deal with the theory and detection of dark stars, at the GC and in the first dark matter halos.

In \hyperref[papI]{\Fairbairn}, we perform a first numerical investigation of the possible effects of WIMP dark matter upon main sequence stars.  We modify a simple static stellar structure code to show that stable stars supported by the annihilation of dark matter in their cores have surface temperatures and luminosities that make them resemble protostars (Fig.~\ref{fig_I}).  The stable solutions we find to the four coupled differential equations of stellar structure \citep[see e.g.][or any other standard stellar astrophysics text]{KipW,CandO,Stix} track the protostellar cooling trajectory known as the Hayashi track.  The more dark matter is added to the stars, the further their positions are shifted in the Hertzsprung-Russell (HR) diagram, back up the Hayashi track and away from the standard hydrogen-burning solutions of the zero-age main sequence (ZAMS).

We find that the static code is unable to arrive at stable structures across the whole Hayashi track, as it fails to find solutions in the regime where neither fusion nor WIMP annihilation dominates the stellar energy budget.  This can be seen as a series of gaps in the evolutionary tracks of Fig.~\ref{fig_I}.  We investigate these regions further with a full `dark' stellar evolution code \textsf{DarkStars} \citepalias{Scott09b}, and show that the absence of solutions in the intermediate regime is simply due to the numerical limitations of the static code.

We go on to investigate the structure and evolution of main sequence dark stars in detail in \hyperref[papII]{\ScottII}, using \textsf{DarkStars}.  We carefully describe the input physics for the \textsf{DarkStars} code, and proceed to catalogue the changes in evolutionary paths, core temperatures, core densities and convection zones that result from different rates of energy injection by WIMPs, as a function of the stellar mass.  We show that low-mass stars are generally much more strongly affected than high-mass ones, and that WIMP annihilation in stellar cores can significantly extend stars' main sequence lifetimes.

\begin{figure}[tb]
\centering
\includegraphics[width=0.8\linewidth]{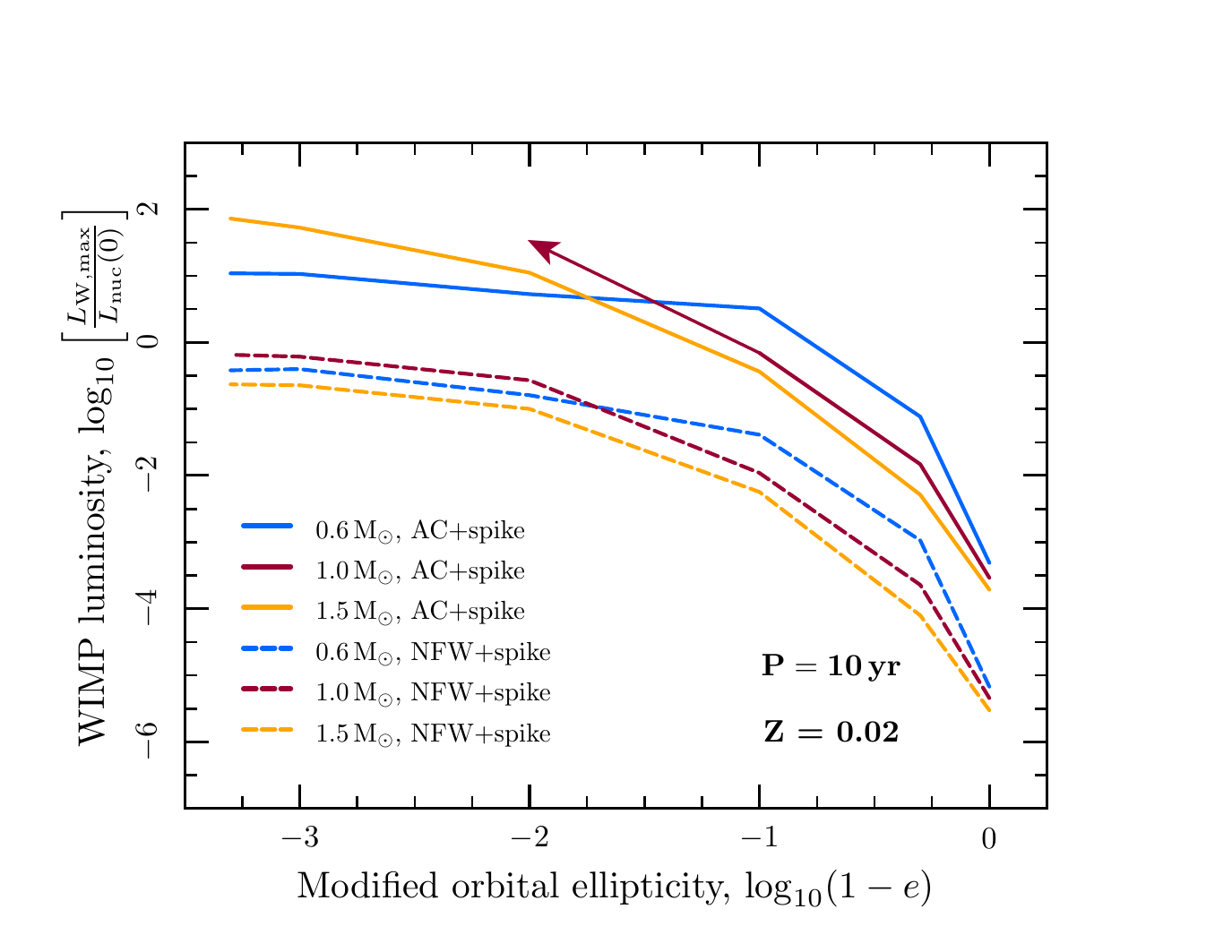}
\caption{WIMP-to-nuclear luminosity ratios achieved by stars on orbits with 10-year periods around the Galactic centre.  Dark matter annihilation can produce up to 100 times as much power as nuclear fusion in stars on realistic orbits in our own Galaxy.  If the Galactic halo has been adiabatically contracted (AC+spike), annihilation can equal nuclear fusion in stars on orbits with eccentricities greater than $e=0.9$, for masses less than about 1.5\,M$_\odot$.  If not (NFW+spike), stars of a solar mass or less require $e\gtrsim0.99$ to approach break-even between annihilation and fusion.  The arrow indicates that the 1\,M$_\odot$, AC+spike curve is expected to continue in this direction, but converging stellar models becomes rather difficult for such high WIMP luminosities.  From \protect{\hyperref[papII]{\ScottII}}.}
\label{fig_II}
\end{figure}

We investigate what injection rates could be realised in stars orbiting close to the supermassive black hole at the GC, paying attention to stellar orbits, dark matter density profiles and velocity distributions.  We derive new expressions for the capture rate from a truncated Maxwellian distribution of velocities, and outline an approximate fit to the distribution of velocities seen in the Via Lactea \citep{vialactea} simulation.  We find that low-mass stars on highly elliptical orbits near the GC could exhibit significant signs of WIMP capture and annihilation (Fig.~\ref{fig_II}).  The orbits required are very similar to the orbits on which more massive stars have already been observed at the GC.  The weakest point in this conclusion is the requirement of an adiabatically-contracted NFW density profile for the most marked effects of WIMP annihilation to be realised.

We show in \hyperref[papI]{\Fairbairn}, and then even more clearly in \hyperref[papII]{\ScottII}, that dark stars resemble protostars because the annihilation of WIMP dark matter in their cores causes them to cool and expand.  This result is in qualitative agreement with earlier analytical estimates \citep{SalatiSilk89}, though our more accurate numerical calculations show that the stars occupy slightly different positions in the HR diagram than simple polytropic models would suggest.  We conclude that the reason for this behaviour is the negative specific heat of a self-gravitating system; when extra energy is injected by WIMP annihilation, the self-gravitating body cools and expands.  Cooling and expansion of a stellar core reduces the rate of nuclear fusion.  Reduced fusion rates mean that stars' core hydrogen lasts longer, leading to the observed increase in \mbox{(quasi-)main} sequence lifetimes.

The resemblance of the `WIMP-burning' tracks to the Hayashi track can be understood in terms of two aspects of the physics of WIMP annihilation in stellar cores.  Our appreciation of these two aspects was not as well developed at the time of writing of either \hyperref[papI]{\Fairbairn} or \hyperref[papII]{\ScottII} as it has since become, so the following discussion is largely absent from those manuscripts.  The first aspect is simply that the energy injected by WIMPs plays a similar role to the gravitational energy released during the contraction of a protostar, as they are both responsible for exactly the same type of term in the stellar luminosity equation (one of the four equations mentioned above).  

The second and most crucial aspect is that the rate of energy production due to dark matter annihilation is not coupled to the gas pressure or temperature in any substantial way; the WIMP distribution, and hence annihilation rate, depend only very weakly upon the stellar structure.  Whereas the fusion rate depends in an essential way on the gas equation of state in the stellar core, the dark matter annihilation rate essentially depends only on the WIMP density, which is mostly decoupled from everything else.  Greater energy input from fusion would cause the core to cool and expand in a similar manner to what is seen to occur due to dark matter annihilation.  In this case however, the expansion would have the stabilising effect of reducing the fusion yield, and causing the core to recontract.  On the other hand, structural changes caused by energy injection from WIMP annihilations have almost no effect upon the net rate of annihilation, so there is no stabilising feedback.  This is very much akin to the energy released in gravitational contraction; the amount of energy released depends only upon the changing gravitational potential, not directly upon the thermodynamic gas properties.  This is also the main reason we see the gaps in tracks in \hyperref[papI]{\Fairbairn} and Fig.~\ref{fig_I}, and see in \hyperref[papII]{\ScottII} that although we can find viable solutions in these regions, such stars are only barely stable.

\begin{figure}[tb]
\centering
\begin{minipage}[c]{0.5\textwidth}
  \textbf{a}\vspace{-8mm}\\\includegraphics[width=\textwidth, trim = 20 150 20 150, clip=true]{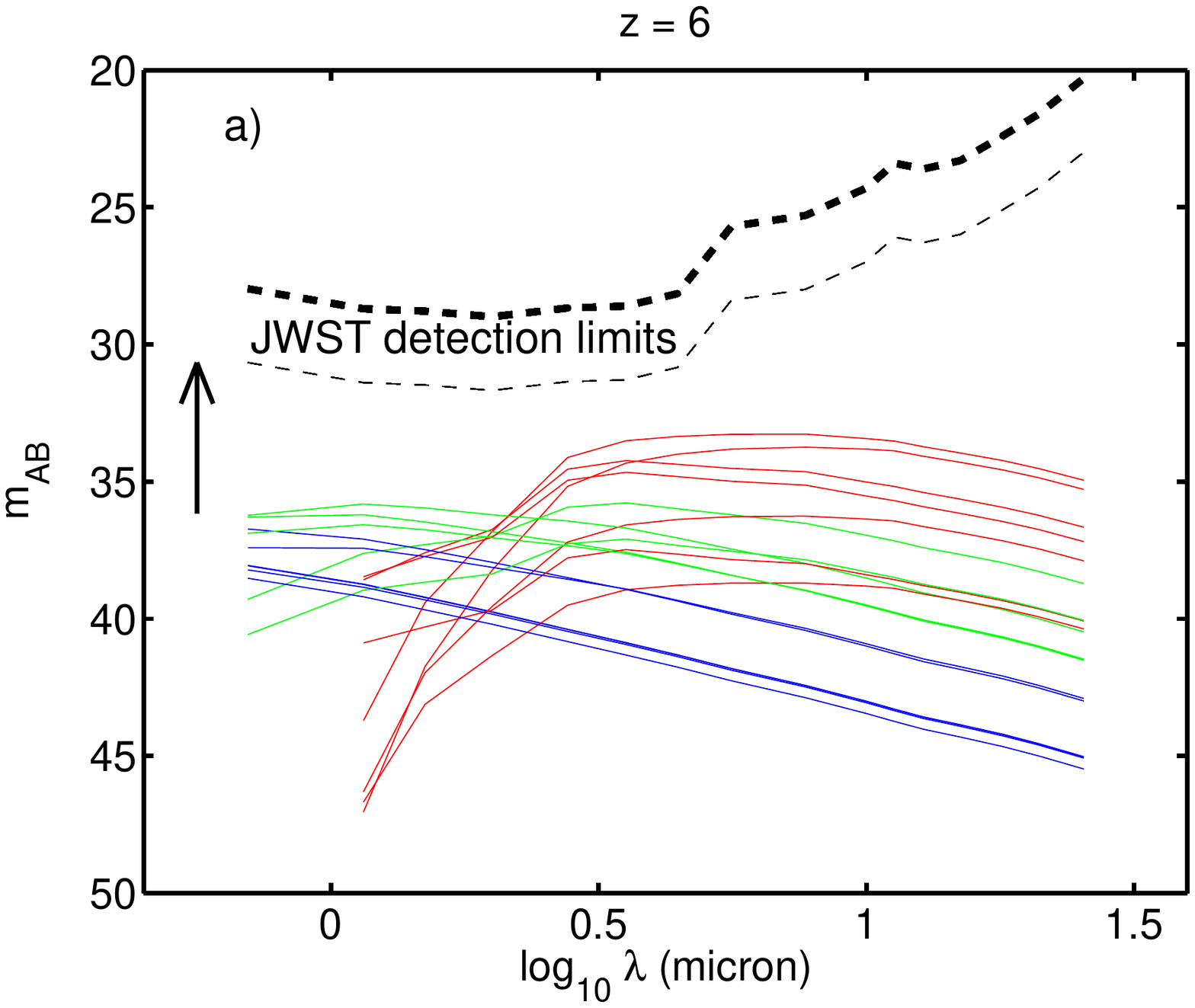}
\end{minipage}%
\begin{minipage}[c]{0.5\textwidth}
  \textbf{b}\vspace{-8mm}\\\includegraphics[width=\textwidth, trim = 20 150 20 150, clip=true]{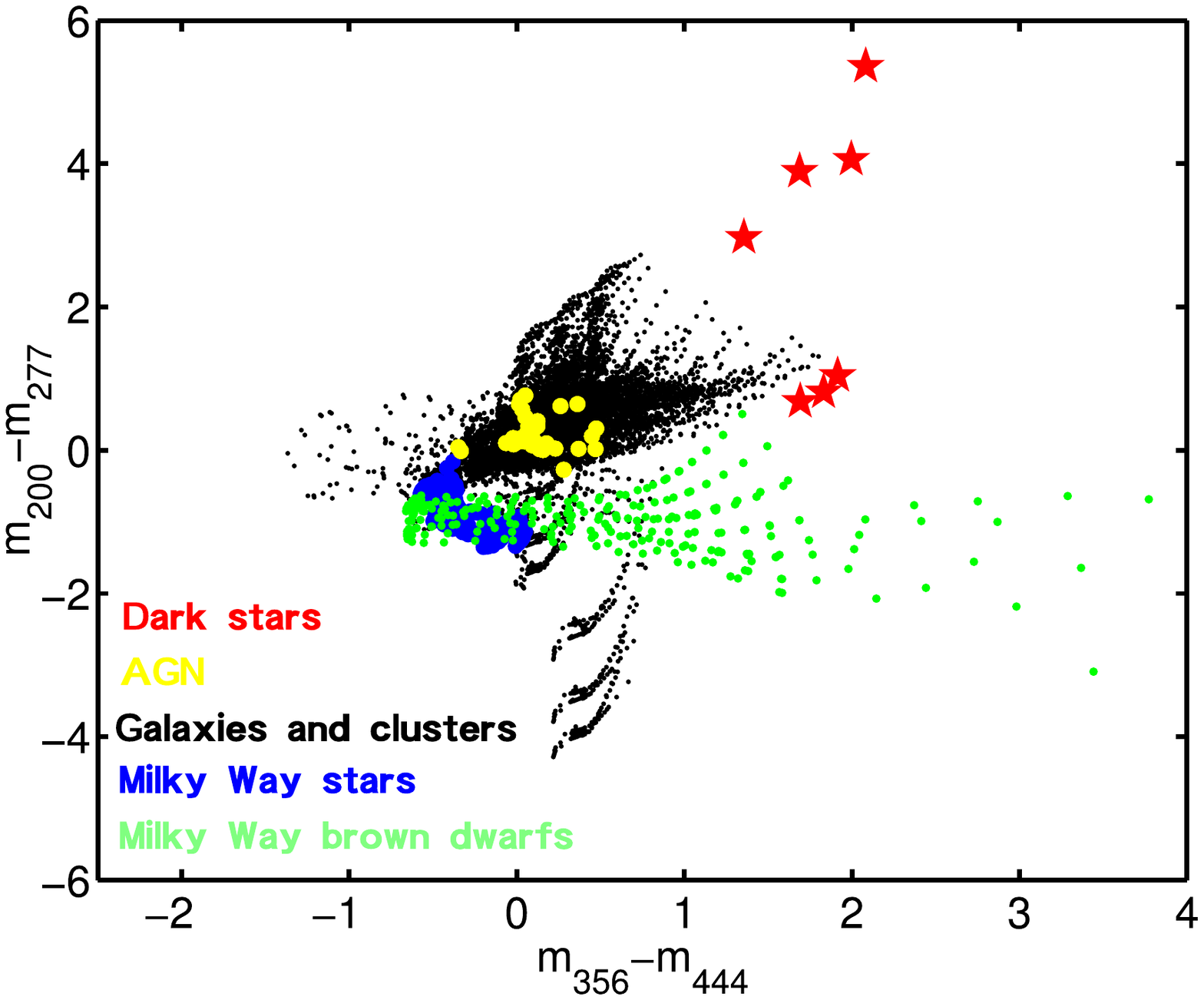}
\end{minipage}
\caption{\emph{Left (a)}: predicted apparent AB magnitudes of dark stars at $z = 6$, as a function the central wavelength observed in JWST broadband filters. Each line corresponds to a different dark star model, where models differ in their lifetimes and masses.  Models differing in stellar mass also necessarily have different intrinsic luminosities, surface temperatures and surface gravities.  Colours refer to surface temperatures $T_\mathrm{eff} \le 8000$\,K (red), $8000$\,K $< T_\mathrm{eff} \le 30\,000$\,K (green) and $T_\mathrm{eff} > 30\,000$\,K (blue).  Dashed lines refer to $5\sigma$ and $10\sigma$ JWST detection limits after $3.6 \times 10^5$\,s (100\,hr) and $10^4$\,s of exposure, respectively.  The arrow indicates the degree to which curves would be shifted upwards if dark stars were observed though the lensing cluster MACS J0717.5+3745.  In this case, the coolest dark stars would become individually detectable with JWST.  \emph{Right (b)}: Colour-colour diagram for dark stars and possible interloper populations, based on apparent magnitude differences in different JWST filters.  The very red spectra of dark stars mean that if they are bright enough to be seen at all, dark stars would occupy a rather unique corner of the colour-colour plot, and therefore be relatively easy to distinguish from other objects.  Both figures from \protect{\hyperref[papVI]{\Zackrisson}}.}
\label{fig_VI}
\end{figure}

We turn to dark stars in the early Universe in \hyperref[papVI]{\Zackrisson}, investigating their observability with the upcoming James Webb Space Telescope (JWST).  We compute model stellar atmospheres and synthetic spectra for the dark stars of \citet{Spolyar09}, and feed them through the JWST filters to ascertain their detectability as single objects.  We find that single dark stars in the early Universe will not be detectable by JWST except when viewed through a substantial gravitational lens, and even then only for certain values of their lifetimes and masses.  The expected fluxes from different dark stars are shown in comparison to JWST detection thresholds in Fig.~\ref{fig_VI}a.  The scale arrow in this figure indicates the amount by which the fluxes would be boosted if viewed through the lensing cluster MACS J0717.5+3745, which provides a magnification of $\mu=160$.  Their very red (i.e.~cool) spectra mean that if they are visible at all, high-redshift dark stars will occupy a rather unique position in the colour-colour diagram of Fig.~\ref{fig_VI}b, so should be clearly distinguishable from interloper objects. 

We also show that if early dark stars live long enough, their longevity will combine with their redder spectra to produce a rather peculiar feature in the integrated spectra of high-redshift galaxies.  The flux contributed by the dark stars will tend to amass in redshift space, producing characteristic red bumps in the spectra and significantly redder overall galaxy colours.  If at least 1\% of stars in early-type galaxies are indeed dark stars, this feature should be detectable in galaxy spectra.

\begin{figure}[tbp]
\centering
\includegraphics[width=0.7\linewidth, trim = 20 74 20 100, clip = true]{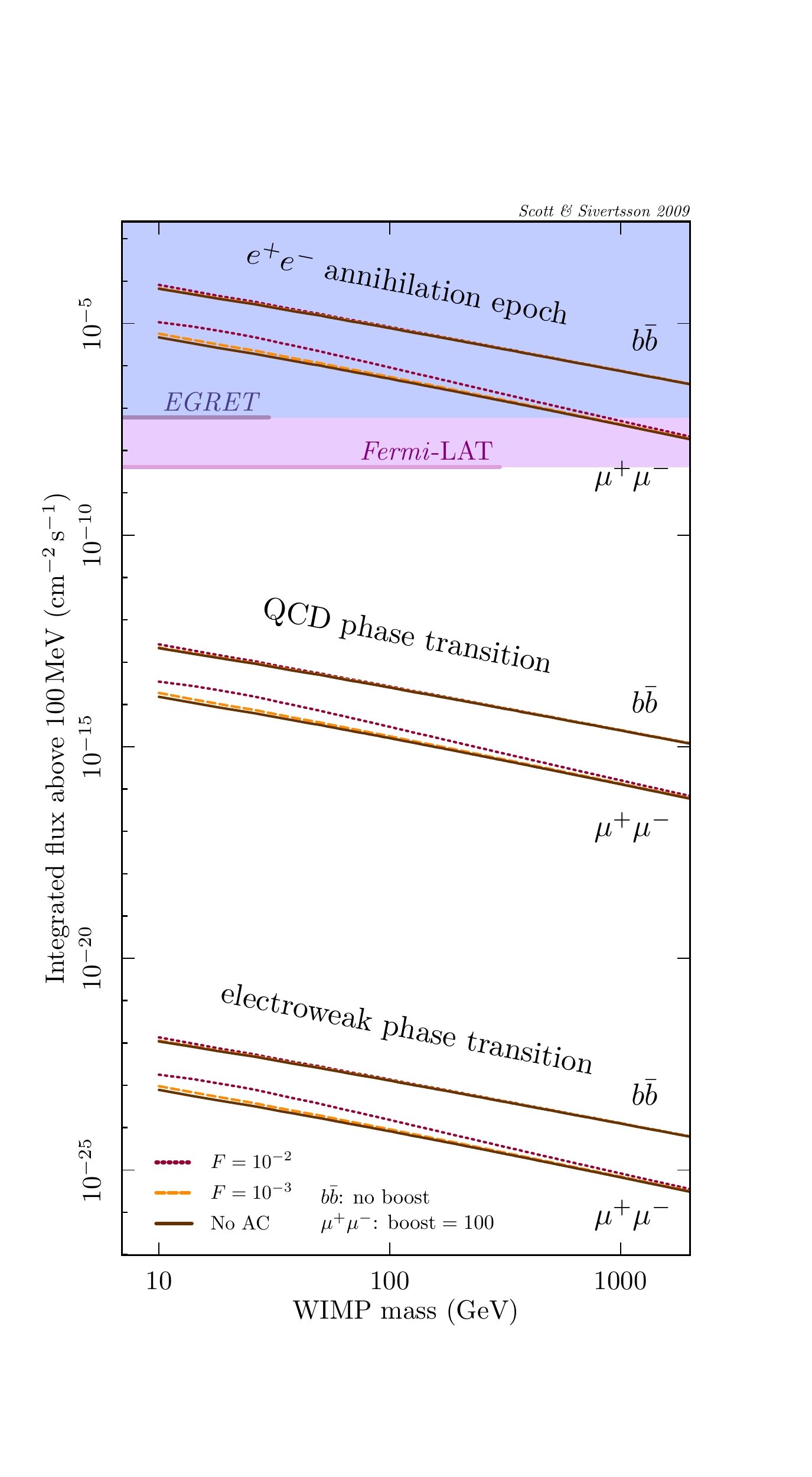}
\caption{Integrated fluxes above 100\,MeV for primordial ultracompact minihalos at a distance of $d=100$\,pc, consisting of WIMPs annihilating into either $b\bar{b}$ or $\mu^+\mu^-$ pairs.  Curves are shown for different phase transitions giving rise to minihalos, and various degrees of adiabatic contraction.  Adiabatically-contracted minihalos are assumed to have a fraction $F$ of their mass collapsed into a constant-density baryonic core of radius $10^{-3}R_\mathrm{h}$.  Also shown are approximate $5\sigma$, power-law, high-latitude, point-source sensitivities for 2 weeks of pointed \emph{EGRET} and one year of all-sky \emph{Fermi}-LAT observations.  Solid limits indicate instruments' nominal energy ranges.  Virtually all types of minihalos produced in the $e^+e^-$ annihilation epoch would be detectable by \emph{Fermi} after 1 year.  Minihalos produced in the QCD confinement phase transition would also be detectable if one or more were closer than $d\sim1$\,pc.  From \protect{\hyperref[papIII]{\SSiv}}.}
\label{fig_III}
\end{figure}

\citet{Akrami10,Scott09c,Zackrisson10,SS09,Fairbairn08,Scott09}

\hyperref[papIII]{\SSiv} and \hyperref[papIV]{\ScottIV} discuss novel aspects of indirect detection of WIMP dark matter using gamma rays, whilst \hyperref[papIV]{\ScottIV} and \hyperref[papV]{\Akrami} both deal with statistical scanning of SUSY parameter spaces.

In \hyperref[papIII]{\SSiv}, we investigate the prospects for indirect detection from a new class of dark matter substructure, proposed by \citet{Ricotti09}.  These ultracompact primordial minihalos would be formed by gravitational collapse of large-amplitude, small-scale density fluctuations in the early Universe.  Such perturbations might have been induced by phase transitions such as the QCD confinement transition, or specific late-time features in the inflaton potential.\footnote{In the submitted version of \hyperref[papIII]{\SSiv} (\protect{\href{http://arxiv.org/abs/0908.4082v1}{arXiv:0908.4082v1}}), we referred to these objects as `PLUMs' (Primordially-Laid Ultracompact Minihalos) so as to distinguish them from minihalos produced by the standard spectrum of inflationary perturbations.  Although that acronym did not ultimately survive to publication, we are pleased to see that it has resurfaced in the literature anyway \protect\citep{Lacki10}.}  The resultant minihalos could consist of PBHs which have later accreted dark matter, or be made almost entirely of dark matter.  The first case requires $\delta\gtrsim30\%$ for the initial PBH formation (cf.\ Sect.~\ref{goodbadugly}), whereas the second only needs $10^{-3}\lesssim\delta\lesssim30\%$, making the formation of PBH-free minihalos considerably more likely.

We compute the gamma-ray fluxes expected from ultracompact minihalos without a central PBH, formed in phase transitions in the early Universe.  We compare the detection prospects for the \emph{Fermi}-LAT and existing ACTs, showing that they provide largely complementary detection capabilities across a range of WIMP masses.  We show that single minihalos from the electron-positron annihilation epoch would be eminently observable today with existing instruments (Fig.~\ref{fig_III}), and in some cases should even have been seen by \emph{Fermi}'s immediate predecessor, \emph{EGRET} (the Energetic Gamma Ray Experiment Telescope).  We find that minihalos from the QCD confinement phase transition, arguably the best candidate transition for producing sufficient density perturbations to form the minihalos, should also be detectable today if any exist within \mbox{$\mathcal{O}$(1\,pc)} of Earth.  If their population is large enough, it is quite likely that at least one such minihalo would be close enough to detect with existing instruments.  We provide an expression for the cosmological density of ultracompact minihalos, which exhibits a strong dependence upon the spectral index $n$ of the density perturbations responsible for their formation.

\begin{figure}[tb]
  \begin{minipage}{0.31\textwidth}
  \centering
  \includegraphics[width=\linewidth, trim = 20 170 0 200, clip=true]{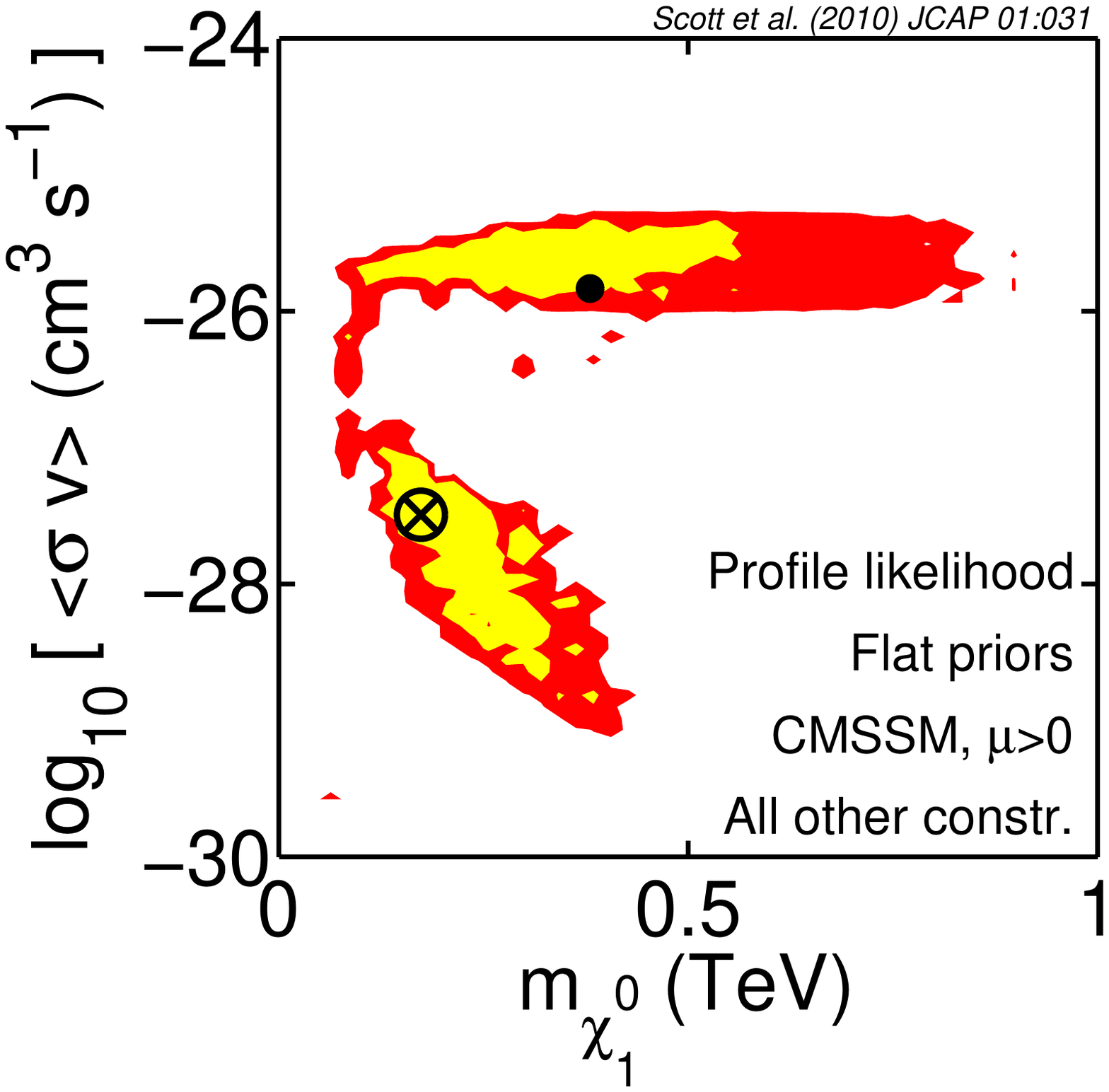}
  \includegraphics[width=\linewidth, trim = 20 170 0 200, clip=true]{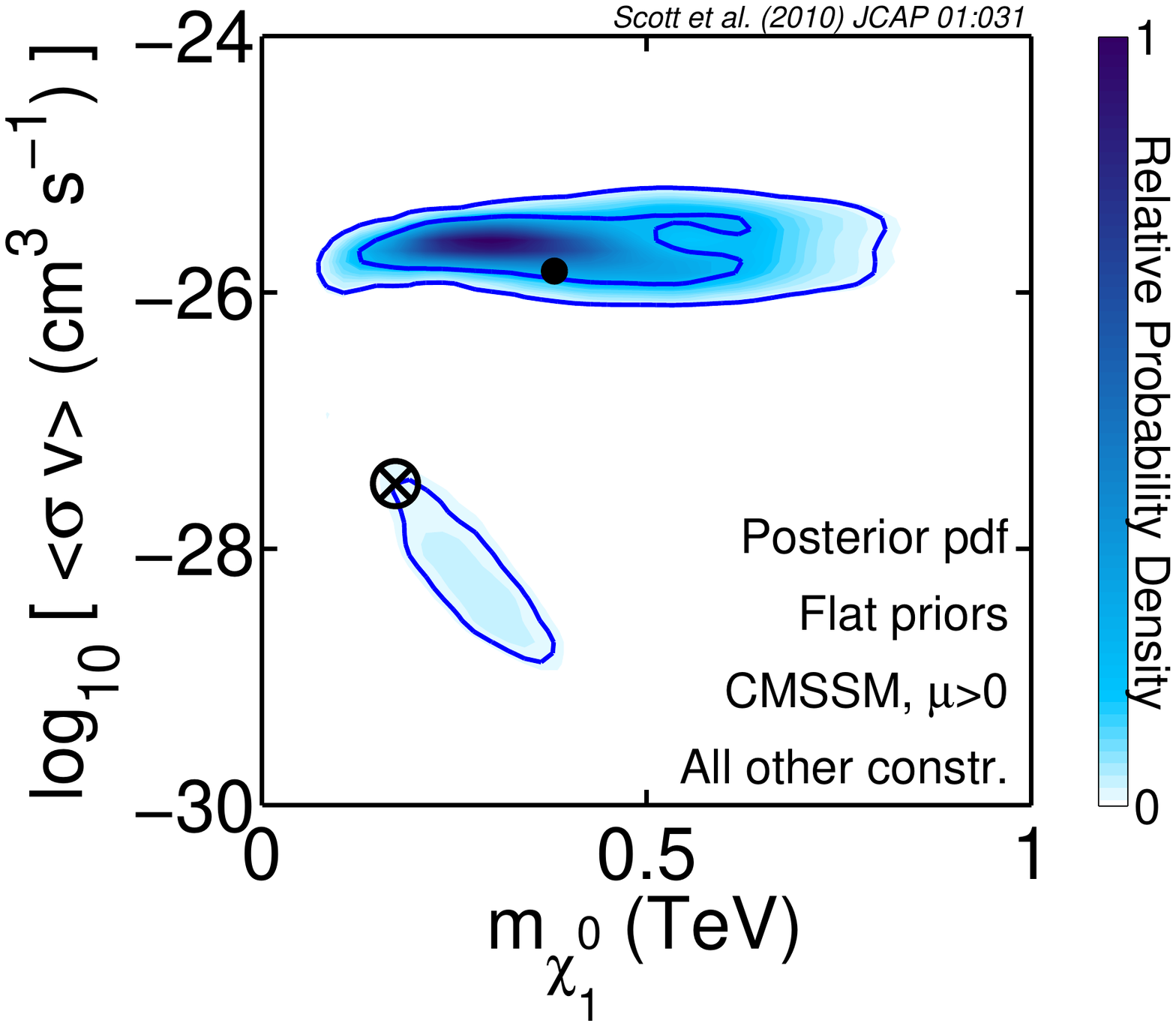}
  \end{minipage}
  \begin{minipage}{0.31\textwidth}
  \includegraphics[width=\linewidth, trim = 20 170 0 200, clip=true]{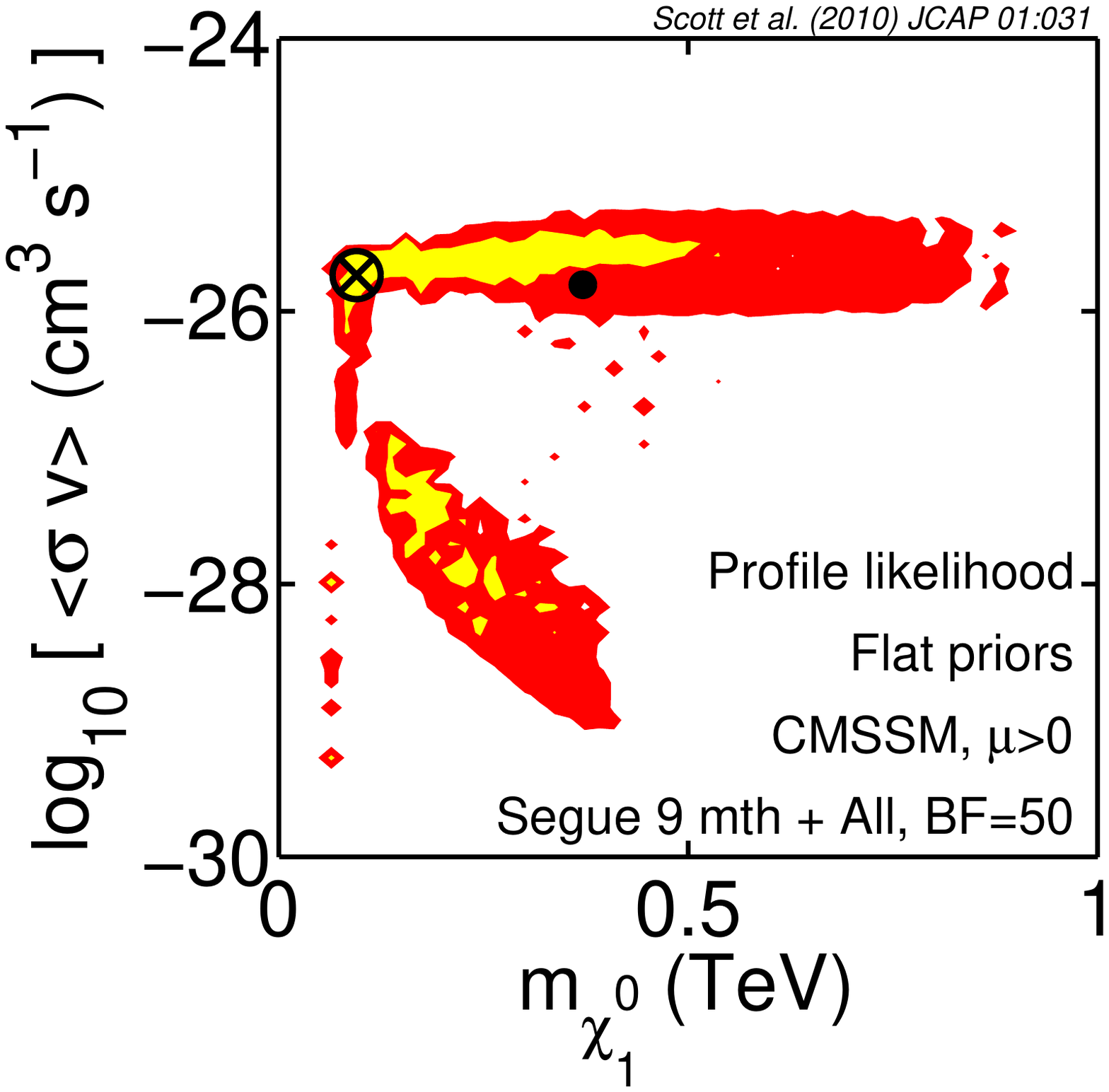}
  \includegraphics[width=\linewidth, trim = 20 170 0 200, clip=true]{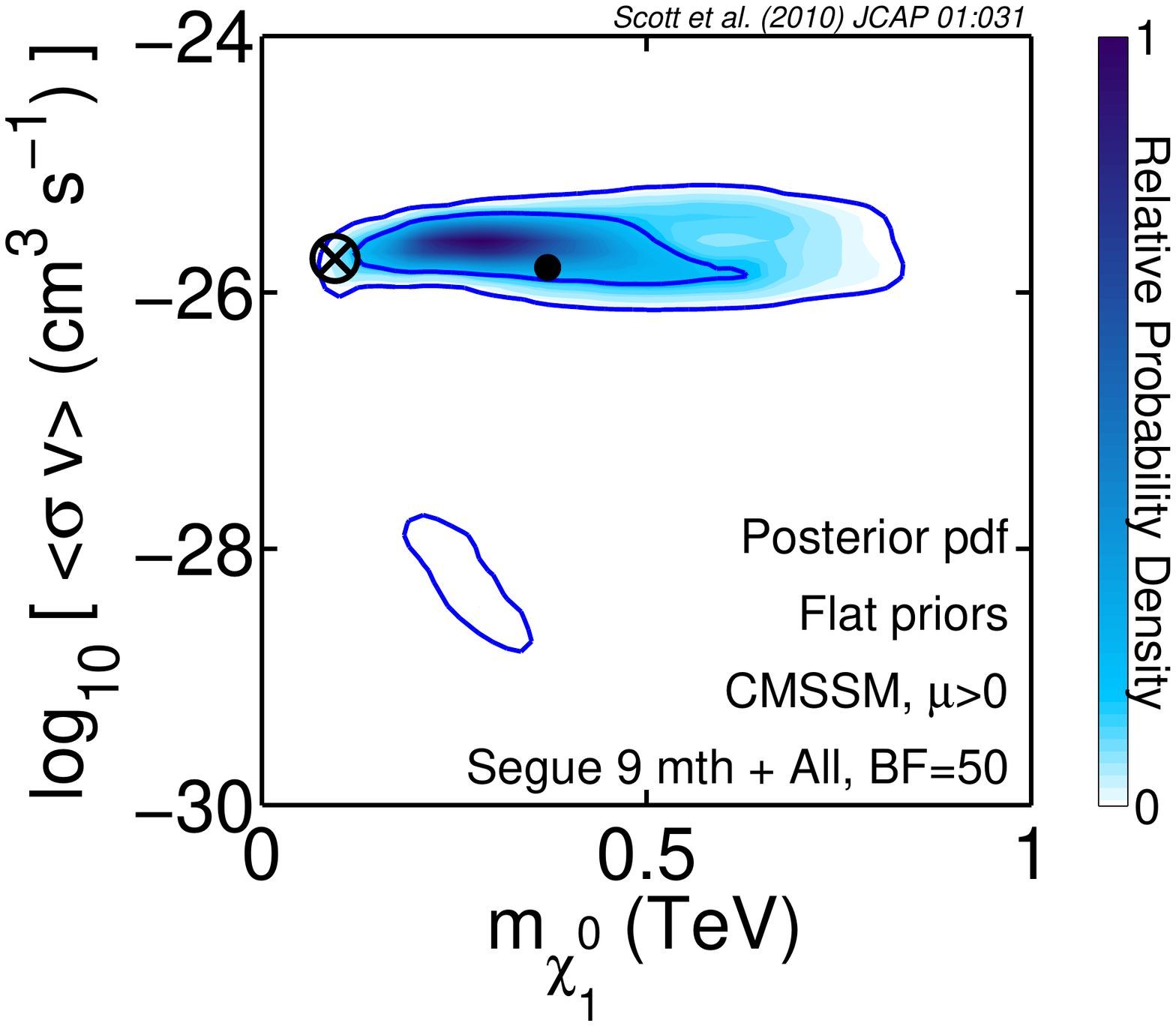}
  \end{minipage}
  \begin{minipage}{0.31\textwidth}
  \includegraphics[width=\linewidth, trim = 20 170 0 200, clip=true]{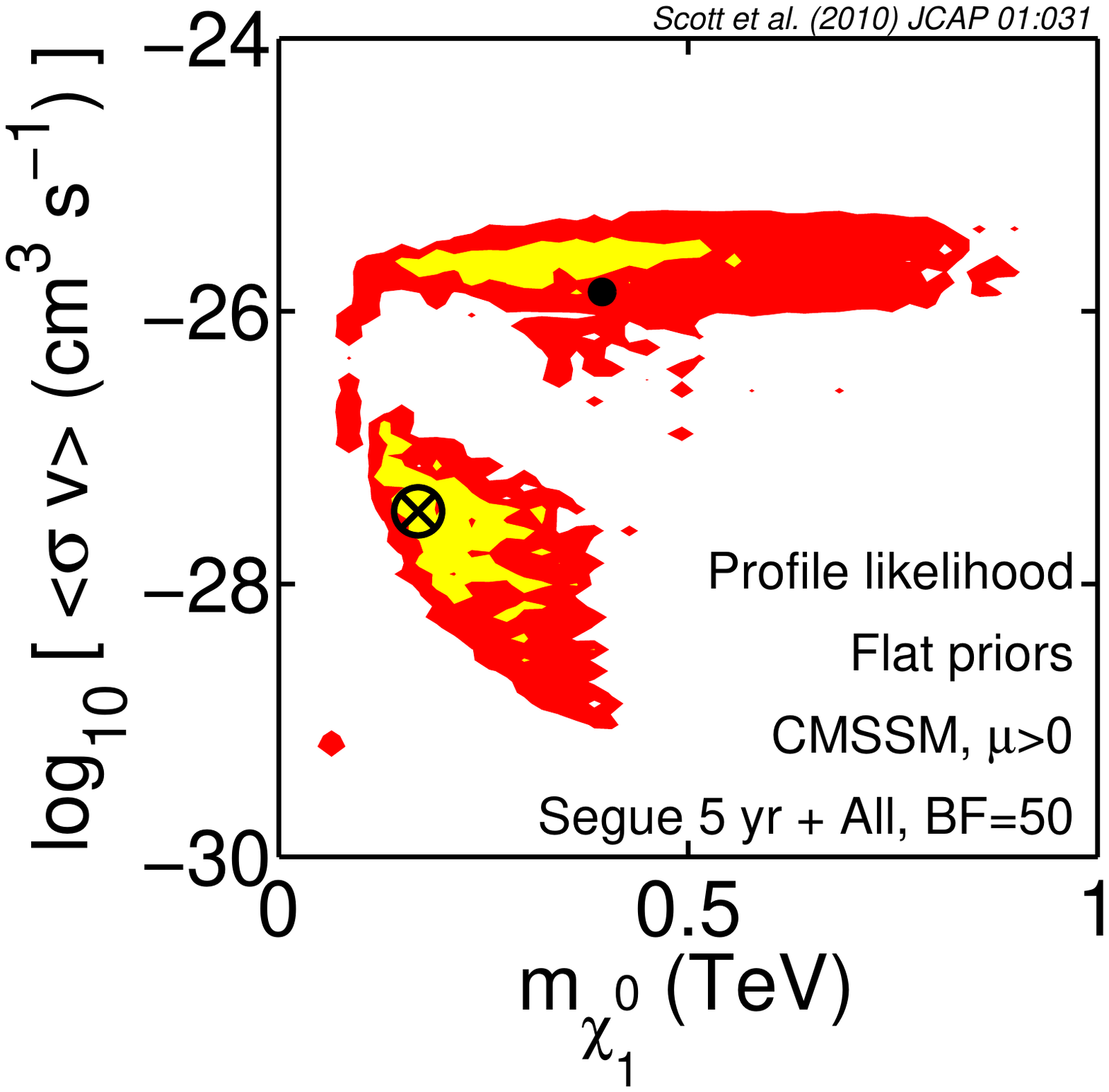}
  \includegraphics[width=\linewidth, trim = 20 170 0 200, clip=true]{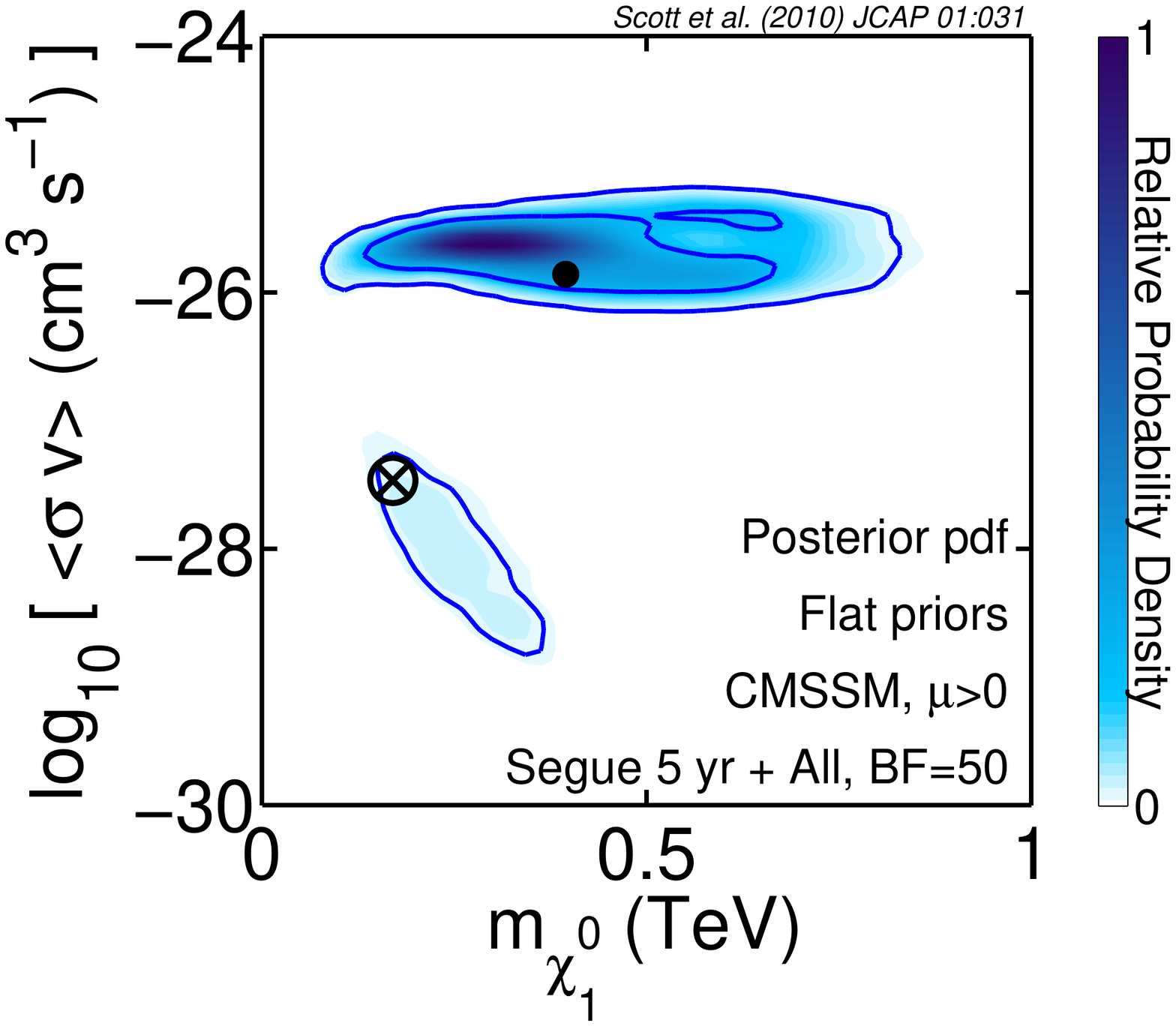}
  \end{minipage}
  \caption{CMSSM annihilation cross-sections consistent with all experimental constraints, with and without additional constraints from \emph{Fermi}-LAT observations of the dwarf galaxy Segue 1.  Scans assume that dark matter consists predominantly of neutralinos.  Plots show regions favoured by existing constraints only (\emph{left}), and with the addition of 9 months of \emph{Fermi} observations towards Segue 1 (\emph{middle}).  Also shown is the expected impact upon the favoured masses and cross sections if no signal is observed from Segue 1 after 5 years (\emph{right}).  Upper subfigures give profile likelihoods (yellow and red indicate 68\% and 95\% confidence regions respectively), whereas lower subfigures provide marginalised Bayesian posteriors (with 68\% and 95\% credible regions given by solid blue contours).  Solid dots give posterior means, and crosses indicate best-fit points.  The bulk of models disfavoured by observations of Segue 1 are already strongly disfavoured by other constraints, such as the relic density.  After 5 years of non-detection, some small inroads will be made into interesting parts of the parameter space at very low masses and high cross-sections.  From \protect{\hyperref[papIV]{\ScottIV}}.}
\label{fig_IV}
\end{figure}

In \hyperref[papIV]{\ScottIV}, we consider the constraints placed upon the CMSSM by \emph{Fermi} observations of the dwarf galaxy Segue 1.  We perform full global SUSY fits, and combine them with likelihoods from \emph{Fermi} data to ascertain which parts of the CMSSM parameter space are most favoured by all current data, and to what degree.  Specifically, we include constraints from the relic density, electroweak precision observables, the anomalous magnetic moment of the muon, rare processes in $B$-physics and accelerator bounds on sparticle and Higgs masses.  We perform full spectral and spatial fits to the \emph{Fermi} data, including a full convolution with the instrumental energy dispersion and point spread function for all $5.5\times10^4$ CMSSM models we compute.  

We show that the lack of any gamma-ray signal from Segue 1 disfavours some CMSSM models with very large annihilation cross-sections and low neutralino masses.  These models all possess cross-sections where the neutralino would not be the dominant component of dark matter, so are already disfavoured by the relic density constraint (Fig.~\ref{fig_IV}).  We also give a topical discussion of the relative merits of the use of the profile likelihood and the Bayesian posterior in SUSY scans.  We give examples of the additional physical insight that can be gained by considering both analysis techniques for any given parameter scan, rather than arguing the (rather futile) case that one is somehow more `correct' than the other.  Despite the weakness of the constraint drawn on the CMSSM parameter space, this paper makes a significant and concrete contribution to the literature, as it presents the first inclusion of constraints from indirect detection in full statistical SUSY scans.  It was also the first dark matter paper to use \emph{Fermi} gamma-ray data, from either within or outside the LAT collaboration.

\begin{figure}[tb]
  \centering
  \begin{minipage}{0.5\textwidth}
  \textbf{a}\vspace{-8mm}\\\includegraphics[width=\linewidth]{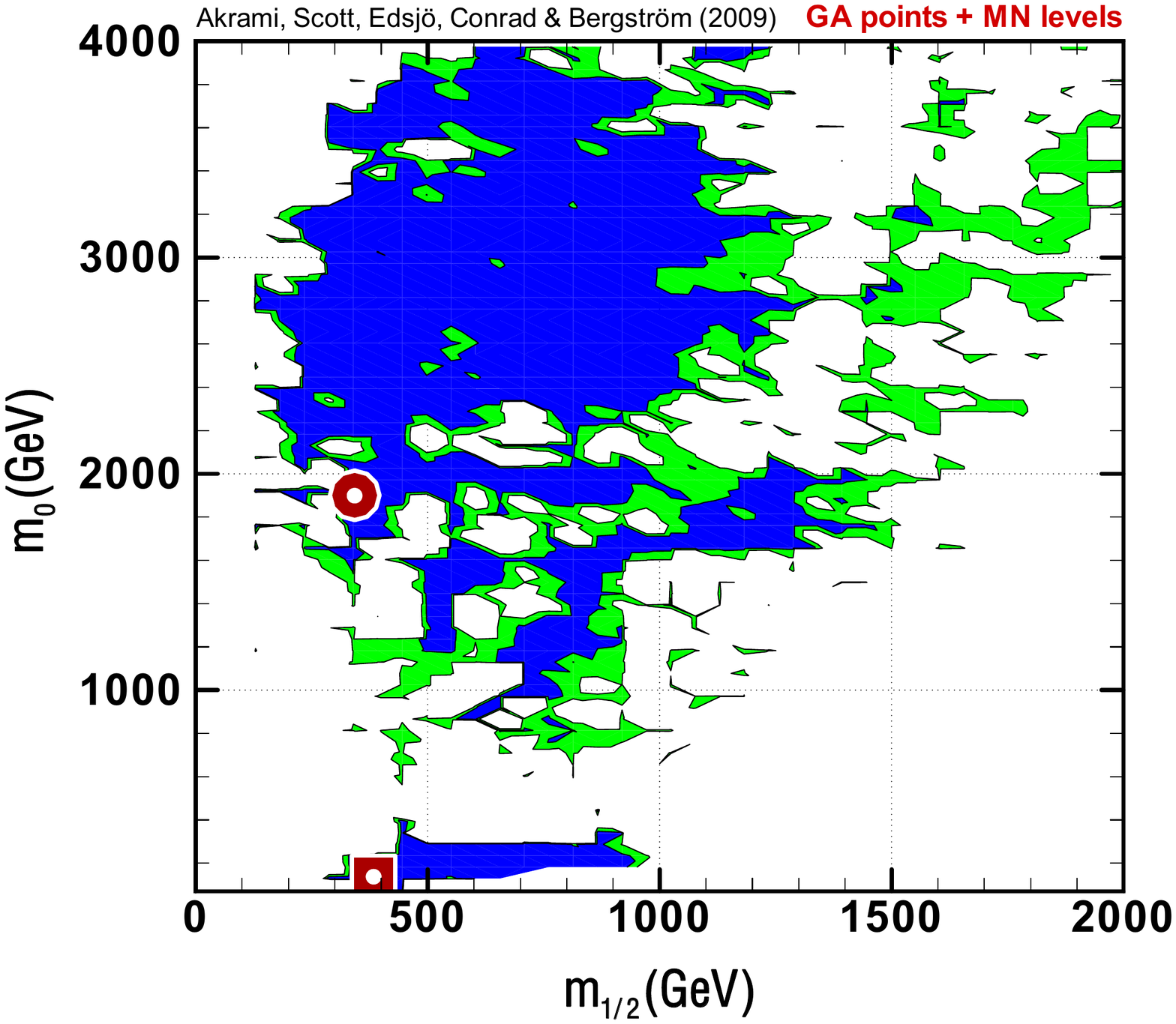}
  \end{minipage}%
  \begin{minipage}{0.5\textwidth}
  \textbf{b}\vspace{-8mm}\\\includegraphics[width=\linewidth]{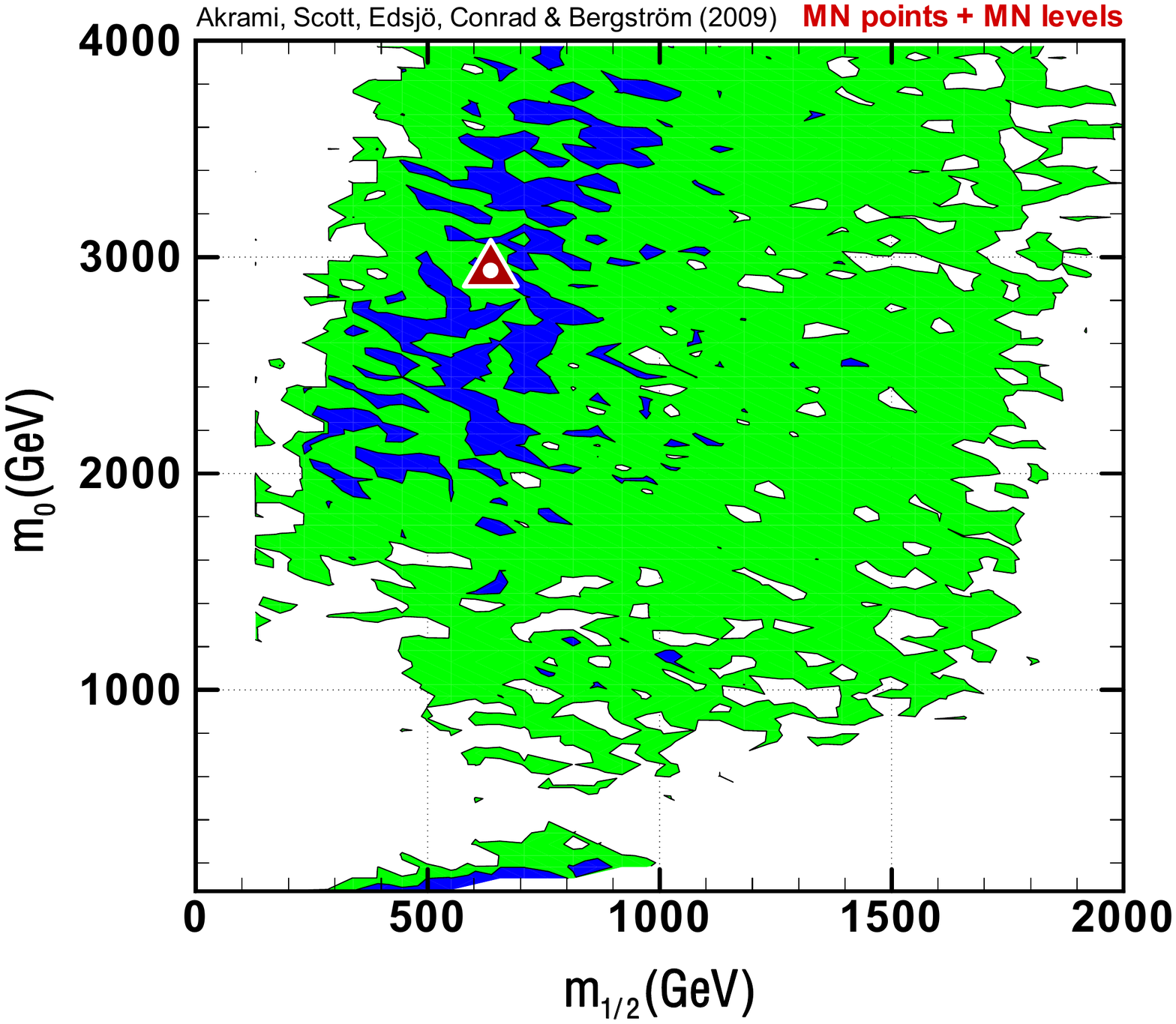}
  \end{minipage}
  \caption{Best-fit scalar ($m_0$) and gaugino ($M_\frac12$) mass parameters found by genetic algorithms (\emph{left}) and nested sampling (\emph{right}) in the CMSSM.  Blue and green shading indicate isolikelihood contours corresponding to 1-- and 2$\sigma$ confidence regions in the nested sampling scan.  Circular, square and triangular markers indicate the best fit in the genetic scan, the best-fit co-annihilation point in the genetic scan, and the best fit in the nested-sampling scan, respectively.  The genetic algorithm finds more very good fits, but nested sampling maps the regions where moderately good fits can be found more completely.  From \protect{\hyperref[papV]{\Akrami}}.}
  \label{fig_V}
\end{figure}

\hyperref[papV]{\Akrami} explores the scanning algorithms used in statistical analyses of SUSY parameter spaces.  In particular, we test whether genetic algorithms might perform more effective scans in the context of the profile likelihood than Bayesian techniques, such as nested sampling and MCMCs.  We implement the genetic algorithm code \textsf{PIKAIA} \citep{Pikaia} in the SUSY-scanning code \textsf{SuperBayeS} \citep{Trotta08}, using the same experimental data as in \hyperref[papIV]{\ScottIV} except for observations of Segue 1.  

Our results are somewhat surprising: we find a substantial improvement in the likelihood of the best-fit point when using genetic algorithms instead of nested sampling to map the CMSSM likelihood surface (Fig.~\ref{fig_V}).  We also uncover a section of the stau co-annihilation region of the CMSSM at large $m_0$, consistent with all constraints.  This sub-region appears to have been entirely missed in previous scans.  Contrary to findings by other authors, often based on Bayesian MCMC methods, our best-fit point occurs in the focus-point region, rather than the stau co-annihilation region.  This particular discrepancy likely has more to do with differences in physics codes than the scanning algorithm itself.  Nonetheless, our direct comparison with nested sampling in \textsf{SuperBayeS} indicates that all groups using Bayesian techniques to perform SUSY scans in the context of the profile likelihood should have significant concerns about whether the convergence of their scans is sufficient for a non-Bayesian analysis.

\section{Outlook}

This thesis presents a number of novel approaches and techniques for discovering or constraining the nature of dark matter.  Its results also provide some initial applications of those techniques to constraining dark matter models, especially SUSY in the form of the CMSSM.  

A myriad of extensions to the studies described here could be imagined, from hybrid scanning algorithms, additional observables and nuisances in SUSY scans, to post-main sequence evolution of dark stars and more detailed modelling of the halos in which they reside.  The scanning techniques of \hyperref[papIV]{\ScottIV} and \hyperref[papV]{\Akrami}, along with the improved solar abundances of \citetalias{AGSS}, \citetalias{Scott09Ni} and \citetalias{ScottVII}, should help sharpen our understanding of the limits on nuclear scattering cross sections provided by neutrino telescopes.  Together with an analysis of unidentified point sources in \emph{Fermi} data, the cosmological density of ultracompact minihalos and flux predictions derived in \hyperref[papIII]{\SSiv} would also be useful for constraining the spectrum of perturbations produced in various phase transitions in the early Universe.

Together, \hyperref[papI]{\Fairbairn}, \hyperref[papII]{\ScottII} and \hyperref[papVI]{\Zackrisson} indicate that the prospects for detecting dark stars or using them to constrain nuclear scattering cross-sections are not fantastic, although results are promising enough not to write off the possibility.  Dark stars in the early universe are probably more likely to exist and show effects of WIMP annihilation, but those in our own Galaxy might be easier to detect and investigate in detail.  Based on the results of \hyperref[papIV]{\ScottIV} and other careful analyses of \emph{Fermi} data \citep{LATdwarfpaper,LATcosmowimp,LATclusterpaper}, it would seem that indirect detection is also unlikely to provide a reliable discovery or exclusion of standard WIMP dark matter in the near future.  This statement would not apply if we happen to be very lucky with respect to the identity or early history of dark matter; if it were of a Sommerfeld-enhanced or non-thermal variety, or the QCD phase transition in the early Universe substantially increased the substructure boost factor through the creation of ultracompact minihalos, the picture for indirect would be rather more optimistic.

Notwithstanding these rather fine-tuned scenarios, it would seem that the best hope for dark matter detection in the next few years is probably at the LHC, and in direct detection experiments.  In this case, analyses like those of \hyperref[papIV]{\ScottIV} and \hyperref[papV]{\Akrami} will become especially important as we try to extract information about dark matter from LHC data, and cross-correlate it with dedicated dark matter probes.  A credible discovery of dark matter is unlikely to to be achieved through a single detection channel, much less a convincing identification in terms of the underlying quantum field theory.

\end{fmffile}

\bibliography{DMbiblio,SUSYbiblio,AbuGen,CObiblio}

\begin{thebibliography}{266}
\expandafter\ifx\csname natexlab\endcsname\relax\def\natexlab#1{#1}\fi
\addcontentsline{toc}{chapter}{\bibname}
\bibfont

\bibitem[{{Aaronson} {et~al.}(1979){Aaronson}, {Huchra} \&
  {Mould}}]{Aaronson79}
{Aaronson}, M., {Huchra}, J. \& {Mould}, J. 1979, \emph{\apj}, \textbf{229}, 1

\bibitem[{{Abazajian} {et~al.}(2010){Abazajian}, {Agrawal}, {Chacko} \&
  {Kilic}}]{Abazajian10}
{Abazajian}, K.~N., {Agrawal}, P., {Chacko}, Z. \& {Kilic}, C. 2010, \emph{\jcap},
  \textbf{11}, 041, 
  \href{http://arxiv.org/abs/1002.3820}{arXiv:1002.3820}

\bibitem[{{Abbasi} {et~al.}(2009){Abbasi}, {Abdou}, {Ackermann},
  {Adams}, {Ahlers}, {Andeen}, {Auffenberg} \& {Bai} {et al.}}]{IceCube09}
{Abbasi}, R., {Abdou}, Y., {Ackermann}, M., {Adams}, J., {Ahlers}, M.,
  {Andeen}, K., {Auffenberg}, J. \& {Bai}, X. {et al.} 2009{\natexlab{a}},
  \emph{\prl}, \textbf{102}, 201302,
  \href{http://arxiv.org/abs/0902.2460}{arXiv:0902.2460}

\bibitem[{{Abbasi} {et~al.}(2010){Abbasi}, {Abdou}, {Ackermann},
  {Adams}, {Ahlers}, {Andeen}, {Auffenberg} \& {Bai} {et al.}}]{IceCube09_KK}
---. 2010, \emph{\prd}, \textbf{81}, 057101, 
  \href{http://arxiv.org/abs/0910.4480}{arXiv:0910.4480}

\bibitem[{{Abdo} {et~al.}(2009){Abdo}, {Ackermann}, {Ajello}, {Atwood},
  {Axelsson}, {Baldini}, {Ballet} \& {Barbiellini} {et al.}}]{Fermielectron}
{Abdo}, A.~A., {Ackermann}, M., {Ajello}, M., {Atwood}, W.~B., {Axelsson}, M.,
  {Baldini}, L., {Ballet}, J. \& {Barbiellini}, G. {et al.} 2009, \emph{\prl},
  \textbf{102}, 181101, \href{http://arxiv.org/abs/0905.0025}{arXiv:0905.0025}

\bibitem[{{Abdo} {et~al.}(2010){Abdo}, {Ackermann}, {Ajello}, {Atwood},
  {Axelsson}, {Baldini}, {Ballet} \& {Barbiellini} {et al.}}]{LATstarburst}
---. 2010, \emph{\apjl}, \textbf{709}, L152,
  \href{http://arxiv.org/abs/arXiv:0911.5327}{arXiv:0911.5327}

\bibitem[{Abdo {et~al.}(2010)Abdo, Ackermann, Ajello, Atwood, Baldini, Ballet,
  Barbiellini \& Bastieri {et al.}}]{LATEGBGpaper}
Abdo, A.~A., Ackermann, M., Ajello, M., Atwood, W.~B., Baldini, L., Ballet, J.,
  Barbiellini, G. \& Bastieri, D. {et al.} 2010, \emph{\prl}, \textbf{104},
  101101, \href{http://arxiv.org/abs/1002.3603}{arXiv:1002.3603}

\bibitem[{{Abdo} {et~al.}(2010{\natexlab{a}}){Abdo}, {Ackermann}, {Ajello},
  {Atwood}, {Baldini}, {Ballet}, {Barbiellini} \& {Bastieri} {et
  al.}}]{LATlinepaper}
{Abdo}, A.~A., {Ackermann}, M., {Ajello}, M., {Atwood}, W.~B., {Baldini}, L.,
  {Ballet}, J., {Barbiellini}, G. \& {Bastieri}, D. {et al.}
  2010{\natexlab{a}}, \emph{\prl}, \textbf{104}, 091302,
  \href{http://arxiv.org/abs/1001.4836}{arXiv:1001.4836}

\bibitem[{{Abdo} {et~al.}(2010{\natexlab{b}}){Abdo}, {Ackermann}, {Ajello},
  {Atwood}, {Baldini}, {Ballet}, {Barbiellini} \& {Bastieri} {et
  al.}}]{LATdwarfpaper}
---. 2010{\natexlab{b}}, \emph{\apj}, \textbf{712}, 147,
  \href{http://arxiv.org/abs/1001.4531}{arXiv:1001.4531}

\bibitem[{{Abdo} {et~al.}(2010{\natexlab{c}}){Abdo}, {Ackermann}, {Ajello},
  {Baldini}, {Ballet}, {Barbiellini}, {Bastieri} \& {Bechtol} {et
  al.}}]{LATcosmowimp}
{Abdo}, A.~A., {Ackermann}, M., {Ajello}, M., {Baldini}, L., {Ballet}, J.,
  {Barbiellini}, G., {Bastieri}, D. \& {Bechtol}, K. {et al.}
  2010{\natexlab{c}}, \emph{\jcap}, \textbf{4}, 14,
  \href{http://arxiv.org/abs/1002.4415}{arXiv:1002.4415}

\bibitem[{{Acero} {et~al.}(2010){Acero}, {Aharonian}, {Akhperjanian}, {Anton},
  {Barres de Almeida}, {Bazer-Bachi}, {Becherini} \& {Behera} {et
  al.}}]{HESS10_GC}
{Acero}, F., {Aharonian}, F., {Akhperjanian}, A.~G., {Anton}, G., {Barres de
  Almeida}, U., {Bazer-Bachi}, A.~R., {Becherini}, Y. \& {Behera}, B. {et al.}
  2010, \emph{\mnras}, \textbf{402}, 1877,
  \href{http://arxiv.org/abs/arXiv:0911.1912}{arXiv:0911.1912}

\bibitem[{{Ackermann} {et~al.}(2006){Ackermann}, {Ahrens}, {Bai}, {Bartelt},
  {Barwick}, {Bay}, {Becka} \& {Becker} {et al.}}]{AMANDA06}
{Ackermann}, M., {Ahrens}, J., {Bai}, X., {Bartelt}, M., {Barwick}, S.~W.,
  {Bay}, R.~C., {Becka}, T. \& {Becker}, K. {et al.} 2006, \emph{\app},
  \textbf{24}, 459,
  \href{http://arxiv.org/abs/arXiv:astro-ph/0508518}{arXiv:astro-ph/0508518}

\bibitem[{{Ackermann} {et~al.}(2010){Ackermann}, {Ajello}, {Allafort},
  {Baldini}, {Ballet}, {Barbiellini}, {Bastieri} \& {Bechtol} {et
  al.}}]{LATclusterpaper}
{Ackermann}, M., {Ajello}, M., {Allafort}, A., {Baldini}, L., {Ballet}, J.,
  {Barbiellini}, G., {Bastieri}, D. \& {Bechtol}, K. {et al.} 2010,
  \emph{\jcap}, \textbf{5}, 25,
  \href{http://arxiv.org/abs/1002.2239}{arXiv:1002.2239}

\bibitem[{{Adriani} {et~al.}(2009{\natexlab{a}}){Adriani}, {Barbarino},
  {Bazilevskaya}, {Bellotti}, {Boezio}, {Bogomolov}, {Bonechi} \& {Bongi} {et
  al.}}]{pamelapositron}
{Adriani}, O., {Barbarino}, G.~C., {Bazilevskaya}, G.~A., {Bellotti}, R.,
  {Boezio}, M., {Bogomolov}, E.~A., {Bonechi}, L. \& {Bongi}, M. {et al.}
  2009{\natexlab{a}}, \emph{\nat}, \textbf{458}, 607,
  \href{http://arxiv.org/abs/0810.4995}{arXiv:0810.4995}

\bibitem[{{Adriani} {et~al.}(2009{\natexlab{b}}){Adriani}, {Barbarino},
  {Bazilevskaya}, {Bellotti}, {Boezio}, {Bogomolov}, {Bonechi} \& {Bongi} {et
  al.}}]{PAMELAantiproton}
---. 2009{\natexlab{b}}, \emph{\prl}, \textbf{102}, 051101,
  \href{http://arxiv.org/abs/0810.4994}{arXiv:0810.4994}

\bibitem[{{Aharonian} {et~al.}(2008){Aharonian}, {Akhperjanian}, {Barres de
  Almeida}, {Bazer-Bachi}, {Becherini}, {Behera}, {Benbow} \& {Bernl{\"o}hr}
  {et al.}}]{HESSelectron}
{Aharonian}, F., {Akhperjanian}, A.~G., {Barres de Almeida}, U., {Bazer-Bachi},
  A.~R., {Becherini}, Y., {Behera}, B., {Benbow}, W. \& {Bernl{\"o}hr}, K. {et
  al.} 2008, \emph{\prl}, \textbf{101}, 261104,
  \href{http://arxiv.org/abs/0811.3894}{arXiv:0811.3894}

\bibitem[{{Ahmad} {et~al.}(2001){Ahmad}, {Allen}, {Andersen}, {Anglin},
  {B{\"u}hler}, {Barton}, {Beier} \& {Bercovitch} {et al.}}]{Ahmad01}
{Ahmad}, Q.~R., {Allen}, R.~C., {Andersen}, T.~C., {Anglin}, J.~D.,
  {B{\"u}hler}, G., {Barton}, J.~C., {Beier}, E.~W. \& {Bercovitch}, M. {et
  al.} 2001, \emph{\prl}, \textbf{87}, 071301,
  \href{http://arxiv.org/abs/arXiv:nucl-ex/0106015}{arXiv:nucl-ex/0106015}

\bibitem[{{Ahmed} {et~al.}(2010){Ahmed}, {Akerib}, {Arrenberg}, {Bailey},
  {Balakishiyeva}, {Baudis}, {Bauer} \& {Brink} {et al.}}]{cdms2event}
{Ahmed}, Z., {Akerib}, D.~S., {Arrenberg}, S., {Bailey}, C.~N.,
  {Balakishiyeva}, D., {Baudis}, L., {Bauer}, D.~A. \& {Brink}, P.~L. {et al.}
  2010, \emph{Science}, \textbf{327}, 1619,
  \href{http://arxiv.org/abs/0912.3592}{arXiv:0912.3592}

\bibitem[{{Aitchison}(2007)}]{Aitchison}
{Aitchison}, I.~J.~R. 2007, \emph{{Supersymmetry in Particle Physics: An
  Elementary Introduction}} (Cambridge University Press)

\bibitem[{{Akerib} {et~al.}(2006){Akerib}, {Attisha}, {Bailey}, {Baudis},
  {Bauer}, {Brink}, {Brusov} \& {Bunker} {et al.}}]{supercdms}
{Akerib}, D.~S., {Attisha}, M.~J., {Bailey}, C.~N., {Baudis}, L., {Bauer},
  D.~A., {Brink}, P.~L., {Brusov}, P.~P. \& {Bunker}, R. {et al.} 2006,
  \emph{\nima}, \textbf{559}, 411

\bibitem[{{Akrami} {et~al.}(2010){Akrami}, {Scott}, Edsj{\"o}, {Conrad} \&
  {Bergstr{\"o}m}}]{Akrami10}
{Akrami}, Y., {Scott}, P., Edsj{\"o}, J., {Conrad}, J. \& {Bergstr{\"o}m}, L.
  2010, \emph{\jhep}, \textbf{4}, 57,
  \href{http://arxiv.org/abs/0910.3950}{arXiv:0910.3950}, \citepalias{Akrami10}

\bibitem[{{Alcock} {et~al.}(2000){Alcock}, {Allsman}, {Alves}, {Axelrod},
  {Becker}, {Bennett}, {Cook} \& {Dalal} {et al.}}]{Alcock00}
{Alcock}, C., {Allsman}, R.~A., {Alves}, D.~R., {Axelrod}, T.~S., {Becker},
  A.~C., {Bennett}, D.~P., {Cook}, K.~H. \& {Dalal}, N. {et al.} 2000,
  \emph{\apj}, \textbf{542}, 281,
  \href{http://arxiv.org/abs/arXiv:astro-ph/0001272}{arXiv:astro-ph/0001272}

\bibitem[{{Allanach} \& {Lester}(2006)}]{Allanach06}
{Allanach}, B.~C. \& {Lester}, C.~G. 2006, \emph{\prd}, \textbf{73}, 015013,
  \href{http://arxiv.org/abs/arXiv:hep-ph/0507283}{arXiv:hep-ph/0507283}

\bibitem[{{Amsler} {et~al.}(2008){Amsler}, {Doser}, {Antonelli}, {Asner},
  {Babu}, {Baer}, {Band} \& {Barnett} {et al.}}]{PDG}
{Amsler}, C., {Doser}, M., {Antonelli}, M., {Asner}, D.~M., {Babu}, K.~S.,
  {Baer}, H., {Band}, H.~R. \& {Barnett}, R.~M. {et al.} 2008, \emph{\plb},
  \textbf{667}, 1

\bibitem[{{Angle} {et~al.}(2008){Angle}, {Aprile}, {Arneodo}, {Baudis},
  {Bernstein}, {Bolozdynya}, {Brusov} \& {Coelho} {et al.}}]{Angle08a}
{Angle}, J., {Aprile}, E., {Arneodo}, F., {Baudis}, L., {Bernstein}, A.,
  {Bolozdynya}, A., {Brusov}, P. \& {Coelho}, L.~C.~C. {et al.} 2008,
  \emph{\prl}, \textbf{100}, 021303,
  \href{http://arxiv.org/abs/0706.0039}{arXiv:0706.0039}

\bibitem[{{Angulo} {et~al.}(1999){Angulo}, {Arnould}, {Rayet}, {Descouvemont},
  {Baye}, {Leclercq-Willain}, {Coc} \& {Barhoumi} {et al.}}]{NACRE}
{Angulo}, C., {Arnould}, M., {Rayet}, M., {Descouvemont}, P., {Baye}, D.,
  {Leclercq-Willain}, C., {Coc}, A. \& {Barhoumi}, S. {et al.} 1999,
  \emph{\nphysa}, \textbf{656}, 3

\bibitem[{{Arik} {et~al.}(2009){Arik}, {Aune}, {Autiero}, {Barth}, {Belov},
  {Beltr{\'a}n}, {Borghi} \& {Bourlis} {et al.}}]{CAST09}
{Arik}, E., {Aune}, S., {Autiero}, D., {Barth}, K., {Belov}, A., {Beltr{\'a}n},
  B., {Borghi}, S. \& {Bourlis}, G. {et al.} 2009, \emph{\jcap}, \textbf{2}, 8,
  \href{http://arxiv.org/abs/0810.4482}{arXiv:0810.4482}

\bibitem[{{Arkani-Hamed} {et~al.}(2009){Arkani-Hamed}, {Finkbeiner}, {Slatyer}
  \& {Weiner}}]{AHDM}
{Arkani-Hamed}, N., {Finkbeiner}, D.~P., {Slatyer}, T.~R. \& {Weiner}, N. 2009,
  \emph{\prd}, \textbf{79}, 015014,
  \href{http://arxiv.org/abs/0810.0713}{arXiv:0810.0713}

\bibitem[{{Asplund} {et~al.}(2009){Asplund}, {Grevesse}, {Sauval} \&
  {Scott}}]{AGSS}
{Asplund}, M., {Grevesse}, N., {Sauval}, A.~J. \& {Scott}, P. 2009,
  \emph{\araa}, \textbf{47}, 481,
  \href{http://arxiv.org/abs/arXiv:0909.0948}{arXiv:0909.0948},
  \citepalias{AGSS}

\bibitem[{{Asplund} {et~al.}(2006){Asplund}, {Lambert}, {Nissen}, {Primas} \&
  {Smith}}]{AsplundLi6}
{Asplund}, M., {Lambert}, D.~L., {Nissen}, P.~E., {Primas}, F. \& {Smith},
  V.~V. 2006, \emph{\apj}, \textbf{644}, 229,
  \href{http://arxiv.org/abs/arXiv:astro-ph/0510636}{arXiv:astro-ph/0510636}

\bibitem[{{Asztalos} {et~al.}(2010){Asztalos}, {Carosi}, {Hagmann}, {Kinion},
  {van Bibber}, {Hotz}, {Rosenberg} \& {Rybka} {et al.}}]{ADMX10}
{Asztalos}, S.~J., {Carosi}, G., {Hagmann}, C., {Kinion}, D., {van Bibber}, K.,
  {Hotz}, M., {Rosenberg}, L.~J. \& {Rybka}, G. {et al.} 2010, \emph{\prl},
  \textbf{104}, 041301,
  \href{http://arxiv.org/abs/arXiv:0910.5914}{arXiv:0910.5914}

\bibitem[{{Atwood} {et~al.}(2009){Atwood}, {Abdo}, {Ackermann}, {Althouse},
  {Anderson}, {Axelsson}, {Baldini} \& {Ballet} {et al.}}]{Atwood09}
{Atwood}, W.~B., {Abdo}, A.~A., {Ackermann}, M., {Althouse}, W., {Anderson},
  B., {Axelsson}, M., {Baldini}, L. \& {Ballet}, J. {et al.} 2009, \emph{\apj},
  \textbf{697}, 1071, \href{http://arxiv.org/abs/0902.1089}{arXiv:0902.1089}

\bibitem[{{Baer} \& {Profumo}(2005)}]{Baer05}
{Baer}, H. \& {Profumo}, S. 2005, \emph{\jcap}, \textbf{12}, 8,
  \href{http://arxiv.org/abs/arXiv:astro-ph/0510722}{arXiv:astro-ph/0510722}

\bibitem[{{Baer} \& {Tata}(2006)}]{BaerTata}
{Baer}, H. \& {Tata}, X. 2006, \emph{{Weak Scale Supersymmetry}} (Cambridge
  University Press)

\bibitem[{{Baltz} {et~al.}(2006){Baltz}, {Battaglia}, {Peskin} \&
  {Wizansky}}]{Baltz06}
{Baltz}, E.~A., {Battaglia}, M., {Peskin}, M.~E. \& {Wizansky}, T. 2006,
  \emph{\prd}, \textbf{74}, 103521,
  \href{http://arxiv.org/abs/arXiv:hep-ph/0602187}{arXiv:hep-ph/0602187}

\bibitem[{{Barbieri} {et~al.}(2006){Barbieri}, {Hall} \&
  {Rychkov}}]{Barbieri06}
{Barbieri}, R., {Hall}, L.~J. \& {Rychkov}, V.~S. 2006, \emph{\prd},
  \textbf{74}, 015007,
  \href{http://arxiv.org/abs/arXiv:hep-ph/0603188}{arXiv:hep-ph/0603188}

\bibitem[{{Battaglia}(2009)}]{Battaglia09}
{Battaglia}, M. 2009, \emph{\njp}, \textbf{11}, 105025

\bibitem[{{Battaglia} {et~al.}(2004){Battaglia}, {Hinchliffe} \&
  {Tovey}}]{Battaglia04}
{Battaglia}, M., {Hinchliffe}, I. \& {Tovey}, D. 2004, \emph{\jpg},
  \textbf{30}, 217

\bibitem[{{Bekenstein}(2004)}]{Bekenstein04}
{Bekenstein}, J.~D. 2004, \emph{\prd}, \textbf{70}, 083509,
  \href{http://arxiv.org/abs/arXiv:astro-ph/0403694}{arXiv:astro-ph/0403694}

\bibitem[{{B{\'e}langer} {et~al.}(2007){B{\'e}langer}, {Boudjema}, {Pukhov} \&
  {Semenov}}]{MicrOmegas}
{B{\'e}langer}, G., {Boudjema}, F., {Pukhov}, A. \& {Semenov}, A. 2007,
  \emph{\cpc}, \textbf{176}, 367,
  \href{http://arxiv.org/abs/arXiv:hep-ph/0607059}{arXiv:hep-ph/0607059}

\bibitem[{{Belikov} \& {Hooper}(2010)}]{Belikov10}
{Belikov}, A.~V. \& {Hooper}, D. 2010, \emph{\prd}, \textbf{81}, 043505,
  \href{http://arxiv.org/abs/0906.2251}{arXiv:0906.2251}

\bibitem[{{Belli} {et~al.}(2002){Belli}, {Cerulli}, {Fornengo} \&
  {Scopel}}]{Belli02}
{Belli}, P., {Cerulli}, R., {Fornengo}, N. \& {Scopel}, S. 2002, \emph{\prd},
  \textbf{66}, 043503,
  \href{http://arxiv.org/abs/arXiv:hep-ph/0203242}{arXiv:hep-ph/0203242}

\bibitem[{{Bergstr{\"o}m}(1989)}]{Bergstrom89}
{Bergstr{\"o}m}, L. 1989, \emph{\plb}, \textbf{225}, 372

\bibitem[{{Bergstr\"om}(2000)}]{Bergstrom00}
{Bergstr\"om}, L. 2000, \emph{\repprogphys}, \textbf{63}, 793,
  \href{http://arxiv.org/abs/arXiv:hep-ph/0002126}{arXiv:hep-ph/0002126}

\bibitem[{{Bergstr{\"o}m}(2009)}]{Bergstrom09}
{Bergstr{\"o}m}, L. 2009, \emph{\njp}, \textbf{11}, 105006,
  \href{http://arxiv.org/abs/0903.4849}{arXiv:0903.4849}

\bibitem[{{Bergstr{\"o}m} {et~al.}(2009{\natexlab{a}}){Bergstr{\"o}m},
  {Bertone}, {Bringmann}, {Edsj{\"o}} \& {Taoso}}]{Bergstrom09_MW}
{Bergstr{\"o}m}, L., {Bertone}, G., {Bringmann}, T., {Edsj{\"o}}, J. \&
  {Taoso}, M. 2009{\natexlab{a}}, \emph{\prd}, \textbf{79}, 081303,
  \href{http://arxiv.org/abs/0812.3895}{arXiv:0812.3895}

\bibitem[{{Bergstr{\"o}m} {et~al.}(2008){Bergstr{\"o}m}, {Bringmann} \&
  {Edsj{\"o}}}]{Bergstrom08}
{Bergstr{\"o}m}, L., {Bringmann}, T. \& {Edsj{\"o}}, J. 2008, \emph{\prd},
  \textbf{78}, 103520, \href{http://arxiv.org/abs/0808.3725}{arXiv:0808.3725}

\bibitem[{{Bergstr{\"o}m} {et~al.}(2009{\natexlab{b}}){Bergstr{\"o}m},
  {Edsj{\"o}} \& {Zaharijas}}]{Bergstrom09b}
{Bergstr{\"o}m}, L., {Edsj{\"o}}, J. \& {Zaharijas}, G. 2009{\natexlab{b}},
  \emph{\prl}, \textbf{103}, 031103,
  \href{http://arxiv.org/abs/0905.0333}{arXiv:0905.0333}

\bibitem[{{Bergstr{\"o}m} \& {Snellman}(1988)}]{Bergstrom88}
{Bergstr{\"o}m}, L. \& {Snellman}, H. 1988, \emph{\prd}, \textbf{37}, 3737

\bibitem[{{Bergstr{\"o}m} \& {Ullio}(1997)}]{Bergstrom97}
{Bergstr{\"o}m}, L. \& {Ullio}, P. 1997, \emph{\nphysb}, \textbf{504}, 27,
  \href{http://arxiv.org/abs/arXiv:hep-ph/9706232}{arXiv:hep-ph/9706232}

\bibitem[{{Bergstr{\"o}m} {et~al.}(1998){Bergstr{\"o}m}, {Ullio} \&
  {Buckley}}]{Bergstrom98}
{Bergstr{\"o}m}, L., {Ullio}, P. \& {Buckley}, J.~H. 1998, \emph{\app},
  \textbf{9}, 137,
  \href{http://arxiv.org/abs/arXiv:astro-ph/9712318}{arXiv:astro-ph/9712318}

\bibitem[{{Bern} {et~al.}(2009){Bern}, {Carrasco}, {Dixon}, {Johansson} \&
  {Roiban}}]{Bern09}
{Bern}, Z., {Carrasco}, J.~J.~M., {Dixon}, L.~J., {Johansson}, H. \& {Roiban},
  R. 2009, \emph{\prl}, \textbf{103}, 081301,
  \href{http://arxiv.org/abs/0905.2326}{arXiv:0905.2326}

\bibitem[{{Bernabei} {et~al.}(2008){Bernabei}, {Belli}, {Cappella}, {Cerulli},
  {Dai}, {D'Angelo}, {He} \& {Incicchitti} {et al.}}]{Bernabei08}
{Bernabei}, R., {Belli}, P., {Cappella}, F., {Cerulli}, R., {Dai}, C.~J.,
  {D'Angelo}, A., {He}, H.~L. \& {Incicchitti}, A. {et al.} 2008, \emph{\epjc},
  167, \href{http://arxiv.org/abs/0804.2741}{arXiv:0804.2741}

\bibitem[{{Bernabei} {et~al.}(2000){Bernabei}, {Belli}, {Cerulli},
  {Montecchia}, {Amato}, {Ignesti}, {Incicchitti} \& {Prosperi} {et
  al.}}]{Bernabei00}
{Bernabei}, R., {Belli}, P., {Cerulli}, R., {Montecchia}, F., {Amato}, M.,
  {Ignesti}, G., {Incicchitti}, A. \& {Prosperi}, D. {et al.} 2000,
  \emph{\plb}, \textbf{480}, 23

\bibitem[{Bertone(2010)}]{BertoneBook}
Bertone, G., ed. 2010, \emph{{Particle Dark Matter: Observations, Models and
  Searches}} (Cambridge University Press)

\bibitem[{{Bertone} {et~al.}(2009){Bertone}, {Cirelli}, {Strumia} \&
  {Taoso}}]{Bertone09}
{Bertone}, G., {Cirelli}, M., {Strumia}, A. \& {Taoso}, M. 2009, \emph{\jcap},
  \textbf{3}, 9, \href{http://arxiv.org/abs/0811.3744}{arXiv:0811.3744}

\bibitem[{{Bertone} \& {Fairbairn}(2008)}]{Bertone07}
{Bertone}, G. \& {Fairbairn}, M. 2008, \emph{\prd}, \textbf{77}, 043515,
  \href{http://arxiv.org/abs/arXiv:0709.1485}{arXiv:0709.1485}

\bibitem[{{Bertone} {et~al.}(2005){Bertone}, {Hooper} \& {Silk}}]{Bertone05}
{Bertone}, G., {Hooper}, D. \& {Silk}, J. 2005, \emph{\physrep}, \textbf{405},
  279, \href{http://arxiv.org/abs/arXiv:hep-ph/0404175}{arXiv:hep-ph/0404175}

\bibitem[{{Bertone} {et~al.}(2004){Bertone}, {Nezri}, {Orloff} \&
  {Silk}}]{Bertone04}
{Bertone}, G., {Nezri}, E., {Orloff}, J. \& {Silk}, J. 2004, \emph{\prd},
  \textbf{70}, 063503,
  \href{http://arxiv.org/abs/arXiv:astro-ph/0403322}{arXiv:astro-ph/0403322}

\bibitem[{{Birkedal} {et~al.}(2004){Birkedal}, {Matchev} \&
  {Perelstein}}]{Birkedal04}
{Birkedal}, A., {Matchev}, K. \& {Perelstein}, M. 2004, \emph{\prd},
  \textbf{70}, 077701,
  \href{http://arxiv.org/abs/arXiv:hep-ph/0403004}{arXiv:hep-ph/0403004}

\bibitem[{{Blasi} \& {Serpico}(2009)}]{Blasi09}
{Blasi}, P. \& {Serpico}, P.~D. 2009, \emph{\prl}, \textbf{103}, 081103,
  \href{http://arxiv.org/abs/0904.0871}{arXiv:0904.0871}

\bibitem[{{Blennow} {et~al.}(2008){Blennow}, {Edsj{\"o}} \&
  {Ohlsson}}]{Blennow08}
{Blennow}, M., {Edsj{\"o}}, J. \& {Ohlsson}, T. 2008, \emph{\jcap}, \textbf{1},
  21, \href{http://arxiv.org/abs/arXiv:0709.3898}{arXiv:0709.3898}

\bibitem[{{Blennow} {et~al.}(2010){Blennow}, {Melb{\'e}us} \&
  {Ohlsson}}]{Blennow10}
{Blennow}, M., {Melb{\'e}us}, H. \& {Ohlsson}, T. 2010, \emph{\jcap},
  \textbf{1}, 18, \href{http://arxiv.org/abs/0910.1588}{arXiv:0910.1588}

\bibitem[{{Bottino} {et~al.}(2002){Bottino}, {Fiorentini}, {Fornengo}, {Ricci},
  {Scopel} \& {Villante}}]{Bottino02}
{Bottino}, A., {Fiorentini}, G., {Fornengo}, N., {Ricci}, B., {Scopel}, S. \&
  {Villante}, F.~L. 2002, \emph{\prd}, \textbf{66}, 053005,
  \href{http://arxiv.org/abs/arXiv:hep-ph/0206211}{arXiv:hep-ph/0206211}

\bibitem[{{Bouquet} \& {Salati}(1989)}]{BouquetSalati89a}
{Bouquet}, A. \& {Salati}, P. 1989, \emph{\aap}, \textbf{217}, 270

\bibitem[{{Bringmann}(2009)}]{Bringmann09_clumps}
{Bringmann}, T. 2009, \emph{\njp}, \textbf{11}, 105027,
  \href{http://arxiv.org/abs/0903.0189}{arXiv:0903.0189}

\bibitem[{{Bringmann} {et~al.}(2008){Bringmann}, {Bergstr{\"o}m} \&
  {Edsj{\"o}}}]{Bringmann08}
{Bringmann}, T., {Bergstr{\"o}m}, L. \& {Edsj{\"o}}, J. 2008, \emph{\jhep},
  \textbf{1}, 49, \href{http://arxiv.org/abs/arXiv:0710.3169}{arXiv:0710.3169}

\bibitem[{{Bringmann} {et~al.}(2009){Bringmann}, {Doro} \&
  {Fornasa}}]{Bringmann09}
{Bringmann}, T., {Doro}, M. \& {Fornasa}, M. 2009, \emph{\jcap}, \textbf{1},
  16, \href{http://arxiv.org/abs/arXiv:0809.2269}{arXiv:0809.2269}

\bibitem[{{Bruch} {et~al.}(2009){Bruch}, {Peter}, {Read}, {Baudis} \&
  {Lake}}]{Bruch09}
{Bruch}, T., {Peter}, A.~H.~G., {Read}, J., {Baudis}, L. \& {Lake}, G. 2009,
  \emph{\plb}, \textbf{674}, 250,
  \href{http://arxiv.org/abs/0902.4001}{arXiv:0902.4001}

\bibitem[{{Buckley} {et~al.}(2010){Buckley}, {Spolyar}, {Freese}, {Hooper} \&
  {Murayama}}]{Buckley10}
{Buckley}, M.~R., {Spolyar}, D., {Freese}, K., {Hooper}, D. \& {Murayama}, H.
  2010, \emph{\prd}, \textbf{81}, 016006

\bibitem[{Carrol \& Ostlie(1996)}]{CandO}
Carrol, B.~W. \& Ostlie, D.~A. 1996, \emph{An Introduction to Modern
  Astrophysics} (Addison-Wesley, Reading, MA)

\bibitem[{{Casanellas} \& {Lopes}(2009)}]{Casanellas09}
{Casanellas}, J. \& {Lopes}, I. 2009, \emph{\apj}, \textbf{705}, 135,
  \href{http://arxiv.org/abs/0909.1971}{arXiv:0909.1971}

\bibitem[{{Cerde\~no} \& {Green}(2010)}]{CerdenoGreen10}
{Cerde\~no}, D.~G. \& {Green}, A.~M. 2010, in \emph{Particle Dark Matter:
  Observations, Models and Searches}, ed. G.~Bertone (Cambridge University
  Press), 347--369,
  \href{http://arxiv.org/abs/arXiv:1002.1912}{arXiv:1002.1912}

\bibitem[{{Chamseddine} {et~al.}(1982){Chamseddine}, {Arnowitt} \&
  {Nath}}]{Chamseddine82}
{Chamseddine}, A.~H., {Arnowitt}, R. \& {Nath}, P. 1982, \emph{\prl},
  \textbf{49}, 970

\bibitem[{{Chang} {et~al.}(2008){Chang}, {Adams}, {Ahn}, {Bashindzhagyan},
  {Christl}, {Ganel}, {Guzik} \& {Isbert} {et al.}}]{Chang08}
{Chang}, J., {Adams}, J.~H., {Ahn}, H.~S., {Bashindzhagyan}, G.~L., {Christl},
  M., {Ganel}, O., {Guzik}, T.~G. \& {Isbert}, J. {et al.} 2008, \emph{\nat},
  \textbf{456}, 362

\bibitem[{{Charbonneau}(1995)}]{Pikaia}
{Charbonneau}, P. 1995, \emph{\apjs}, \textbf{101}, 309

\bibitem[{{Chen} {et~al.}(2009){Chen}, {Mohapatra}, {Nussinov} \&
  {Zhang}}]{Chen09}
{Chen}, S., {Mohapatra}, R.~N., {Nussinov}, S. \& {Zhang}, Y. 2009,
  \emph{\plb}, \textbf{677}, 311,
  \href{http://arxiv.org/abs/0903.2562}{arXiv:0903.2562}

\bibitem[{{Cholis} {et~al.}(2009){Cholis}, {Goodenough}, {Hooper}, {Simet} \&
  {Weiner}}]{Cholis09}
{Cholis}, I., {Goodenough}, L., {Hooper}, D., {Simet}, M. \& {Weiner}, N. 2009,
  \emph{\prd}, \textbf{80}, 123511,
  \href{http://arxiv.org/abs/0809.1683}{arXiv:0809.1683}

\bibitem[{{Choutko} \& {Giovacchini}(2008)}]{Choutko08}
{Choutko}, V. \& {Giovacchini}, F. 2008, in \emph{Proceedings of the
  30$^\mathrm{th}$ International Cosmic Ray Conference, M\'erida, Mexico},
  765--768

\bibitem[{{Cirelli} {et~al.}(2006){Cirelli}, {Fornengo} \&
  {Strumia}}]{Cirelli06}
{Cirelli}, M., {Fornengo}, N. \& {Strumia}, A. 2006, \emph{\nphysb},
  \textbf{753}, 178,
  \href{http://arxiv.org/abs/arXiv:hep-ph/0512090}{arXiv:hep-ph/0512090}

\bibitem[{{Cirelli} \& {Strumia}(2009)}]{Cirelli09}
{Cirelli}, M. \& {Strumia}, A. 2009, \emph{\njp}, \textbf{11}, 105005,
  \href{http://arxiv.org/abs/0903.3381}{arXiv:0903.3381}

\bibitem[{{Clowe} {et~al.}(2006){Clowe}, {Brada{\v c}}, {Gonzalez},
  {Markevitch}, {Randall}, {Jones} \& {Zaritsky}}]{bullet}
{Clowe}, D., {Brada{\v c}}, M., {Gonzalez}, A.~H., {Markevitch}, M., {Randall},
  S.~W., {Jones}, C. \& {Zaritsky}, D. 2006, \emph{\apjl}, \textbf{648}, L109,
  \href{http://arxiv.org/abs/arXiv:astro-ph/0608407}{arXiv:astro-ph/0608407}

\bibitem[{{Coc} {et~al.}(2004){Coc}, {Vangioni-Flam}, {Descouvemont},
  {Adahchour} \& {Angulo}}]{Coc04}
{Coc}, A., {Vangioni-Flam}, E., {Descouvemont}, P., {Adahchour}, A. \&
  {Angulo}, C. 2004, \emph{\apj}, \textbf{600}, 544,
  \href{http://arxiv.org/abs/arXiv:astro-ph/0309480}{arXiv:astro-ph/0309480}

\bibitem[{{Cole} {et~al.}(2005){Cole}, {Percival}, {Peacock}, {Norberg},
  {Baugh}, {Frenk}, {Baldry} \& {Bland-Hawthorn} {et al.}}]{Cole05}
{Cole}, S., {Percival}, W.~J., {Peacock}, J.~A., {Norberg}, P., {Baugh}, C.~M.,
  {Frenk}, C.~S., {Baldry}, I. \& {Bland-Hawthorn}, J. {et al.} 2005,
  \emph{\mnras}, \textbf{362}, 505,
  \href{http://arxiv.org/abs/arXiv:astro-ph/0501174}{arXiv:astro-ph/0501174}

\bibitem[{{Coleman} \& {Mandula}(1967)}]{Coleman67}
{Coleman}, S. \& {Mandula}, J. 1967, \emph{\physrev}, \textbf{159}, 1251

\bibitem[{{Colless} {et~al.}(2001){Colless}, {Dalton}, {Maddox}, {Sutherland},
  {Norberg}, {Cole}, {Bland-Hawthorn} \& {Bridges} {et al.}}]{Colless01}
{Colless}, M., {Dalton}, G., {Maddox}, S., {Sutherland}, W., {Norberg}, P.,
  {Cole}, S., {Bland-Hawthorn}, J. \& {Bridges}, T. {et al.} 2001,
  \emph{\mnras}, \textbf{328}, 1039,
  \href{http://arxiv.org/abs/arXiv:astro-ph/0106498}{arXiv:astro-ph/0106498}

\bibitem[{{Covi} \& {Kim}(2009)}]{CoviKim}
{Covi}, L. \& {Kim}, J.~E. 2009, \emph{\njp}, \textbf{11}, 105003,
  \href{http://arxiv.org/abs/0902.0769}{arXiv:0902.0769}

\bibitem[{{Davidson} {et~al.}(2000){Davidson}, {Hannestad} \&
  {Raffelt}}]{Davidson00}
{Davidson}, S., {Hannestad}, S. \& {Raffelt}, G. 2000, \emph{\jhep},
  \textbf{5}, 3,
  \href{http://arxiv.org/abs/arXiv:hep-ph/0001179}{arXiv:hep-ph/0001179}

\bibitem[{{de Blok} {et~al.}(2001){de Blok}, {McGaugh}, {Bosma} \&
  {Rubin}}]{deBlok01}
{de Blok}, W.~J.~G., {McGaugh}, S.~S., {Bosma}, A. \& {Rubin}, V.~C. 2001,
  \emph{\apjl}, \textbf{552}, L23,
  \href{http://arxiv.org/abs/arXiv:astro-ph/0103102}{arXiv:astro-ph/0103102}

\bibitem[{{Del Popolo} \& {Kroupa}(2009)}]{DelPopolo09}
{Del Popolo}, A. \& {Kroupa}, P. 2009, \emph{\aap}, \textbf{502}, 733,
  \href{http://arxiv.org/abs/0906.1146}{arXiv:0906.1146}

\bibitem[{{Desai} {et~al.}(2004){Desai}, {Ashie}, {Fukuda}, {Fukuda},
  {Ishihara}, {Itow}, {Koshio} \& {Minamino} {et al.}}]{Desai04}
{Desai}, S., {Ashie}, Y., {Fukuda}, S., {Fukuda}, Y., {Ishihara}, K., {Itow},
  Y., {Koshio}, Y. \& {Minamino}, A. {et al.} 2004, \emph{\prd}, \textbf{70},
  083523,
  \href{http://arxiv.org/abs/arXiv:hep-ex/0404025}{arXiv:hep-ex/0404025}

\bibitem[{{Diemand} {et~al.}(2007){Diemand}, {Kuhlen} \& {Madau}}]{vialactea}
{Diemand}, J., {Kuhlen}, M. \& {Madau}, P. 2007, \emph{\apj}, \textbf{657},
  262,
  \href{http://arxiv.org/abs/arXiv:astro-ph/0611370}{arXiv:astro-ph/0611370}

\bibitem[{{Diemand} {et~al.}(2008){Diemand}, {Kuhlen}, {Madau}, {Zemp},
  {Moore}, {Potter} \& {Stadel}}]{vialacteaII}
{Diemand}, J., {Kuhlen}, M., {Madau}, P., {Zemp}, M., {Moore}, B., {Potter}, D.
  \& {Stadel}, J. 2008, \emph{\nat}, \textbf{454}, 735,
  \href{http://arxiv.org/abs/0805.1244}{arXiv:0805.1244}

\bibitem[{{Dobler} {et~al.}(2010){Dobler}, {Finkbeiner}, {Cholis}, {Slatyer} \&
  {Weiner}}]{Dobler09}
{Dobler}, G., {Finkbeiner}, D.~P., {Cholis}, I., {Slatyer}, T.~R. \& {Weiner},
  N. 2010, \emph{\apj}, \textbf{717}, 825, \href{http://arxiv.org/abs/arXiv:0910.4583}{arXiv:0910.4583}

\bibitem[{{Donato} {et~al.}(2008){Donato}, {Fornengo} \& {Maurin}}]{Donato08}
{Donato}, F., {Fornengo}, N. \& {Maurin}, D. 2008, \emph{\prd}, \textbf{78},
  043506, \href{http://arxiv.org/abs/0803.2640}{arXiv:0803.2640}

\bibitem[{{Donato} {et~al.}(2000){Donato}, {Fornengo} \& {Salati}}]{Donato00}
{Donato}, F., {Fornengo}, N. \& {Salati}, P. 2000, \emph{\prd}, \textbf{62},
  043003,
  \href{http://arxiv.org/abs/arXiv:hep-ph/9904481}{arXiv:hep-ph/9904481}

\bibitem[{{Donato} {et~al.}(2009){Donato}, {Maurin}, {Brun}, {Delahaye} \&
  {Salati}}]{Donato09}
{Donato}, F., {Maurin}, D., {Brun}, P., {Delahaye}, T. \& {Salati}, P. 2009,
  \emph{\prl}, \textbf{102}, 071301,
  \href{http://arxiv.org/abs/0810.5292}{arXiv:0810.5292}

\bibitem[{{Ellis} {et~al.}(2010){Ellis}, {Olive}, {Savage} \&
  {Spanos}}]{Ellis09}
{Ellis}, J., {Olive}, K.~A., {Savage}, C. \& {Spanos}, V.~C. 2010, \emph{\prd}, \textbf{81}, 085004,
  \href{http://arxiv.org/abs/0912.3137}{arXiv:0912.3137}

\bibitem[{Essig {et~al.}(2009)Essig, Sehgal \& Strigari}]{Essig09}
Essig, R., Sehgal, N. \& Strigari, L.~E. 2009, \emph{\prd}, \textbf{80},
  023506, \href{http://arxiv.org/abs/0902.4750}{arXiv:0902.4750}

\bibitem[{{Faber} \& {Jackson}(1976)}]{FaberJackson}
{Faber}, S.~M. \& {Jackson}, R.~E. 1976, \emph{\apj}, \textbf{204}, 668

\bibitem[{{Fairbairn} {et~al.}(2007){Fairbairn}, {Rashba} \&
  {Troitsky}}]{Fairbairn07}
{Fairbairn}, M., {Rashba}, T. \& {Troitsky}, S. 2007, \emph{\prl}, \textbf{98},
  201801,
  \href{http://arxiv.org/abs/arXiv:astro-ph/0610844}{arXiv:astro-ph/0610844}

\bibitem[{{Fairbairn} \& {Schwetz}(2009)}]{Fairbairn09}
{Fairbairn}, M. \& {Schwetz}, T. 2009, \emph{\jcap}, \textbf{1}, 37,
  \href{http://arxiv.org/abs/arXiv:0808.0704}{arXiv:0808.0704}

\bibitem[{{Fairbairn} {et~al.}(2008){Fairbairn}, {Scott} \&
  {Edsj{\"o}}}]{Fairbairn08}
{Fairbairn}, M., {Scott}, P. \& {Edsj{\"o}}, J. 2008, \emph{\prd}, \textbf{77},
  047301, \href{http://arxiv.org/abs/arXiv:0710.3396}{arXiv:0710.3396}, \citepalias{Fairbairn08}

\bibitem[{{Falk} {et~al.}(1994){Falk}, {Olive} \& {Srednicki}}]{Falk94}
{Falk}, T., {Olive}, K.~A. \& {Srednicki}, M. 1994, \emph{\plb}, \textbf{339},
  248, \href{http://arxiv.org/abs/arXiv:hep-ph/9409270}{arXiv:hep-ph/9409270}

\bibitem[{{Feng}(2010)}]{Feng10}
{Feng}, J.~L. 2010, in \emph{Particle Dark Matter: Observations, Models and
  Searches}, ed. G.~Bertone (Cambridge University Press), 190--203,
  \href{http://arxiv.org/abs/arXiv:1002.3828}{arXiv:1002.3828}

\bibitem[{{Feng} \& {Kumar}(2008)}]{FengKumar08}
{Feng}, J.~L. \& {Kumar}, J. 2008, \emph{\prl}, \textbf{101}, 231301,
  \href{http://arxiv.org/abs/0803.4196}{arXiv:0803.4196}

\bibitem[{{Finkbeiner} \& {Weiner}(2007)}]{FinkbeinerWeiner07}
{Finkbeiner}, D.~P. \& {Weiner}, N. 2007, \emph{\prd}, \textbf{76}, 083519,
  \href{http://arxiv.org/abs/arXiv:astro-ph/0702587}{arXiv:astro-ph/0702587}

\bibitem[{{Flacke} {et~al.}(2009){Flacke}, {Menon}, {Hooper} \&
  {Freese}}]{Flacke09}
{Flacke}, T., {Menon}, A., {Hooper}, D. \& {Freese}, K. 2009,
  \href{http://arxiv.org/abs/0908.0899}{arXiv:0908.0899}

\bibitem[{{Freedman} {et~al.}(2001){Freedman}, {Madore}, {Gibson}, {Ferrarese},
  {Kelson}, {Sakai}, {Mould} \& {Kennicutt} {et al.}}]{Freedman01}
{Freedman}, W.~L., {Madore}, B.~F., {Gibson}, B.~K., {Ferrarese}, L., {Kelson},
  D.~D., {Sakai}, S., {Mould}, J.~R. \& {Kennicutt}, Jr., R.~C. {et al.} 2001,
  \emph{\apj}, \textbf{553}, 47,
  \href{http://arxiv.org/abs/arXiv:astro-ph/0012376}{arXiv:astro-ph/0012376}

\bibitem[{{Freese} {et~al.}(2008{\natexlab{a}}){Freese}, {Bodenheimer},
  {Spolyar} \& {Gondolo}}]{Freese08c}
{Freese}, K., {Bodenheimer}, P., {Spolyar}, D. \& {Gondolo}, P.
  2008{\natexlab{a}}, \emph{\apjl}, \textbf{685}, L101,
  \href{http://arxiv.org/abs/arXiv:0806.0617}{arXiv:0806.0617}

\bibitem[{{Freese} {et~al.}(2009){Freese}, {Gondolo}, {Sellwood} \&
  {Spolyar}}]{Freese08b}
{Freese}, K., {Gondolo}, P., {Sellwood}, J.~A. \& {Spolyar}, D. 2009,
  \emph{\apj}, \textbf{693}, 1563,
  \href{http://arxiv.org/abs/0805.3540}{arXiv:0805.3540}

\bibitem[{{Freese} {et~al.}(2010){Freese}, {Ilie}, {Spolyar}, {Valluri} \&
  {Bodenheimer}}]{Freese10}
{Freese}, K., {Ilie}, C., {Spolyar}, D., {Valluri}, M. \& {Bodenheimer}, P.
  2010, \emph{\apj}, \textbf{716}, 1397, 
  \href{http://arxiv.org/abs/arXiv:1002.2233}{arXiv:1002.2233}

\bibitem[{{Freese} {et~al.}(2008{\natexlab{b}}){Freese}, {Spolyar} \&
  {Aguirre}}]{Freese08a}
{Freese}, K., {Spolyar}, D. \& {Aguirre}, A. 2008{\natexlab{b}}, \emph{\jcap},
  \textbf{11}, 14, \href{http://arxiv.org/abs/arXiv:0802.1724}{arXiv:0802.1724}

\bibitem[{{Fujita} {et~al.}(2009){Fujita}, {Kohri}, {Yamazaki} \&
  {Ioka}}]{Fujita09}
{Fujita}, Y., {Kohri}, K., {Yamazaki}, R. \& {Ioka}, K. 2009, \emph{\prd},
  \textbf{80}, 063003, \href{http://arxiv.org/abs/0903.5298}{arXiv:0903.5298}

\bibitem[{{Gaitskell}(2004)}]{Gaitskell04}
{Gaitskell}, R.~J. 2004, \emph{\arnps}, \textbf{54}, 315

\bibitem[{{Galli} {et~al.}(2009){Galli}, {Iocco}, {Bertone} \&
  {Melchiorri}}]{Galli09}
{Galli}, S., {Iocco}, F., {Bertone}, G. \& {Melchiorri}, A. 2009, \emph{\prd},
  \textbf{80}, 023505, \href{http://arxiv.org/abs/0905.0003}{arXiv:0905.0003}

\bibitem[{{Geha} {et~al.}(2009){Geha}, {Willman}, {Simon}, {Strigari}, {Kirby},
  {Law} \& {Strader}}]{Geha09}
{Geha}, M., {Willman}, B., {Simon}, J.~D., {Strigari}, L.~E., {Kirby}, E.~N.,
  {Law}, D.~R. \& {Strader}, J. 2009, \emph{\apj}, \textbf{692}, 1464,
  \href{http://arxiv.org/abs/arXiv:0809.2781}{arXiv:0809.2781}

\bibitem[{{Gentile} {et~al.}(2004){Gentile}, {Salucci}, {Klein}, {Vergani} \&
  {Kalberla}}]{Gentile04}
{Gentile}, G., {Salucci}, P., {Klein}, U., {Vergani}, D. \& {Kalberla}, P.
  2004, \emph{\mnras}, \textbf{351}, 903,
  \href{http://arxiv.org/abs/arXiv:astro-ph/0403154}{arXiv:astro-ph/0403154}

\bibitem[{{Giudice} \& {Rattazzi}(1999)}]{Giudice99}
{Giudice}, G.~F. \& {Rattazzi}, R. 1999, \emph{\physrep}, \textbf{322}, 419,
  \href{http://arxiv.org/abs/arXiv:hep-ph/9801271}{arXiv:hep-ph/9801271}

\bibitem[{{Gondolo} {et~al.}(2004){Gondolo}, {Edsj{\"o}}, {Ullio},
  {Bergstr{\"o}m}, {Schelke} \& {Baltz}}]{DarkSUSY}
{Gondolo}, P., {Edsj{\"o}}, J., {Ullio}, P., {Bergstr{\"o}m}, L., {Schelke}, M.
  \& {Baltz}, E.~A. 2004, \emph{\jcap}, \textbf{7}, 8,
  \href{http://arxiv.org/abs/arXiv:astro-ph/0406204}{arXiv:astro-ph/0406204}

\bibitem[{{Goodenough} \& {Hooper}(2009)}]{Goodenough09}
{Goodenough}, L. \& {Hooper}, D. 2009,
  \href{http://arxiv.org/abs/arXiv:0910.2998}{arXiv:0910.2998}

\bibitem[{{Goodman} \& {Witten}(1985)}]{GoodmanWitten85}
{Goodman}, M.~W. \& {Witten}, E. 1985, \emph{\prd}, \textbf{31}, 3059

\bibitem[{{Gould}(1987)}]{Gould87b}
{Gould}, A. 1987, \emph{\apj}, \textbf{321}, 571

\bibitem[{{Gould}(1991)}]{Gould91}
---. 1991, \emph{\apj}, \textbf{368}, 610

\bibitem[{{Governato} {et~al.}(2010){Governato}, {Brook}, {Mayer}, {Brooks},
  {Rhee}, {Wadsley}, {Jonsson} \& {Willman} {et al.}}]{Governato10}
{Governato}, F., {Brook}, C., {Mayer}, L., {Brooks}, A., {Rhee}, G., {Wadsley},
  J., {Jonsson}, P. \& {Willman}, B. {et al.} 2010, \emph{\nat}, \textbf{463},
  203, \href{http://arxiv.org/abs/arXiv:0911.2237}{arXiv:0911.2237}

\bibitem[{{Green}(2007)}]{Green07}
{Green}, A.~M. 2007, \emph{\jcap}, \textbf{8}, 22,
  \href{http://arxiv.org/abs/arXiv:hep-ph/0703217}{arXiv:hep-ph/0703217}

\bibitem[{{Green}(2008)}]{Green08}
---. 2008, \emph{\jcap}, \textbf{7}, 5,
  \href{http://arxiv.org/abs/0805.1704}{arXiv:0805.1704}

\bibitem[{{Green} {et~al.}(2004{\natexlab{a}}){Green}, {Hofmann} \&
  {Schwarz}}]{Green04b}
{Green}, A.~M., {Hofmann}, S. \& {Schwarz}, D.~J. 2004{\natexlab{a}},
  \emph{\mnras}, \textbf{353}, L23,
  \href{http://arxiv.org/abs/arXiv:astro-ph/0309621}{arXiv:astro-ph/0309621}

\bibitem[{{Green} {et~al.}(2005){Green}, {Hofmann} \& {Schwarz}}]{Green05}
---. 2005, \emph{\jcap}, \textbf{8}, 3,
  \href{http://arxiv.org/abs/arXiv:astro-ph/0503387}{arXiv:astro-ph/0503387}

\bibitem[{{Green} \& {Liddle}(1997)}]{GreenLiddle}
{Green}, A.~M. \& {Liddle}, A.~R. 1997, \emph{\prd}, \textbf{56}, 6166,
  \href{http://arxiv.org/abs/arXiv:astro-ph/9704251}{arXiv:astro-ph/9704251}

\bibitem[{{Green} {et~al.}(2004{\natexlab{b}}){Green}, {Liddle}, {Malik} \&
  {Sasaki}}]{Green04c}
{Green}, A.~M., {Liddle}, A.~R., {Malik}, K.~A. \& {Sasaki}, M.
  2004{\natexlab{b}}, \emph{\prd}, \textbf{70}, 041502,
  \href{http://arxiv.org/abs/arXiv:astro-ph/0403181}{arXiv:astro-ph/0403181}

\bibitem[{{Haag} {et~al.}(1975){Haag}, {Lopusza{\'n}ski} \& {Sohnius}}]{Haag75}
{Haag}, R., {Lopusza{\'n}ski}, J.~T. \& {Sohnius}, M. 1975, \emph{\nphysb},
  \textbf{88}, 257

\bibitem[{{Hailey}(2009)}]{Hailey09}
{Hailey}, C.~J. 2009, \emph{\njp}, \textbf{11}, 105022

\bibitem[{Hansen {et~al.}(2006)Hansen, Moore, Zemp \& Stadel}]{Hansen:2005yj}
Hansen, S.~H., Moore, B., Zemp, M. \& Stadel, J. 2006, \emph{\jcap},
  \textbf{0601}, 014,
  \href{http://arxiv.org/abs/astro-ph/0505420}{arXiv:astro-ph/0505420}

\bibitem[{{Hisano} {et~al.}(2009){Hisano}, {Kawasaki}, {Kohri} \&
  {Nakayama}}]{Hisano09}
{Hisano}, J., {Kawasaki}, M., {Kohri}, K. \& {Nakayama}, K. 2009, \emph{\prd},
  \textbf{79}, 043516, \href{http://arxiv.org/abs/0812.0219}{arXiv:0812.0219}

\bibitem[{{Hofmann} {et~al.}(2001){Hofmann}, {Schwarz} \&
  {St{\"o}cker}}]{Hofman01}
{Hofmann}, S., {Schwarz}, D.~J. \& {St{\"o}cker}, H. 2001, \emph{\prd},
  \textbf{64}, 083507,
  \href{http://arxiv.org/abs/arXiv:astro-ph/0104173}{arXiv:astro-ph/0104173}

\bibitem[{{Holdom}(1986)}]{Holdom86}
{Holdom}, B. 1986, \emph{\plb}, \textbf{166}, 196

\bibitem[{{Hooper} {et~al.}(2009){Hooper}, {Blasi} \& {Dario
  Serpico}}]{Hooper09}
{Hooper}, D., {Blasi}, P. \& {Dario Serpico}, P. 2009, \emph{\jcap},
  \textbf{1}, 25, \href{http://arxiv.org/abs/0810.1527}{arXiv:0810.1527}

\bibitem[{{Hooper} {et~al.}(2010){Hooper}, {Spolyar}, {Vallinotto} \&
  {Gnedin}}]{Hooper10}
{Hooper}, D., {Spolyar}, D., {Vallinotto}, A. \& {Gnedin}, N.~Y. 2010, \emph{\prd},
  \textbf{81}, 103531,
  \href{http://arxiv.org/abs/arXiv:1002.0005}{arXiv:1002.0005}

\bibitem[{{Hu} \& {Dodelson}(2002)}]{HuDodelson}
{Hu}, W. \& {Dodelson}, S. 2002, \emph{\araa}, \textbf{40}, 171,
  \href{http://arxiv.org/abs/arXiv:astro-ph/0110414}{arXiv:astro-ph/0110414}

\bibitem[{{Ibarra} \& {Tran}(2009)}]{Ibarra09}
{Ibarra}, A. \& {Tran}, D. 2009, \emph{\jcap}, \textbf{6}, 4,
  \href{http://arxiv.org/abs/0904.1410}{arXiv:0904.1410}

\bibitem[{{Ingelman} \& {Thunman}(1996)}]{Ingelman96}
{Ingelman}, G. \& {Thunman}, M. 1996, \emph{\prd}, \textbf{54}, 4385,
  \href{http://arxiv.org/abs/arXiv:hep-ph/9604288}{arXiv:hep-ph/9604288}

\bibitem[{{Iocco}(2008)}]{Iocco08a}
{Iocco}, F. 2008, \emph{\apjl}, \textbf{677}, L1,
  \href{http://arxiv.org/abs/arXiv:0802.0941}{arXiv:0802.0941}

\bibitem[{{Iocco} {et~al.}(2008){Iocco}, {Bressan}, {Ripamonti}, {Schneider},
  {Ferrara} \& {Marigo}}]{Iocco08b}
{Iocco}, F., {Bressan}, A., {Ripamonti}, E., {Schneider}, R., {Ferrara}, A. \&
  {Marigo}, P. 2008, \emph{\mnras}, \textbf{390}, 1655,
  \href{http://arxiv.org/abs/arXiv:0805.4016}{arXiv:0805.4016}

\bibitem[{{Iocco} {et~al.}(2009){Iocco}, {Mangano}, {Miele}, {Pisanti} \&
  {Serpico}}]{IoccoBBN}
{Iocco}, F., {Mangano}, G., {Miele}, G., {Pisanti}, O. \& {Serpico}, P.~D.
  2009, \emph{\physrep}, \textbf{472}, 1,
  \href{http://arxiv.org/abs/arXiv:0809.0631}{arXiv:0809.0631}

\bibitem[{{Jedamzik} \& {Pospelov}(2009)}]{Jedamzik09}
{Jedamzik}, K. \& {Pospelov}, M. 2009, \emph{\njp}, \textbf{11}, 105028,
  \href{http://arxiv.org/abs/0906.2087}{arXiv:0906.2087}

\bibitem[{{Josan} {et~al.}(2009){Josan}, {Green} \& {Malik}}]{Josan09}
{Josan}, A.~S., {Green}, A.~M. \& {Malik}, K.~A. 2009, \emph{\prd},
  \textbf{79}, 103520,
  \href{http://arxiv.org/abs/arXiv:0903.3184}{arXiv:0903.3184}

\bibitem[{{Jungman} {et~al.}(1996){Jungman}, {Kamionkowski} \&
  {Griest}}]{Jungman96}
{Jungman}, G., {Kamionkowski}, M. \& {Griest}, K. 1996, \emph{\physrep},
  \textbf{267}, 195,
  \href{http://arxiv.org/abs/arXiv:hep-ph/9506380}{arXiv:hep-ph/9506380}

\bibitem[{{Kang} {et~al.}(2008){Kang}, {Luty} \& {Nasri}}]{Kang08}
{Kang}, J., {Luty}, M.~A. \& {Nasri}, S. 2008, \emph{\jhep}, \textbf{9}, 86,
  \href{http://arxiv.org/abs/arXiv:hep-ph/0611322}{arXiv:hep-ph/0611322}

\bibitem[{{Katz} {et~al.}(2010){Katz}, {Blum} \& {Waxman}}]{Katz09}
{Katz}, B., {Blum}, K. \& {Waxman}, E. 2010, \emph{\mnras}, \textbf{405}, 1458,
  \href{http://arxiv.org/abs/arXiv:0907.1686}{arXiv:0907.1686}

\bibitem[{{Kim} \& {Carosi}(2010)}]{Kim08}
{Kim}, J.~E. \& {Carosi}, G. 2010, \emph{\rmp}, \textbf{82}, 557, 
  \href{http://arxiv.org/abs/arXiv:0807.3125}{arXiv:0807.3125}

\bibitem[{Kippenhahn \& Weigert(1991)}]{KipW}
Kippenhahn, R. \& Weigert, A. 1991, \emph{Stellar Structure and Evolution},
  $1^{\mathrm{st}}$ edn. (Springer-Verlag, Berlin)

\bibitem[{{Kolb} {et~al.}(1999){Kolb}, {Chung} \& {Riotto}}]{Kolb99}
{Kolb}, E.~W., {Chung}, D.~J.~H. \& {Riotto}, A. 1999, in \emph{Proceedings of
  Dark matter in Astrophysics and Particle Physics: Dark 98}, ed.
  {H.~V.~Klapdor-Kleingrothaus \& L.~Baudis}, 592,
  \href{http://arxiv.org/abs/arXiv:hep-ph/9810361}{arXiv:hep-ph/9810361}

\bibitem[{{Kolb} \& {Turner}(1990)}]{KolbTurner}
{Kolb}, E.~W. \& {Turner}, M.~S. 1990, \emph{{The Early Universe}} (Frontiers
  in Physics, Addison-Wesley, Reading, MA)

\bibitem[{{Komatsu} {et~al.}(2009){Komatsu}, {Dunkley}, {Nolta}, {Bennett},
  {Gold}, {Hinshaw}, {Jarosik} \& {Larson} {et al.}}]{WMAP5}
{Komatsu}, E., {Dunkley}, J., {Nolta}, M.~R., {Bennett}, C.~L., {Gold}, B.,
  {Hinshaw}, G., {Jarosik}, N. \& {Larson}, D. {et al.} 2009, \emph{\apjs},
  \textbf{180}, 330, \href{http://arxiv.org/abs/0803.0547}{arXiv:0803.0547}

\bibitem[{{Komatsu} {et~al.}(2011){Komatsu}, {Smith}, {Dunkley}, {Bennett},
  {Gold}, {Hinshaw}, {Jarosik} \& {Larson} {et al.}}]{wmap7}
{Komatsu}, E., {Smith}, K.~M., {Dunkley}, J., {Bennett}, C.~L., {Gold}, B.,
  {Hinshaw}, G., {Jarosik}, N. \& {Larson}, D. {et al.} 2011, \emph{\apjs},
  \textbf{192}, 18, \href{http://arxiv.org/abs/1001.4538}{arXiv:1001.4538}

\bibitem[{{Kowalski} {et~al.}(2008){Kowalski}, {Rubin}, {Aldering},
  {Agostinho}, {Amadon}, {Amanullah}, {Balland} \& {Barbary} {et
  al.}}]{Kowalski08}
{Kowalski}, M., {Rubin}, D., {Aldering}, G., {Agostinho}, R.~J., {Amadon}, A.,
  {Amanullah}, R., {Balland}, C. \& {Barbary}, K. {et al.} 2008, \emph{\apj},
  \textbf{686}, 749, \href{http://arxiv.org/abs/0804.4142}{arXiv:0804.4142}

\bibitem[{{Kuhlen} {et~al.}(2008){Kuhlen}, {Diemand} \& {Madau}}]{Kuhlen08}
{Kuhlen}, M., {Diemand}, J. \& {Madau}, P. 2008, \emph{\apj}, \textbf{686},
  262, \href{http://arxiv.org/abs/0805.4416}{arXiv:0805.4416}

\bibitem[{{Kusenko}(2009)}]{Kusenko09}
{Kusenko}, A. 2009, \emph{\physrep}, \textbf{481}, 1,
  \href{http://arxiv.org/abs/0906.2968}{arXiv:0906.2968}

\bibitem[{{Kuster} {et~al.}(2008){Kuster}, {Raffelt} \&
  Beltr{\'a}n}]{LNPaxions}
{Kuster}, M., {Raffelt}, G. \& Beltr{\'a}n, B., eds. 2008, \emph{\lnp}, Vol.
  741, \emph{Axions: Theory, Cosmology, and Experimental Searches}
  (Springer-Verlag, Berlin)

\bibitem[{{Lacki} \& {Beacom}(2010)}]{Lacki10}
{Lacki}, B.~C. \& {Beacom}, J.~F. 2010, \emph{\apj}, \textbf{720}, L67, 
  \href{http://arxiv.org/abs/1003.3466}{arXiv:1003.3466}

\bibitem[{{Larson} {et~al.}(2011){Larson}, {Dunkley}, {Hinshaw}, {Komatsu},
  {Nolta}, {Bennett}, {Gold} \& {Halpern} {et al.}}]{WMAP7Larson}
{Larson}, D., {Dunkley}, J., {Hinshaw}, G., {Komatsu}, E., {Nolta}, M.~R.,
  {Bennett}, C.~L., {Gold}, B. \& {Halpern}, M. {et al.} 2011, \emph{\apjs}, \textbf{192}, 16,
  \href{http://arxiv.org/abs/arXiv:1001.4635}{arXiv:1001.4635}

\bibitem[{{Lebedenko} {et~al.}(2009){Lebedenko}, {Ara{\'u}jo}, {Barnes},
  {Bewick}, {Cashmore}, {Chepel}, {Currie} \& {Davidge} {et al.}}]{zeplinDec08}
{Lebedenko}, V.~N., {Ara{\'u}jo}, H.~M., {Barnes}, E.~J., {Bewick}, A.,
  {Cashmore}, R., {Chepel}, V., {Currie}, A. \& {Davidge}, D. {et al.} 2009,
  \emph{\prd}, \textbf{80}, 052010,
  \href{http://arxiv.org/abs/0812.1150}{arXiv:0812.1150}

\bibitem[{{Lee} \& {Mohapatra}(1995)}]{LeeMohapatra}
{Lee}, D. \& {Mohapatra}, R.~N. 1995, \emph{\prd}, \textbf{51}, 1353,
  \href{http://arxiv.org/abs/arXiv:hep-ph/9406328}{arXiv:hep-ph/9406328}

\bibitem[{{Lim} {et~al.}(2009)}]{Lim09}
{Lim}, G. {et~al.} 2009, in \emph{Proceedings of the 31$^\mathrm{st}$
  International Cosmic Ray Conference, Lodz, Poland},
  \href{http://arxiv.org/abs/0905.2316}{arXiv:0905.2316}

\bibitem[{{Linden} \& {Profumo}(2010)}]{Linden10}
{Linden}, T. \& {Profumo}, S. 2010, \emph{\apj}, \textbf{714}, L228,
  \href{http://arxiv.org/abs/1003.0002}{arXiv:1003.0002}

\bibitem[{{Liu} {et~al.}(2009){Liu}, {Yin} \& {Zhu}}]{Liu09}
{Liu}, J., {Yin}, P. \& {Zhu}, S. 2009, \emph{\prd}, \textbf{79}, 063522,
  \href{http://arxiv.org/abs/0812.0964}{arXiv:0812.0964}

\bibitem[{{Loewenstein} \& {Kusenko}(2010)}]{Lowenstein10}
{Loewenstein}, M. \& {Kusenko}, A. 2010, \emph{\apj}, \textbf{714}, 652, 
  \href{http://arxiv.org/abs/arXiv:0912.0552}{arXiv:0912.0552}

\bibitem[{{Lombardi} {et~al.}(2009){Lombardi}, {Aleksic}, {Barrio}, {Biland},
  {Doro}, {Elsaesser}, {Gaug} \& {Mannheim} {et al.}}]{Lombardi09}
{Lombardi}, S., {Aleksic}, J., {Barrio}, J.~A., {Biland}, A., {Doro}, M.,
  {Elsaesser}, D., {Gaug}, M. \& {Mannheim}, K. {et al.} 2009, in
  \emph{Proceedings of the 31$^\mathrm{st}$ International Cosmic Ray
  Conference, Lodz, Poland},
  \href{http://arxiv.org/abs/arXiv:0907.0738}{arXiv:0907.0738}

\bibitem[{{Lundberg} \& {Edsj{\"o}}(2004)}]{Lundberg04}
{Lundberg}, J. \& {Edsj{\"o}}, J. 2004, \emph{\prd}, \textbf{69}, 123505,
  \href{http://arxiv.org/abs/arXiv:astro-ph/0401113}{arXiv:astro-ph/0401113}

\bibitem[{{Luty}(2005)}]{LutyTASI}
{Luty}, M.~A. 2005, \emph{{2004 TASI Lectures on Supersymmetry Breaking}},
  \href{http://arxiv.org/abs/arXiv:hep-th/0509029}{arXiv:hep-th/0509029}

\bibitem[{{Mandal} {et~al.}(2010){Mandal}, {Buckley}, {Freese}, {Spolyar} \&
  {Murayama}}]{Mandal10}
{Mandal}, S.~K., {Buckley}, M.~R., {Freese}, K., {Spolyar}, D. \& {Murayama},
  H. 2010, \emph{\prd}, \textbf{81}, 043508,
  \href{http://arxiv.org/abs/0911.5188}{arXiv:0911.5188}

\bibitem[{{Markevitch}(2006)}]{Markevitch06}
{Markevitch}, M. 2006, in \emph{ESA Special Publication}, Vol. 604,
  \emph{Proceedings of The X-ray Universe 2005}, ed. {A.~Wilson}, 723,
  \href{http://arxiv.org/abs/arXiv:astro-ph/0511345}{arXiv:astro-ph/0511345}

\bibitem[{Martin(1997)}]{Martin}
Martin, S.~P. 1997, in \emph{Perspectives on supersymmetry}, ed. G.~L. Kane,
  1--98, \href{http://arxiv.org/abs/hep-ph/9709356}{arXiv:hep-ph/9709356}

\bibitem[{{Martinez} {et~al.}(2009){Martinez}, {Bullock}, {Kaplinghat},
  {Strigari} \& {Trotta}}]{Martinez09}
{Martinez}, G.~D., {Bullock}, J.~S., {Kaplinghat}, M., {Strigari}, L.~E. \&
  {Trotta}, R. 2009, \emph{\jcap}, \textbf{6}, 14,
  \href{http://arxiv.org/abs/arXiv:0902.4715}{arXiv:0902.4715}

\bibitem[{{Massey} {et~al.}(2007){Massey}, {Rhodes}, {Ellis}, {Scoville},
  {Leauthaud}, {Finoguenov}, {Capak} \& {Bacon} {et al.}}]{Massey07}
{Massey}, R., {Rhodes}, J., {Ellis}, R., {Scoville}, N., {Leauthaud}, A.,
  {Finoguenov}, A., {Capak}, P. \& {Bacon}, D. {et al.} 2007, \emph{\nat},
  \textbf{445}, 286,
  \href{http://arxiv.org/abs/arXiv:astro-ph/0701594}{arXiv:astro-ph/0701594}

\bibitem[{{Maurin} {et~al.}(2001){Maurin}, {Donato}, {Taillet} \&
  {Salati}}]{Maurin01}
{Maurin}, D., {Donato}, F., {Taillet}, R. \& {Salati}, P. 2001, \emph{\apj},
  \textbf{555}, 585,
  \href{http://arxiv.org/abs/arXiv:astro-ph/0101231}{arXiv:astro-ph/0101231}

\bibitem[{{McCullough} \& {Fairbairn}(2010)}]{McCullough10}
{McCullough}, M. \& {Fairbairn}, M. 2010, \emph{\prd},
  \textbf{81}, 083520,
  \href{http://arxiv.org/abs/arXiv:1001.2737}{arXiv:1001.2737}

\bibitem[{{Merritt}(2010)}]{Merritt10}
{Merritt}, D. 2010, in \emph{Particle Dark Matter: Observations, Models and
  Searches}, ed. G.~Bertone (Cambridge University Press), 83--98,
  \href{http://arxiv.org/abs/arXiv:1001.3706}{arXiv:1001.3706}

\bibitem[{{Milgrom}(1983)}]{Milgrom83}
{Milgrom}, M. 1983, \emph{\apj}, \textbf{270}, 365

\bibitem[{{Morrissey} {et~al.}(2005){Morrissey}, {Tait} \&
  {Wagner}}]{Morrissey05}
{Morrissey}, D.~E., {Tait}, T.~M.~P. \& {Wagner}, C.~E.~M. 2005, \emph{\prd},
  \textbf{72}, 095003,
  \href{http://arxiv.org/abs/hep-ph/0508123}{arXiv:hep-ph/0508123}

\bibitem[{{Moskalenko} \& {Strong}(1998)}]{Moskalenko98}
{Moskalenko}, I.~V. \& {Strong}, A.~W. 1998, \emph{\apj}, \textbf{493}, 694,
  \href{http://arxiv.org/abs/arXiv:astro-ph/9710124}{arXiv:astro-ph/9710124}

\bibitem[{{Moskalenko} \& {Wai}(2007)}]{Moskalenko07}
{Moskalenko}, I.~V. \& {Wai}, L.~L. 2007, \emph{\apjl}, \textbf{659}, L29,
  \href{http://arxiv.org/abs/arXiv:astro-ph/0702654}{arXiv:astro-ph/0702654}

\bibitem[{{Natarajan} \& {Schwarz}(2008)}]{Natarajan08a}
{Natarajan}, A. \& {Schwarz}, D.~J. 2008, \emph{\prd}, \textbf{78}, 103524,
  \href{http://arxiv.org/abs/arXiv:0805.3945}{arXiv:0805.3945}

\bibitem[{{Natarajan} {et~al.}(2009){Natarajan}, {Tan} \&
  {O'Shea}}]{Natarajan08b}
{Natarajan}, A., {Tan}, J.~C. \& {O'Shea}, B.~W. 2009, \emph{\apj},
  \textbf{692}, 574, \href{http://arxiv.org/abs/0807.3769}{arXiv:0807.3769}

\bibitem[{{Navarro} {et~al.}(1996){Navarro}, {Frenk} \& {White}}]{NFW}
{Navarro}, J.~F., {Frenk}, C.~S. \& {White}, S.~D.~M. 1996, \emph{\apj},
  \textbf{462}, 563,
  \href{http://arxiv.org/abs/arXiv:astro-ph/9508025}{arXiv:astro-ph/9508025}

\bibitem[{{Navarro} {et~al.}(2004){Navarro}, {Hayashi}, {Power}, {Jenkins},
  {Frenk}, {White}, {Springel} \& {Stadel} {et al.}}]{NFWsmooth}
{Navarro}, J.~F., {Hayashi}, E., {Power}, C., {Jenkins}, A.~R., {Frenk}, C.~S.,
  {White}, S.~D.~M., {Springel}, V. \& {Stadel}, J. {et al.} 2004,
  \emph{\mnras}, \textbf{349}, 1039,
  \href{http://arxiv.org/abs/arXiv:astro-ph/0311231}{arXiv:astro-ph/0311231}

\bibitem[{{Nomura} \& {Thaler}(2009)}]{NTDM}
{Nomura}, Y. \& {Thaler}, J. 2009, \emph{\prd}, \textbf{79}, 075008,
  \href{http://arxiv.org/abs/0810.5397}{arXiv:0810.5397}

\bibitem[{{Pacaud} {et~al.}(2007){Pacaud}, {Pierre}, {Adami}, {Altieri},
  {Andreon}, {Chiappetti}, {Detal} \& {Duc} {et al.}}]{XMM}
{Pacaud}, F., {Pierre}, M., {Adami}, C., {Altieri}, B., {Andreon}, S.,
  {Chiappetti}, L., {Detal}, A. \& {Duc}, P. {et al.} 2007, \emph{\mnras},
  \textbf{382}, 1289, \href{http://arxiv.org/abs/0709.1950}{arXiv:0709.1950}

\bibitem[{{Paczynski}(1986)}]{Paczynski86}
{Paczynski}, B. 1986, \emph{\apj}, \textbf{304}, 1

\bibitem[{{Peccei}(2008)}]{Peccei08}
{Peccei}, R.~D. 2008, in \emph{\lnp}, Vol. 741, \emph{Axions: Theory,
  Cosmology, and Experimental Searches}, ed. M.~{Kuster}, G.~{Raffelt} \&
  B.~Beltr{\'a}n (Springer-Verlag, Berlin), 3

\bibitem[{{Peccei} \& {Quinn}(1977)}]{PQ}
{Peccei}, R.~D. \& {Quinn}, H.~R. 1977, \emph{\prl}, \textbf{38}, 1440

\bibitem[{{Percival} {et~al.}(2010){Percival}, {Reid}, {Eisenstein}, {Bahcall},
  {Budavari}, {Frieman}, {Fukugita} \& {Gunn} {et al.}}]{Percival10}
{Percival}, W.~J., {Reid}, B.~A., {Eisenstein}, D.~J., {Bahcall}, N.~A.,
  {Budavari}, T., {Frieman}, J.~A., {Fukugita}, M. \& {Gunn}, J.~E. {et al.}
  2010, \emph{\mnras}, \textbf{401}, 2148,
  \href{http://arxiv.org/abs/0907.1660}{arXiv:0907.1660}

\bibitem[{{Perlmutter} {et~al.}(1999){Perlmutter}, {Aldering}, {Goldhaber},
  {Knop}, {Nugent}, {Castro}, {Deustua} \& {Fabbro} {et al.}}]{Perlmutter99}
{Perlmutter}, S., {Aldering}, G., {Goldhaber}, G., {Knop}, R.~A., {Nugent}, P.,
  {Castro}, P.~G., {Deustua}, S. \& {Fabbro}, S. {et al.} 1999, \emph{\apj},
  \textbf{517}, 565,
  \href{http://arxiv.org/abs/arXiv:astro-ph/9812133}{arXiv:astro-ph/9812133}

\bibitem[{{Peskin} \& {Schroeder}(1995)}]{Peskin}
{Peskin}, M.~E. \& {Schroeder}, D.~V. 1995, \emph{{An Introduction to Quantum
  Field Theory}} (Westview Press)

\bibitem[{{Peter}(2009)}]{Peter09}
{Peter}, A.~H.~G. 2009, \emph{\prd}, \textbf{79}, 103532,
  \href{http://arxiv.org/abs/0902.1347}{arXiv:0902.1347}

\bibitem[{{Pieri} {et~al.}(2009){Pieri}, {Pizzella}, {Corsini}, {Dalla
  Bont{\`a}} \& {Bertola}}]{Pieri09}
{Pieri}, L., {Pizzella}, A., {Corsini}, E.~M., {Dalla Bont{\`a}}, E. \&
  {Bertola}, F. 2009, \emph{\aap}, \textbf{496}, 351,
  \href{http://arxiv.org/abs/0812.1494}{arXiv:0812.1494}

\bibitem[{{Piran} {et~al.}(2009){Piran}, {Shaviv} \& {Nakar}}]{Piran09}
{Piran}, T., {Shaviv}, N.~J. \& {Nakar}, E. 2009, in \emph{Proceedings of
  Rencontres de Moriond 2009 -- Very High Energy Phenomena in the Universe},
  \href{http://arxiv.org/abs/arXiv:0905.0904}{arXiv:0905.0904}

\bibitem[{{Press} \& {Spergel}(1985)}]{Press85}
{Press}, W.~H. \& {Spergel}, D.~N. 1985, \emph{\apj}, \textbf{296}, 679

\bibitem[{{Profumo}(2008)}]{Profumo08}
{Profumo}, S. 2008,
  \href{http://arxiv.org/abs/arXiv:0812.4457}{arXiv:0812.4457}

\bibitem[{{Profumo} \& {Jeltema}(2009)}]{Profumo09}
{Profumo}, S. \& {Jeltema}, T.~E. 2009, \emph{\jcap}, \textbf{7}, 20,
  \href{http://arxiv.org/abs/0906.0001}{arXiv:0906.0001}

\bibitem[{{Pugnat} {et~al.}(2008){Pugnat}, {Duvillaret}, {Jost}, {Vitrant},
  {Romanini}, {Siemko}, {Ballou} \& {Barbara} {et al.}}]{Pugnat08}
{Pugnat}, P., {Duvillaret}, L., {Jost}, R., {Vitrant}, G., {Romanini}, D.,
  {Siemko}, A., {Ballou}, R. \& {Barbara}, B. {et al.} 2008, \emph{\prd},
  \textbf{78}, 092003, \href{http://arxiv.org/abs/0712.3362}{arXiv:0712.3362}

\bibitem[{{Putze} {et~al.}(2010){Putze}, {Derome} \& {Maurin}}]{Putze10}
{Putze}, A., {Derome}, L. \& {Maurin}, D. 2010, \emph{\aap}, \textbf{516}, A66,
  \href{http://arxiv.org/abs/1001.0551}{arXiv:1001.0551}

\bibitem[{{Putze} {et~al.}(2009){Putze}, {Derome}, {Maurin}, {Perotto} \&
  {Taillet}}]{Putze09}
{Putze}, A., {Derome}, L., {Maurin}, D., {Perotto}, L. \& {Taillet}, R. 2009,
  \emph{\aap}, \textbf{497}, 991,
  \href{http://arxiv.org/abs/0808.2437}{arXiv:0808.2437}

\bibitem[{{Raffelt}(2008)}]{Raffelt08}
{Raffelt}, G.~G. 2008, in \emph{\lnp}, Vol. 741, \emph{Axions: Theory,
  Cosmology, and Experimental Searches}, ed. M.~{Kuster}, G.~{Raffelt} \&
  B.~Beltr{\'a}n (Springer-Verlag, Berlin), 51

\bibitem[{{Randall} \& {Sundrum}(1999)}]{Randall99}
{Randall}, L. \& {Sundrum}, R. 1999, \emph{\nphysb}, \textbf{557}, 79,
  \href{http://arxiv.org/abs/arXiv:hep-th/9810155}{arXiv:hep-th/9810155}

\bibitem[{{Read} {et~al.}(2009){Read}, {Mayer}, {Brooks}, {Governato} \&
  {Lake}}]{Read09}
{Read}, J.~I., {Mayer}, L., {Brooks}, A.~M., {Governato}, F. \& {Lake}, G.
  2009, \emph{\mnras}, \textbf{397}, 44,
  \href{http://arxiv.org/abs/0902.0009}{arXiv:0902.0009}

\bibitem[{{Regis} \& {Ullio}(2008)}]{Regis08}
{Regis}, M. \& {Ullio}, P. 2008, \emph{\prd}, \textbf{78}, 043505,
  \href{http://arxiv.org/abs/0802.0234}{arXiv:0802.0234}

\bibitem[{{Ricotti} \& {Gould}(2009)}]{Ricotti09}
{Ricotti}, M. \& {Gould}, A. 2009, \emph{\apj}, \textbf{707}, 979,
  \href{http://arxiv.org/abs/0908.0735}{arXiv:0908.0735}

\bibitem[{{Riess} {et~al.}(1998){Riess}, {Filippenko}, {Challis},
  {Clocchiatti}, {Diercks}, {Garnavich}, {Gilliland} \& {Hogan} {et
  al.}}]{Riess98}
{Riess}, A.~G., {Filippenko}, A.~V., {Challis}, P., {Clocchiatti}, A.,
  {Diercks}, A., {Garnavich}, P.~M., {Gilliland}, R.~L. \& {Hogan}, C.~J. {et
  al.} 1998, \emph{\aj}, \textbf{116}, 1009,
  \href{http://arxiv.org/abs/arXiv:astro-ph/9805201}{arXiv:astro-ph/9805201}

\bibitem[{{Ripamonti} {et~al.}(2007){Ripamonti}, {Mapelli} \&
  {Ferrara}}]{Ripamonti07}
{Ripamonti}, E., {Mapelli}, M. \& {Ferrara}, A. 2007, \emph{\mnras},
  \textbf{374}, 1067,
  \href{http://arxiv.org/abs/arXiv:astro-ph/0606482}{arXiv:astro-ph/0606482}

\bibitem[{{Robilliard} {et~al.}(2007){Robilliard}, {Battesti}, {Fouch{\'e}},
  {Mauchain}, {Sautivet}, {Amiranoff} \& {Rizzo}}]{Robilliard07}
{Robilliard}, C., {Battesti}, R., {Fouch{\'e}}, M., {Mauchain}, J., {Sautivet},
  A., {Amiranoff}, F. \& {Rizzo}, C. 2007, \emph{\prl}, \textbf{99}, 190403,
  \href{http://arxiv.org/abs/0707.1296}{arXiv:0707.1296}

\bibitem[{{Rubin} {et~al.}(1985){Rubin}, {Burstein}, {Ford} \&
  {Thonnard}}]{Rubin85}
{Rubin}, V.~C., {Burstein}, D., {Ford}, Jr., W.~K. \& {Thonnard}, N. 1985,
  \emph{\apj}, \textbf{289}, 81

\bibitem[{{Rubin} \& {Ford}(1970)}]{Rubin70}
{Rubin}, V.~C. \& {Ford}, Jr., W.~K. 1970, \emph{\apj}, \textbf{159}, 379

\bibitem[{{Rubin} {et~al.}(1978){Rubin}, {Thonnard} \& {Ford}}]{Rubin78}
{Rubin}, V.~C., {Thonnard}, N. \& {Ford}, Jr., W.~K. 1978, \emph{\apjl},
  \textbf{225}, L107

\bibitem[{{Ruiz de Austri} {et~al.}(2006){Ruiz de Austri}, {Trotta} \&
  {Roszkowski}}]{Ruiz06}
{Ruiz de Austri}, R., {Trotta}, R. \& {Roszkowski}, L. 2006, \emph{\jhep},
  \textbf{5}, 2,
  \href{http://arxiv.org/abs/arXiv:hep-ph/0602028}{arXiv:hep-ph/0602028}

\bibitem[{{Salati} \& {Silk}(1989)}]{SalatiSilk89}
{Salati}, P. \& {Silk}, J. 1989, \emph{\apj}, \textbf{338}, 24

\bibitem[{{Samtleben} {et~al.}(2007){Samtleben}, {Staggs} \&
  {Winstein}}]{CMBpedestrian}
{Samtleben}, D., {Staggs}, S. \& {Winstein}, B. 2007, \emph{\arnps},
  \textbf{57}, 245, \href{http://arxiv.org/abs/0803.0834}{arXiv:0803.0834}

\bibitem[{{Sandick} {et~al.}(2010){Sandick}, {Spolyar}, {Buckley}, {Freese} \&
  {Hooper}}]{Sandick09}
{Sandick}, P., {Spolyar}, D., {Buckley}, M., {Freese}, K. \& {Hooper}, D. 2010,
  \emph{\prd}, \textbf{81}, 083506, 
  \href{http://arxiv.org/abs/arXiv:0912.0513}{arXiv:0912.0513}

\bibitem[{{Savage} {et~al.}(2009{\natexlab{a}}){Savage}, {Freese}, {Gondolo} \&
  {Spolyar}}]{Savage09a}
{Savage}, C., {Freese}, K., {Gondolo}, P. \& {Spolyar}, D. 2009{\natexlab{a}},
  \emph{\jcap}, \textbf{9}, 36,
  \href{http://arxiv.org/abs/0901.2713}{arXiv:0901.2713}

\bibitem[{{Savage} {et~al.}(2009{\natexlab{b}}){Savage}, {Gelmini}, {Gondolo}
  \& {Freese}}]{Savage09b}
{Savage}, C., {Gelmini}, G., {Gondolo}, P. \& {Freese}, K. 2009{\natexlab{b}},
  \emph{\jcap}, \textbf{4}, 10,
  \href{http://arxiv.org/abs/0808.3607}{arXiv:0808.3607}

\bibitem[{{Schwetz} {et~al.}(2008){Schwetz}, {T{\'o}rtola} \&
  {Valle}}]{Schwetz08}
{Schwetz}, T., {T{\'o}rtola}, M. \& {Valle}, J.~W.~F. 2008, \emph{\njp},
  \textbf{10}, 113011, \href{http://arxiv.org/abs/0808.2016}{arXiv:0808.2016}

\bibitem[{{Scott} {et~al.}(2006){Scott}, {Asplund}, {Grevesse} \&
  {Sauval}}]{ScottVII}
{Scott}, P., {Asplund}, M., {Grevesse}, N. \& {Sauval}, A.~J. 2006,
  \emph{\aap}, \textbf{456}, 675,
  \href{http://arxiv.org/abs/arXiv:astro-ph/0605116}{arXiv:astro-ph/0605116},
  \citepalias{ScottVII}

\bibitem[{{Scott} {et~al.}(2009{\natexlab{a}}){Scott}, {Asplund}, {Grevesse} \&
  {Sauval}}]{Scott09Ni}
---. 2009{\natexlab{a}}, \emph{\apjl}, \textbf{691}, L119,
  \href{http://arxiv.org/abs/0811.0815}{arXiv:0811.0815},
  \citepalias{Scott09Ni}

\bibitem[{{Scott} {et~al.}(2010{\natexlab{a}}){Scott}, {Conrad}, {Edsj{\"o}},
  {Bergstr{\"o}m}, {Farnier} \& {Akrami}}]{Scott09c}
{Scott}, P., {Conrad}, J., {Edsj{\"o}}, J., {Bergstr{\"o}m}, L., {Farnier}, C.
  \& {Akrami}, Y. 2010{\natexlab{a}}, \emph{\jcap}, \textbf{1}, 31,
  \href{http://arxiv.org/abs/0909.3300}{arXiv:0909.3300},
  \citepalias{Scott09c}

\bibitem[{{Scott} {et~al.}(2008){Scott}, {Edsj{\"o}} \& {Fairbairn}}]{Scott08a}
{Scott}, P., {Edsj{\"o}}, J. \& {Fairbairn}, M. 2008, in \emph{Proceedings of Dark Matter in Astroparticle and Particle Physics: Dark 2007}, ed. H.~K.
  {Klapdor-Kleingrothaus} \& G.~F. {Lewis} (World Scientific, Singapore),
  387--392, \href{http://arxiv.org/abs/arXiv:0711.0991}{arXiv:0711.0991},
  (\citetalias{Scott08a})

\bibitem[{{Scott} {et~al.}(2010{\natexlab{b}}){Scott}, {Edsj{\"o}} \&
  {Fairbairn}}]{Scott09b}
{Scott}, P., {Edsj{\"o}}, J. \& {Fairbairn}, M. 2010{\natexlab{b}}, in
  \emph{{Proceedings of Dark Matter in Astroparticle and Particle Physics: Dark 2009}}, ed.
  {H.~V.~Klapdor-Kleingrothaus \& I.~V.~Krivosheina} (World Scientific,
  Singapore), 320--327,
  \href{http://arxiv.org/abs/arXiv:0904.2395}{arXiv:0904.2395},
  \citepalias{Scott09b}

\bibitem[{{Scott} {et~al.}(2009{\natexlab{b}}){Scott}, {Fairbairn} \&
  {Edsj{\"o}}}]{Scott09}
{Scott}, P., {Fairbairn}, M. \& {Edsj{\"o}}, J. 2009{\natexlab{b}},
  \emph{\mnras}, \textbf{394}, 82,
  \href{http://arxiv.org/abs/arXiv:0809.1871}{arXiv:0809.1871},
  \citepalias{Scott09}

\bibitem[{{Scott} \& {Sivertsson}(2009)}]{SS09}
{Scott}, P. \& {Sivertsson}, S. 2009, \emph{\prl}, \textbf{103}, 211301,
  \href{http://arxiv.org/abs/0908.4082}{arXiv:0908.4082},
  \citepalias{SS09}

\bibitem[{{Seckel} {et~al.}(1991){Seckel}, {Stanev} \& {Gaisser}}]{Seckel91}
{Seckel}, D., {Stanev}, T. \& {Gaisser}, T.~K. 1991, \emph{\apj}, \textbf{382},
  652

\bibitem[{{Serpico} \& {Zaharijas}(2008)}]{Serpico08}
{Serpico}, P.~D. \& {Zaharijas}, G. 2008, \emph{\app}, \textbf{29}, 380,
  \href{http://arxiv.org/abs/0802.3245}{arXiv:0802.3245}

\bibitem[{{Servant} \& {Tait}(2003)}]{ServantTait}
{Servant}, G. \& {Tait}, T.~M.~P. 2003, \emph{\nphysb}, \textbf{650}, 391,
  \href{http://arxiv.org/abs/arXiv:hep-ph/0206071}{arXiv:hep-ph/0206071}

\bibitem[{{Sikivie}(2008)}]{Sikivie08}
{Sikivie}, P. 2008, in \emph{\lnp}, Vol. 741, \emph{Axions: Theory, Cosmology,
  and Experimental Searches}, ed. M.~{Kuster}, G.~{Raffelt} \& B.~Beltr{\'a}n
  (Springer-Verlag, Berlin), 19

\bibitem[{{Silk}(1968)}]{Silk68}
{Silk}, J. 1968, \emph{\apj}, \textbf{151}, 459

\bibitem[{{Slatyer} {et~al.}(2009){Slatyer}, {Padmanabhan} \&
  {Finkbeiner}}]{Slatyer09}
{Slatyer}, T.~R., {Padmanabhan}, N. \& {Finkbeiner}, D.~P. 2009, \emph{\prd},
  \textbf{80}, 043526, \href{http://arxiv.org/abs/0906.1197}{arXiv:0906.1197}

\bibitem[{{Smith} \& {Weiner}(2001)}]{SmithWeiner01}
{Smith}, D. \& {Weiner}, N. 2001, \emph{\prd}, \textbf{64}, 043502,
  \href{http://arxiv.org/abs/arXiv:hep-ph/0101138}{arXiv:hep-ph/0101138}

\bibitem[{{Spite} \& {Spite}(1982)}]{SpiteSpite}
{Spite}, F. \& {Spite}, M. 1982, \emph{\aap}, \textbf{115}, 357

\bibitem[{{Spolyar} {et~al.}(2009){Spolyar}, {Bodenheimer}, {Freese} \&
  {Gondolo}}]{Spolyar09}
{Spolyar}, D., {Bodenheimer}, P., {Freese}, K. \& {Gondolo}, P. 2009,
  \emph{\apj}, \textbf{705}, 1031,
  \href{http://arxiv.org/abs/0903.3070}{arXiv:0903.3070}

\bibitem[{{Spolyar} {et~al.}(2008){Spolyar}, {Freese} \& {Gondolo}}]{Spolyar08}
{Spolyar}, D., {Freese}, K. \& {Gondolo}, P. 2008, \emph{\prl}, \textbf{100},
  051101, \href{http://arxiv.org/abs/arXiv:0705.0521}{arXiv:0705.0521}

\bibitem[{{Springel} {et~al.}(2008{\natexlab{a}}){Springel}, {Wang},
  {Vogelsberger}, {Ludlow}, {Jenkins}, {Helmi}, {Navarro} \& {Frenk} {et
  al.}}]{Aquarius}
{Springel}, V., {Wang}, J., {Vogelsberger}, M., {Ludlow}, A., {Jenkins}, A.,
  {Helmi}, A., {Navarro}, J.~F. \& {Frenk}, C.~S. {et al.} 2008{\natexlab{a}},
  \emph{\mnras}, \textbf{391}, 1685,
  \href{http://arxiv.org/abs/0809.0898}{arXiv:0809.0898}

\bibitem[{{Springel} {et~al.}(2008{\natexlab{b}}){Springel}, {White}, {Frenk},
  {Navarro}, {Jenkins}, {Vogelsberger}, {Wang} \& {Ludlow} {et
  al.}}]{Springel08}
{Springel}, V., {White}, S.~D.~M., {Frenk}, C.~S., {Navarro}, J.~F., {Jenkins},
  A., {Vogelsberger}, M., {Wang}, J. \& {Ludlow}, A. {et al.}
  2008{\natexlab{b}}, \emph{\nat}, \textbf{456}, 73,
  \href{http://arxiv.org/abs/0809.0894}{arXiv:0809.0894}

\bibitem[{{Springel} {et~al.}(2005){Springel}, {White}, {Jenkins}, {Frenk},
  {Yoshida}, {Gao}, {Navarro} \& {Thacker} {et al.}}]{Millenium}
{Springel}, V., {White}, S.~D.~M., {Jenkins}, A., {Frenk}, C.~S., {Yoshida},
  N., {Gao}, L., {Navarro}, J. \& {Thacker}, R. {et al.} 2005, \emph{\nat},
  \textbf{435}, 629,
  \href{http://arxiv.org/abs/arXiv:astro-ph/0504097}{arXiv:astro-ph/0504097}

\bibitem[{{Starkman} {et~al.}(1990){Starkman}, {Gould}, {Esmailzadeh} \&
  {Dimopoulos}}]{Starkman90}
{Starkman}, G.~D., {Gould}, A., {Esmailzadeh}, R. \& {Dimopoulos}, S. 1990,
  \emph{\prd}, \textbf{41}, 3594

\bibitem[{{Steigman}(2007)}]{Steigman07}
{Steigman}, G. 2007, \emph{\arnps}, \textbf{57}, 463,
  \href{http://arxiv.org/abs/0712.1100}{arXiv:0712.1100}

\bibitem[{{Steigman} {et~al.}(1978){Steigman}, {Quintana}, {Sarazin} \&
  {Faulkner}}]{Steigman78}
{Steigman}, G., {Quintana}, H., {Sarazin}, C.~L. \& {Faulkner}, J. 1978,
  \emph{\aj}, \textbf{83}, 1050

\bibitem[{{Stiff} \& {Widrow}(2003)}]{Stiff03}
{Stiff}, D. \& {Widrow}, L.~M. 2003, \emph{\prl}, \textbf{90}, 211301,
  \href{http://arxiv.org/abs/arXiv:astro-ph/0301301}{arXiv:astro-ph/0301301}

\bibitem[{Stix(2002)}]{Stix}
Stix, M. 2002, \emph{The Sun: An Introduction}, $2^{\mathrm{nd}}$ edn.
  (Springer-Verlag, Berlin)

\bibitem[{{Stoehr} {et~al.}(2003){Stoehr}, {White}, {Springel}, {Tormen} \&
  {Yoshida}}]{Stoehr03}
{Stoehr}, F., {White}, S.~D.~M., {Springel}, V., {Tormen}, G. \& {Yoshida}, N.
  2003, \emph{\mnras}, \textbf{345}, 1313,
  \href{http://arxiv.org/abs/arXiv:astro-ph/0307026}{arXiv:astro-ph/0307026}

\bibitem[{{Strong} \& {Moskalenko}(1998)}]{Strong98}
{Strong}, A.~W. \& {Moskalenko}, I.~V. 1998, \emph{\apj}, \textbf{509}, 212,
  \href{http://arxiv.org/abs/arXiv:astro-ph/9807150}{arXiv:astro-ph/9807150}

\bibitem[{{Strong} {et~al.}(2007){Strong}, {Moskalenko} \&
  {Ptuskin}}]{Strong07}
{Strong}, A.~W., {Moskalenko}, I.~V. \& {Ptuskin}, V.~S. 2007, \emph{Annual
  Review of Nuclear and Particle Science}, \textbf{57}, 285,
  \href{http://arxiv.org/abs/arXiv:astro-ph/0701517}{arXiv:astro-ph/0701517}

\bibitem[{{Strong} {et~al.}(2000){Strong}, {Moskalenko} \& {Reimer}}]{Strong00}
{Strong}, A.~W., {Moskalenko}, I.~V. \& {Reimer}, O. 2000, \emph{\apj},
  \textbf{537}, 763,
  \href{http://arxiv.org/abs/arXiv:astro-ph/9811296}{arXiv:astro-ph/9811296}

\bibitem[{{'t Hooft}(1976{\natexlab{a}})}]{t'Hooft76a}
{'t Hooft}, G. 1976{\natexlab{a}}, \emph{\prd}, \textbf{14}, 3432

\bibitem[{{'t Hooft}(1976{\natexlab{b}})}]{t'Hooft76b}
---. 1976{\natexlab{b}}, \emph{\prd}, \textbf{14}, 3432

\bibitem[{{Taoso} {et~al.}(2008){Taoso}, {Bertone} \& {Masiero}}]{Taoso08_10pt}
{Taoso}, M., {Bertone}, G. \& {Masiero}, A. 2008, \emph{\jcap}, \textbf{3}, 22,
  \href{http://arxiv.org/abs/0711.4996}{arXiv:0711.4996}

\bibitem[{{Tegmark} {et~al.}(2004){Tegmark}, {Blanton}, {Strauss}, {Hoyle},
  {Schlegel}, {Scoccimarro}, {Vogeley} \& {Weinberg} {et al.}}]{Tegmark04}
{Tegmark}, M., {Blanton}, M.~R., {Strauss}, M.~A., {Hoyle}, F., {Schlegel}, D.,
  {Scoccimarro}, R., {Vogeley}, M.~S. \& {Weinberg}, D.~H. {et al.} 2004,
  \emph{\apj}, \textbf{606}, 702,
  \href{http://arxiv.org/abs/arXiv:astro-ph/0310725}{arXiv:astro-ph/0310725}

\bibitem[{{Tegmark} {et~al.}(2001){Tegmark}, {Zaldarriaga} \&
  {Hamilton}}]{Tegmark01}
{Tegmark}, M., {Zaldarriaga}, M. \& {Hamilton}, A.~J. 2001, \emph{\prd},
  \textbf{63}, 043007,
  \href{http://arxiv.org/abs/arXiv:astro-ph/0008167}{arXiv:astro-ph/0008167}

\bibitem[{{Tisserand} {et~al.}(2007){Tisserand}, {Le Guillou}, {Afonso},
  {Albert}, {Andersen}, {Ansari}, {Aubourg} \& {Bareyre} {et
  al.}}]{Tisserand06}
{Tisserand}, P., {Le Guillou}, L., {Afonso}, C., {Albert}, J.~N., {Andersen},
  J., {Ansari}, R., {Aubourg}, {\'E}. \& {Bareyre}, P. {et al.} 2007,
  \emph{\aap}, \textbf{469}, 387,
  \href{http://arxiv.org/abs/arXiv:astro-ph/0607207}{arXiv:astro-ph/0607207}

\bibitem[{{Tremaine} \& {Gunn}(1979)}]{TremaineGunn}
{Tremaine}, S. \& {Gunn}, J.~E. 1979, \emph{\prl}, \textbf{42}, 407

\bibitem[{{Trotta} {et~al.}(2008){Trotta}, {Feroz}, {Hobson}, {Roszkowski} \&
  {Ruiz de Austri}}]{Trotta08}
{Trotta}, R., {Feroz}, F., {Hobson}, M., {Roszkowski}, L. \& {Ruiz de Austri},
  R. 2008, \emph{\jhep}, \textbf{12}, 24,
  \href{http://arxiv.org/abs/arXiv:0809.3792}{arXiv:0809.3792}

\bibitem[{{Tucker-Smith} \& {Weiner}(2005)}]{TuckerSmithWeiner05}
{Tucker-Smith}, D. \& {Weiner}, N. 2005, \emph{\prd}, \textbf{72}, 063509,
  \href{http://arxiv.org/abs/arXiv:hep-ph/0402065}{arXiv:hep-ph/0402065}

\bibitem[{{Tully} \& {Fisher}(1977)}]{TullyFisher}
{Tully}, R.~B. \& {Fisher}, J.~R. 1977, \emph{\aap}, \textbf{54}, 661

\bibitem[{{Tyson} {et~al.}(1998){Tyson}, {Kochanski} \&
  {Dell'Antonio}}]{Tyson98}
{Tyson}, J.~A., {Kochanski}, G.~P. \& {Dell'Antonio}, I.~P. 1998, \emph{\apjl},
  \textbf{498}, L107,
  \href{http://arxiv.org/abs/arXiv:astro-ph/9801193}{arXiv:astro-ph/9801193}

\bibitem[{{Ullio} \& {Bergstr{\"o}m}(1998)}]{Ullio98}
{Ullio}, P. \& {Bergstr{\"o}m}, L. 1998, \emph{\prd}, \textbf{57}, 1962,
  \href{http://arxiv.org/abs/arXiv:hep-ph/9707333}{arXiv:hep-ph/9707333}

\bibitem[{{Vikhlinin} {et~al.}(2006){Vikhlinin}, {Kravtsov}, {Forman}, {Jones},
  {Markevitch}, {Murray} \& {Van Speybroeck}}]{Chandra}
{Vikhlinin}, A., {Kravtsov}, A., {Forman}, W., {Jones}, C., {Markevitch}, M.,
  {Murray}, S.~S. \& {Van Speybroeck}, L. 2006, \emph{\apj}, \textbf{640}, 691,
  \href{http://arxiv.org/abs/arXiv:astro-ph/0507092}{arXiv:astro-ph/0507092}

\bibitem[{{Visinelli} \& {Gondolo}(2009)}]{Visinelli09}
{Visinelli}, L. \& {Gondolo}, P. 2009, \emph{\prd}, \textbf{80}, 035024,
  \href{http://arxiv.org/abs/0903.4377}{arXiv:0903.4377}

\bibitem[{{Vitale} {et~al.}(2009){Vitale}, {Morselli} \& {The \emph{Fermi}-LAT
  Collaboration}}]{Vitale09}
{Vitale}, V., {Morselli}, A. \& {The \emph{Fermi}-LAT Collaboration}. 2009, in
  \emph{Proceedings of the 2009 Fermi Symposium, eConf Proceedings}, Vol.
  C091122, \href{http://arxiv.org/abs/arXiv:0912.3828}{arXiv:0912.3828}

\bibitem[{{White}(2007)}]{White07}
{White}, M.~J. 2007, \emph{Proceedings of Rencontres de Moriond 2006 --
  Electroweak Interactions And Unified Theories, \ijmpa}, \textbf{22}, 5771,
  \href{http://arxiv.org/abs/arXiv:hep-ph/0605065}{arXiv:hep-ph/0605065}

\bibitem[{{Wikstr{\"o}m} \& {Edsj{\"o}}(2009)}]{Wikstrom09}
{Wikstr{\"o}m}, G. \& {Edsj{\"o}}, J. 2009, \emph{\jcap}, \textbf{4}, 9,
  \href{http://arxiv.org/abs/0903.2986}{arXiv:0903.2986}

\bibitem[{{Wyrzykowski} {et~al.}(2009){Wyrzykowski}, {Koz{\l}owski}, {Skowron},
  {Belokurov}, {Smith}, {Udalski}, {Szyma{\'n}ski} \& {Kubiak} {et
  al.}}]{OGLE09}
{Wyrzykowski}, {\L}., {Koz{\l}owski}, S., {Skowron}, J., {Belokurov}, V.,
  {Smith}, M.~C., {Udalski}, A., {Szyma{\'n}ski}, M.~K. \& {Kubiak}, M. {et
  al.} 2009, \emph{\mnras}, \textbf{397}, 1228,
  \href{http://arxiv.org/abs/arXiv:0905.2044}{arXiv:0905.2044}

\bibitem[{{Y{\"u}ksel} {et~al.}(2007){Y{\"u}ksel}, {Horiuchi}, {Beacom} \&
  {Ando}}]{Yuksel07}
{Y{\"u}ksel}, H., {Horiuchi}, S., {Beacom}, J.~F. \& {Ando}, S. 2007,
  \emph{\prd}, \textbf{76}, 123506,
  \href{http://arxiv.org/abs/0707.0196}{arXiv:0707.0196}

\bibitem[{{Y{\"u}ksel} {et~al.}(2009){Y{\"u}ksel}, {Kistler} \&
  {Stanev}}]{Yuksel09}
{Y{\"u}ksel}, H., {Kistler}, M.~D. \& {Stanev}, T. 2009, \emph{\prl},
  \textbf{103}, 051101, \href{http://arxiv.org/abs/0810.2784}{arXiv:0810.2784}

\bibitem[{{Zackrisson} {et~al.}(2010){Zackrisson}, {Scott}, {Rydberg}, {Iocco},
  {Edvardsson}, {{\"O}stlin}, {Sivertsson} \& {Zitrin} {et
  al.}}]{Zackrisson10}
{Zackrisson}, E., {Scott}, P., {Rydberg}, C.-E., {Iocco}, F., {Edvardsson}, B.,
  {{\"O}stlin}, G., {Sivertsson}, S. \& {Zitrin}, A. {et al.} 2010,
  \emph{\apj}, \textbf{717}, 257,
  \href{http://arxiv.org/abs/arXiv:1002.3368}{arXiv:1002.3368}, \citepalias{Zackrisson10}

\bibitem[{{Zwicky}(1933)}]{Zwicky33}
{Zwicky}, F. 1933, \emph{Helvetica Physica Acta}, \textbf{6}, 110

\small
\end{thebibliography}

\part{Papers}
\label{papers}

\pagestyle{plain}

\paper{Malcolm Fairbairn, Pat Scott \& Joakim Edsj\"o}{The zero age main sequence of WIMP burners}{\prd}{77}{047301}{2008}{arXiv:0710.3396}{}{papI}
\paper{Pat Scott, Malcolm Fairbairn \& Joakim Edsj\"o}{Dark stars at the Galactic Centre -- the main sequence}{\mnras}{394}{82--104}{2009}{arXiv:0809.1871}{}{papII}
\paper{Pat Scott \& Sofia Sivertsson}{Gamma-rays from ultracompact primordial dark matter minihalos}{\prl}{103}{211301}{2009}{arXiv:0908.4082}{}{papIII}
\paper{Pat Scott, Jan Conrad, Joakim Edsj\"o, Lars Bergstr\"om, Christian Farnier \& Yashar Akrami}{Direct constraints on minimal supersymmetry from Fermi-LAT observations of the dwarf galaxy Segue 1}{\jcap}{01}{031}{2010}{arXiv:0909.3300}{}{papIV}
\paper{Yashar Akrami, Pat Scott, Joakim Edsj\"o, Jan Conrad \& Lars Bergstr\"om}{A profile likelihood analysis of the Constrained MSSM with genetic algorithms}{\jhep}{04}{057}{2010}{arXiv:0910.3950}{}{papV}
\paper{Erik Zackrisson, Pat Scott, Claes-Erik Rydberg, Fabio Iocco, Bengt Edvardsson, G\"oran \"Ostlin, Sofia Sivertsson, Adi Zitrin, Tom Broadhurst \& Paolo Gondolo}{Finding high-redshift dark stars with the James Webb Space Telescope}{\apj}{717}{257}{2010}{arXiv:1002.3368}{}{papVI}

\end{document}